\documentclass[aps,prd,amsmath,amssymb,eqsecnum,nofootinbib,notitlepage,superscriptaddress]{revtex4}

\usepackage[retainorgcmds]{IEEEtrantools}
\usepackage{graphicx}
\usepackage{graphics}
\usepackage{color}
\usepackage[normalem]{ulem}

\usepackage{ulem}
\graphicspath{{../Plots/}}

\newtheorem{proposition}{Proposition}

\newtheorem{theorem}{Theorem}

\newcommand{\Glret}{G_{\ell}}
\newcommand{\Gnd}{G_{\text{nd}}}
\newcommand{\Gd}{G_{\text{d}}}

\newcommand{\Mtwo}{\mathcal{M}_2}
\newcommand{\st}{\mathcal{M}}
\newcommand{\sth}{\hat{\mathcal{M}}}
\newcommand{\Gret}{G_R}
\newcommand{\GFret}{G_{R}}

\newcommand{\indmode}{\ell}
\newcommand{\Gld}{G^\text{d}_{\indmode}}

\newcommand{\sTCS}{\sigma}
\newcommand{\PV}{\text{PV}}

\newcommand{\xp}[1]{x_{+,#1}}
\newcommand{\xm}[1]{x_{-,#1}}
\newcommand{\xpm}[1]{x_{\pm,#1}}
\newcommand{\xpt}[1]{\tilde{x}_{+,#1}}
\newcommand{\xmt}[1]{\tilde{x}_{-,#1}}
\newcommand{\xpmt}[1]{\tilde{x}_{\pm,#1}}
\newcommand{\Vnp}[1]{\mathcal{V}_{+,#1}}
\newcommand{\Vnm}[1]{\mathcal{V}_{-,#1}}
\newcommand{\Vnpm}[1]{\mathcal{V}_{\pm,#1}}
\newcommand{\Vnpb}[1]{\mathcal{\tilde{V}}_{+,#1}}
\newcommand{\Vnmb}[1]{\mathcal{\tilde{V}}_{-,#1}}
\newcommand{\Vnpmb}[1]{\mathcal{\tilde{V}}_{\pm,#1}}
\newcommand{\Wn}[1]{\mathcal{W}_{#1}}
\newcommand{\VHads}{\mathcal{V}}

\newcommand{\dt}{\Delta t}
\newcommand{\be}{\begin{equation}}
\newcommand{\ee}{\end{equation}}
\newcommand{\Hv}{\theta}
\newcommand{\rs}{r_*}
\newcommand{\lam}{L}
\newcommand{\pdd}[2]{\frac{\partial^2{#1}}{\partial{#2}^2}}

\newcommand{\pd}[2]{\frac{\partial{#1}}{\partial{#2}}}
\newcommand{\cgl}{{\cal{U}}_\ell}
\newcommand{\Gl}{G_\ell}
\newcommand{\cgo}{{\cal{U}}_0}
\newcommand{\rlam}{\lam}
\newcommand{\grl}{G_R^{\ell\geq1}}
\newcommand{\grlns}{G_{R,NS}^{\ell\geq1}}
\newcommand{\ca}{{\cal{A}}}
\newcommand{\cb}{{\cal{B}}}
\newcommand{\cc}{{\cal{C}}}
\newcommand{\cs}{{\cal{S}}}

\newcommand{\cw}{{\cal{Z}}}

\newcommand{\zo}{\mathbb{Z}_0}

\newcommand{\gam}{\gamma}
\newcommand{\etane}{\eta_{{\rm{e}},n}}
\newcommand{\etano}{\eta_{{\rm{o}},n}}
\newcommand{\rma}{\mathrm{a}}

\newcommand{\U}{U_{4d}}
\newcommand{\V}{V_{4d}}
\newcommand{\s}{\sigma_{4d}}

\begin{document}
\title{
Global Hadamard form for  the Green function in Schwarzschild spacetime
}

\author{Marc Casals}
\email{marc.casals@uni-leipzig.de,mcasals@cbpf.br,marc.casals@ucd.ie}
\affiliation{Institut f\"ur Theoretische Physik, Universit\"at Leipzig,  Br\"uderstra\ss e  16, 04103 Leipzig, Germany}
\affiliation{Centro Brasileiro de Pesquisas F\'isicas (CBPF), Rio de Janeiro, 
CEP 22290-180, 
Brazil.}
\affiliation{School of Mathematics and Statistics, University College Dublin, Belfield, Dublin 4, Ireland.}

\author{Brien C. Nolan}
\email{brien.nolan@dcu.ie}
\affiliation{School of Mathematical Sciences, Dublin City
University, Glasnevin, Dublin 9, Ireland.}

\begin{abstract}
The  retarded Green function of a wave equation on a 4-dimensional curved background spacetime is a (generalized) function of
two spacetime points and diverges when these are connected by a null geodesic.
The Hadamard form
 makes explicit the form of 
 this divergence but only when one of the points is in  a normal neighbourhood of the other point.
In this paper we derive a  representation for the retarded Green function
for a 
scalar field 
in Schwarzschild spacetime which makes explicit
its 
{\it complete} 
singularity structure beyond the normal neighbourhood.
We interpret 
this representation
as a sum of Hadamard forms,
 the summation being taken over
 the number of times the null wavefront has passed through a caustic point: the sum of Hadamard forms applies to the non-smooth contribution  to the full Green function, not only the singular contribution. (The term non-smooth applies modulo the causality-generating step functions that must appear  in the retarded Green function.)  The singularity structure is determined using two independent approaches, one based on a Bessel function expansion of the Green function, and another that exploits a link between the Green functions of Schwarzschild spacetime and  Pleba{\'n}ski-Hacyan spacetime (the latter approach also yields another representation for the {\it full} Schwarzschild Green function, not just for its non-smooth part).
Our  representation 
is  not valid in a neighbourhood of caustic points.
We deal with these points by providing a separate representation for the Green function in Schwarzschild spacetime
which makes explicit its (different) singularity structure at caustics of this spacetime.
\end{abstract}

\maketitle


\section{Introduction}


A fundamental object for the study of linear field perturbations of a curved spacetime is the retarded Green function (GF) of the wave equation
satisfied by the perturbation.
Heuristically, the GF may be viewed as the value of the field at a spacetime point resulting from the propagation of an `impulsive' source at a base point.
The global --not just local-- behaviour of the  GF is useful, for example, for 
determining the evolution of initial data via a Kirchhoff integral~\cite{Leaver:1986}, 
  for determining the self-force acting  on a particle that is moving on a  background spacetime via the MiSaTaQuWa equation~\cite{Poisson:2011nh} and for determining
 the probability of a quantum particle detector
being excited by a field emitted by another detector~\cite{blasco2015violation,JACMK2019}.

Based on the seminal work by Hadamard~\cite{Hadamard}, 
an analytic expression is known for the GF, $G_R(x,x')$, which is valid within a normal neighbourhood
$\mathcal{N}(x)$
 of 
the base spacetime point $x$
(i.e., a 
neighbourhood $\mathcal{N}(x)$ of $x$ such that every $x'\in \mathcal{N}(x)$  is connected to $x$ by a unique geodesic which
lies in $\mathcal{N}(x)$).
In the case of a scalar field on  a $(3+1)$-dimensional spacetime, this Hadamard form is~\cite{friedlander}:
\be\label{eqhada}
\GFret(x, x') =[\U(x,x')\delta(\s)+ \V(x,x')\Hv(-\s)]\Hv_+(x,x'),
\ee
where 
$\delta$ and $\Hv$  are, respectively, the  Dirac-delta and Heaviside distributions,
$\U$ and  $\V$ are 
smooth biscalars, 
and $\Hv_+(x, x')$ equals $1$ if $x'$ lies to the causal future of $x$ and equals $0$ otherwise.
Here,  $\s=\s(x,x')$ is Synge's world-function, i.e., 
one-half of the squared distance along the (unique) geodesic connecting  $x$ and $x'$.
Eq.~(\ref{eqhada})
explicitly shows
that, in a $(3+1)$-dimensional spacetime,
the GF has a Dirac-delta divergence at points $x'\in \mathcal{N}(x)$ that are connected to $x$ via a null geodesic.
The first term, $\U\delta$, and the second term, $\V\Hv$, on the right-hand side of Eq.~(\ref{eqhada}) are usually called the `direct' and `tail' parts, respectively.

It is further known~\cite{Garabedian, KRW:1997, Ikawa} that, outside a normal neighborhood, the GF continues to diverge when the two spacetime 
points are connected via a null geodesic.
The explicit form of the singularity {\it outside} a normal neighborhood, however, was not known within General Relativity until recently.
Using a variety of methods, it has been shown~\cite{Ori1short,CDOWa,Dolan:2011fh,harte2012caustics,Casals:2012px,Zenginoglu:2012xe,Yang:2013shb} that the {\it global} form of the `leading' singularity of the 
GF generally has the following four-fold structure in Schwarzschild, Kerr and other background spacetimes:
\be
\delta(\s)\to\PV(1/\s)\to-\delta(\s)\to-\PV(1/\s)\to\delta(\s)\to\cdots
\label{4-fold}
\ee 
where the first term corresponds to the direct part in Eq.~\eqref{eqhada} and $\PV$ denotes the Cauchy principal value distribution.  
This change in the character of the singularity is essentially due to the null wavefront of the field perturbation passing through a caustic point (i.e., a spacetime
point where neighboring null geodesics are focused). 
This is indicated in Eq.~\eqref{4-fold} by the arrow `$\to$'. Thus the {\it leading} singularity in $\GFret(x,x')$ has the form $\delta(\s(x,x'))$ (respectively, $\PV(1/\s), -\delta(\s), -\PV(1/\s)$) when there is a null geodesic from $x$ to $x'$ that has passed through $4n$ (respectively, $4n+1, 4n+2, 4n+3$) caustics, where $n$ is a non-negative integer.
The four-fold structure of the GF in Schwarzschild spacetime is beautifully illustrated with numerical animations in~\cite{Video:CausticsSchw}
and has been  used in~\cite{CDOWa,CDOW13,PhysRevD.89.084021} in order to provide an insight into the origin of the self-force and in~\cite{JACMK2019} {and~\cite{caribe2023lensing}} in order to account for interesting features in respectively, the communication {and entanglement} between quantum particle detectors.

The above four-fold singularity structure of the GF, however,  only represents its `leading' singularity.
In~\cite{Casals:2012px}, we proved that there is also a `sub-leading' discontinuity in the case of a black hole toy-model spacetime (Pleba{\'n}ski-Hacyan spacetime, $\mathbb{M}_2\times \mathbb{S}_2$, abbreviated as PH below)
which displays another four-fold structure:
\be
\theta(-\s) \to -\ln\left|\s\right| \to -\theta(-\s) \to \ln\left|\s\right| \to \theta(-\s)\to\cdots
\label{4-fold,sublead}
\ee 
 where the first term corresponds to the tail part in Eq.~\eqref{eqhada}.
To the best of our knowledge, such `sub-leading' four-fold structure has not yet been shown on an actual black hole spacetime.
In this paper
we derive
explicit forms for
 both the leading and sub-leading discontinuities 
  of the  GF of the massless scalar wave equation on Schwarzschild spacetime.
 Furthermore, although we have written the above global four-fold structures in terms of a world-function $\s$, this object is only  well-defined in a geodesically convex domain:
 a region in which all pairs of points are connected by a unique geodesic. In this paper we write the four-fold structures in Schwarzschild spacetime
  in terms of a globally well-defined generalization of the world-function $\hat{\sigma}$ of a spacetime that is conformally related to Schwarzschild.
  
Our derivation is underpinned by a simple conformal transformation of the  Schwarzschild metric to a direct product spacetime, $\Mtwo \times \mathbb{S}_2$,
where $\mathbb{S}_2$ is the two-sphere and $\Mtwo$ is a two-dimensional Lorentzian spacetime (containing the time and radial variables).
In a separate paper~\cite{PhysRevD.92.104030} we
have shown that $\Mtwo$ is a causal domain (i.e.\ is geodesically convex, and obeys a certain causality condition)~\cite{friedlander}, which implies that its corresponding world-function is valid globally. 
This allows us to write a {\it global}  representation of the GF in Schwarzschild spacetime. {This involves a sum over angular modes, indexed by {the multipolar number} $\ell$, of \textit{globally well-defined} 2-dimensional Green functions $G_\ell$ {in the two-dimensional spacetime $\Mtwo$} (see Eq.~\eqref{cgl-eqn} below) multiplied by Legendre polynomials. When we use a  series representation of the Riemann function $\mathcal{U}_\ell$ associated with $G_\ell$ ({that is, $\mathcal{U}_\ell$ is} the coefficient of the Heaviside step function in the 2-dimensional Green function) we obtain a form for the GF in Schwarzschild which makes explicit its {\it complete} four-fold singularity structure in terms of 
the distributions in Eqs.~(\ref{4-fold}) and (\ref{4-fold,sublead})
(with the argument $\s$ replaced by its globalized companion). 
The series representation of $\mathcal{U}_\ell$ is obtained by applying a theorem of Zauderer \cite{zauderer1971modification}, and yields an expression in terms of an infinite series involving Bessel functions with coefficients coming from the Hadamard series (i.e.\ a series in powers of the world function) of the Riemann function of a certain `background' 2-dimensional wave equation. The Bessel series converges whenever the Hadamard series converges. We then
expand for large-$\ell$; resumming yields an expression for the GF in Schwarzschild.
This representation for the GF in Schwarzschild is valid whenever the Hadamard series of the 2-dimensional Riemann function $U$ associated with the background 2-dimensional wave equation converges. This method is essentially the same as the one we used in Sec.V~\cite{Casals:2012px}. Our  large-$\ell$ expansion is also  in the same spirit as~\cite{Dolan:2011fh,Yang:2013shb}, although these 
works
used a further separation in the time variable (via a Fourier transform) and expansion of the GF in terms of the so-called quasinormal modes\footnote{
The
contribution from the branch cut that the GF has in the complex-frequency plane (see, e.g.,~\cite{Leaver:1986,CDOW13,PhysRevLett.109.111101})
was thus neglected in~\cite{Dolan:2011fh,Yang:2013shb}.}, instead of the Green function in the two-dimensional spacetime $\Mtwo$ that we employ.}

Furthermore: In any spacetime dimension,
one can write the  biscalars in the Hadamard form  as a  Hadamard series.
The coefficients of the distributions in our representation for the GF in Schwarzschild
 are given 
 in terms of the
 coefficients in the Hadamard series for
the Riemann function $U$  in $\Mtwo$.
  Therefore, our representation of the GF  has a direct geometrical interpretation in terms of geodesics in 
 this two-dimensional causal domain.
 Our representation for the GF naturally takes the form of a sum of expansions about  each of
the null geodesics in Schwarzschild.  
In effect, we give
an extension of the local Hadamard form for the GF in Schwarzschild spacetime, valid beyond normal neighbourhoods, 
which may be characterized as a `sum of Hadamard forms'. 

This representation for the GF is, however, not valid in a neighbourhood of
 caustic points, for which the angle separation $\gamma$ is equal to $0$ or  $\pi$
 (specifically, this global representation is valid for  $\gamma\in (0,\gamma_0)$, where $\gamma_0=2(\sqrt{2}-1)\pi\simeq 0.828\pi$).
 Indeed, it is known~\cite{Casals:2012px,Zenginoglu:2012xe} that on caustic points in a spherically-symmetric spacetime the above four-fold structure does not generally apply and is instead a two-fold structure.
However, to the best of our knowledge, the analytical form of the singularity of the GF on caustics is not known in Schwarzschild spacetime 
and  we derive its precise form here.
We thus provide a {\it complete} description of the singularity structure of the GF.

As an illustration of the usefulness and the properties of the mentioned Bessel function representation, we calculate and plot (in Fig.~\ref{fig:Bessel exp}) its leading order term (that is after carrying out an $\ell$-mode decomposition but before carrying the full expansion for large-$\ell$)
and  compare it
with an `exact' calculation of the GF obtained using an independent method developed in~\cite{CDOW13}.

As mentioned above, the singularity structure and `sum of Hadamard forms' is valid whenever the Hadamard series of a certain wave equation on $\Mtwo$ converges. 
In this paper we further present numerical studies which provide strong evidence that convergence holds on large domains of $\Mtwo$. It is an open question as to whether or not convergence holds throughout $\Mtwo$; convergence would imply global-in-time validity of our results. 

Specifically, we compare our analytical expressions for the global divergences of the GF  against an ``exact"
semi-analytical calculation of the full GF. We find remarkable agreement for points $x'$ timelike-separated from $x$
up to distances which are quite ``far"  from it.
This indicates that the region where the Hadamard form converges uniformly 
 is either equal  to  the whole of   $\Mtwo$ or, at least, that it includes regions that correspond to extending well beyond the normal neighbourhood of a point in the 4-dimensional Schwarzschild spacetime.
 Furthermore, for points $x'$ ``near"  $x$, our expression 
 in terms of geometrical quantities in $\Mtwo$
 for the direct divergence in the Hadamard form \eqref{eqhada}
 has been successfully used  in the works~\cite{JACMK2019, CNOW:2019,OToole:2021} to greatly facilitate the practical calculation
 of the GF. Thus, these works corroborate our useful expression for the direct divergence.

We also provide another, separate representation for the  GF in Schwarzschild spacetime. This other representation also starts from the Bessel series but, instead of carrying out a large-$\ell$ expansion of its summands, we essentially express the full $\ell$-sum of the $\ell$-dependent factor in the summands  as integrals of the GF in PH spacetime. This new representation not only is of value in itself (in that it offers a way of calculating the  GF in Schwarzschild partly via the GF in PH, which is much easier to calculate)
but also we use it to derive the global singularity structure of the GF in Schwarzschild in an alternative way from that mentioned above (namely, via a large-$\ell$ expansion of the modes of the GF in Schwarzschild), thus offering a check of our results.

For the reader who is only interested in our main results rather than the details of the calculation, we here note our main equations.
Eq.~(\ref{eq:sum Hads}) (see also \eqref{eq:GF Schw from PH discont}) is our sum of Hadamard forms for the GF in Schwarzschild spacetime, which explicitly shows the full structure of the
divergences when the points are connected by a null geodesic.
As mentioned, this form is not valid at caustics, where $\gamma=0$ or $\pi$.
We address this in 
Eq.~\eqref{eq:Grl-caustic-0},
 which gives the explicit full form of the divergences of the  GF at caustics in
the case $\gamma=0$;
Eq.~\eqref{eq:Grl-caustic-pi}
 gives the structure in the case of antipodal points, $\gamma=\pi$.
Finally, Eq.~\eqref{eq:GF Schw from GF PH} is the expression for the GF in Schwarzschild in terms of integrals of the GF in PH, and \eqref{eq:GF Schw Had PH} is a version of it after using the Hadamard form for the GF in PH.

The rest of this paper is organized as follows.

In Sec.~\ref{sec:GF} we define the retarded GF in Schwarzschild spacetime and express it in terms
of GFs in the $2$-D conformal spacetime. We introduce the key background 2-dimensional wave equation, and show how all 2-dimensional GFs derive from this seed equation. More precisely, the 2-dimensional Riemann functions mentioned above can be written as a series of Hadamard coefficents of this seed equation, scaled by Bessel functions $J_k((\ell+\frac12)\sigma), k\geq 0$ where $\ell$ is the multipole index and $\sigma$ is the globally defined world function of the 2-dimensional conformal Schwarzschild spacetime. This follows from a theorem of Zauderer (\cite{zauderer1971modification}; see also Theorem 6.4.2 of \cite{friedlander}), and we will refer to this as the Bessel expansion of the Riemann (or Green) function. This global representation of the 4-dimensional GF is then analysed in the following sections, with the aim of determining its global (singularity) structure in a way that relates to the underlying causal structure of the spacetime.

In Sec.~\ref{sec:PH}, we show how the global retarded GF of Schwarzschild spacetime may be derived by a sequence of iterated integrals  of the corresponding GF of PH spacetime. This yields our first representations of the global singularity structure and of the `sum of Hadamard forms' of the retarded GF of Schwarzschild spacetime.

In Sec.~\ref{sec:Sum Hads} we apply a large-$\ell$ expansion of the Bessel functions arising in the Bessel series obtained in Sec.~\ref{sec:GF} to determine the global singularity structure of the retarded GF $G_R$ on Schwarzschild spacetime, and to determine the `sum over Hadamard forms' expansion of the non-smooth contribution to $G_R$. This approach provides more detail on the distributions that contribute to this sum, relative to the approach of the previous section.

In Sec.~\ref{sec:caustics}, we derive the two-fold singularity structure in the case of caustics in Schwarzschild spacetime.

In Sec.~\ref{sec:numerical} we provide numerical evidence for our results.

We conclude in Sec.~\ref{sec:conclusions} with some comments and suggestions for possible applications of our results.

In Appendix \ref{sec:App}, as an illustration of  the multipolar modes of the GF, we 
calculate these in two simple $(3+1)$-D spacetimes: flat spacetime and Nariai spacetime. In Appendix \ref{sec:M=0}, we consider the zero mass ($M=0$)  limit of the results of this paper (taking us from Schwarzschild spacetime to Minkowski spacetime). This allows us to consider the question of convergence of the key 2-dimensional Hadamard series, and to draw some links between GFs on 2-dimensional anti-de Sitter spacetime  $AdS_2$, PH spacetime, and results relating to the representation of Legendre functions in terms of series of Bessel functions.
Finally, in App.~\ref{sec:SF}, we show how
our results for the Schwarzschild GF could be used to calculate the regularized self-field (i.e., the regularized value of the scalar field created by a scalar point charge evaluated on the location of the charge itself) - see \cite{CNOW:2019}. This self-field is relevant to self-force calculations of  radiation reaction.

Throughout this paper we use geometric units $c=G=1$ and metric signature $(-+++)$.


\section{Green function on Schwarzschild spacetime}\label{sec:GF}
The perturbations by a masslesss scalar field $\Phi$ 
of a background spacetime  $\st$ satisfy a scalar (Klein-Gordon) wave equation.
The retarded Green function  $\GFret(x,x')$
satisfies the inhomogeneous  wave equation with a Dirac-delta distribution source, together with a boundary condition ensuring that $\GFret(x,x')$ vanishes if $x'$ is not in the causal future $J^+(x)$ of the point $x$,
where $x,x'\in \st$. Thus
\be \square\GFret(x,x')=-4\pi\delta_4(x,x'),\qquad \GFret = 0 \hbox{ if } x'\notin J^+(x),\label{gfret-eqs}\ee
where $\square=\nabla_\alpha\nabla^\alpha$ is the d'Alembertian operator, $\delta_4(x,x')\equiv \frac{\delta_{4}(x-x')}{\sqrt{-g(x)}}$ and $g$ is the determinant of the metric of
the background spacetime.

In the Introduction we gave the Hadamard form for the  retarded Green function (GF) in a $(3+1)$-dimensional spacetime,
Eq.~(\ref{eqhada}).
As mentioned, the great advantage of the Hadamard form  is that it makes explicit the form of the singularity of the GF.
This follows from the fact that Synge's world function $\s(x, x')$ is  positive/zero/negative if $x$ and $x'$ are, respectively, spacelike/null/timelike separated.
The main disadvantage of the Hadamard form is that it is only valid in a 
normal neighborhood of $x$, i.e., for $x'\in\mathcal{N}(x)$.
In many situations, however, it is very valuable to know the GF globally. 
In order to obtain a global representation for the GF in Schwarzschild spacetime where the form of its singularities becomes explicit, we will make use of the spherical symmetry of
the spacetime.

We shall use  the usual time coordinate of the Schwarzschild exterior (generated by the timelike Killing vector) and the tortoise radial coordinate $\rs\in (-\infty,+\infty)$
 so that the line element has the form 
\be ds^2 = -f(dt^2-d\rs^2) + r^2d\Omega_2^2,\label{sch-lel}\ee
where $f=f(r)\equiv 1-2M/r$ and $d\Omega_2^2$ is the standard line element of the unit $2$-sphere. The area radius $r$ and tortoise coordinate $\rs$ are related by 
\be \frac{dr}{d\rs}= f.\label{tortoise}\ee In these coordinates, the wave equation for $\GFret$ reads 
\be -f^{-1}\pdd{\GFret}{t}+f^{-1}\pdd{\GFret}{\rs}+\frac{2}{r}\pd{\GFret}{\rs}+\frac{1}{r^2}\nabla^2\GFret=-\frac{4\pi}{r^2f}\delta_2(x^A-x^{A'})\delta_{\mathbb{S}_2}(x^a,x^{a'}),\label{gfret-eq1}
\ee
where $\nabla^2$ is the Laplacian operator on the unit 2-sphere, $x^A=(t,\rs)$ are coordinates on the Lorentzian 2-space (i.e.\ the 2-space that arises by factoring the 4-dimensional spacetime by the action of the $SO(3)$ that generates the spherical symmetry) 
and $x^a=(\theta,\phi)$ are coordinates on the unit 2-sphere. 

At this point, it is usual to rescale the field by a factor $r$: this removes the first order derivative from the wave equation. There is also a geometrical interpretation of this step. The appropriate rescaling amounts to making a conformal transformation of the metric:
\be d\hat{s}^2 \equiv r^{-2}ds^2 =
ds_2^2
+d\Omega_2^2\label{con-sch-lel}\ee
where
\be
ds_2^2 \equiv -\frac{f}{r^2}(dt^2-d\rs^2).\label{2d-lel}
\ee
We will refer to the spacetime with line element (\ref{con-sch-lel}) as the \textit{conformal Schwarzschild spacetime}, which we shall denote as $\sth$.
In its turn, we will refer to the 2-dimensional spacetime with line element (\ref{2d-lel}) as the \textit{2-D conformal space}, and denote it by $\Mtwo$.
By general properties of Green functions  in conformally related spacetimes, we can write~\cite{friedlander,Birrell:Davies}
\be \GFret = \frac{1}{r\cdot r'}\hat{G}_R(x,x'),\label{con-gf}\ee
where $\hat{G}_R(x,x')$ is the retarded Green function for the conformally invariant wave equation on $\sth$. The conformally invariant wave equation of a general $4$-dimensional spacetime with metric tensor ${\rm{g}}$ is  
\be \Box_{\rm{g}}\Phi -\xi R\Phi=0,\quad \xi=\frac16,\label{eq:conf-wave-eq}\ee  
where $R$ is the Ricci scalar corresponding to the metric ${\rm{g}}$.  Using this rescaling, we find
\be -\pdd{\hat{G}_R}{t}+\pdd{\hat{G}_R}{\rs}+\frac{f}{r^2}\left(\nabla^2-\frac{2M}{r}\right)\hat{G}_R=-4\pi\delta_2(x^A-x^{A'})\delta_{\mathbb{S}_2}(x^a,x^{a'}),\label{con-gfret-eq1}
\ee
where we have used the value $R= 12M/r$ of the Ricci scalar of $\sth$. 

In addition to rendering the wave equation more tractable, the conformal rescaling in \eqref{con-sch-lel} introduces a very useful simplification of the world function. The direct product structure of the metric induced by (\ref{con-sch-lel}) yields 
\be \hat{\sigma} = \sigma(x^A,x^{A'}) + \frac12\gamma^2.\label{world-con}\ee
Here, 
$\hat{\sigma}$ is the world function of the conformal Schwarzschild spacetime $\sth$, $\sigma(x^A,x^{A'})$ is the world-function of the 2-dimensional Lorentzian spacetime $\Mtwo$
and $\gamma\in[0,\pi]$ is the geodesic distance 
 on the unit 2-sphere:  
\be \gamma(x^a,x^{a'}) = \hbox{proper distance along the shortest path from $x^a$ to $x^{a'}$ on $\mathbb{S}_2$}.\label{gamma-def} \ee
 
In a previous paper \cite{PhysRevD.92.104030}, we proved that $\Mtwo$ is a \textit{causal domain}. This means, in particular, that each pair of points of this $2$-D spacetime are joined by a unique geodesic. As a consequence, $\sigma$ is defined globally on $\Mtwo$, in contrast with the usual situation in four dimensions: e.g., there is no base point $p$ of conformal Schwarszchild spacetime for which the corresponding maximal normal neighbourhood is the whole spacetime, and consequently $ \hat{\sigma}$ is not defined globally on this spacetime. 
This technical point underpins the present paper, where we use a 2+2 approach to determine certain global properties of the GF on Schwarzschild spacetime. 

There is an immediate pay-off in terms of understanding the global causal structure of Schwarzschild spacetime. The world function $\sTCS$  of the $2$-D spacetime $\Mtwo$ satisfies
\be
\nabla_A \sTCS \nabla^A \sTCS=2\sTCS,
\ee
with the initial conditions $\lim_{x'\to x}\sigma(x,x')=0$ and  $\lim_{x'\to x}\nabla_A \nabla_B \sigma(x,x')=g_{AB}(x)$.
As noted, there is a unique geodesic connecting any given pair of points in $\Mtwo$. Furthermore, any geodesic of the $4$-D spacetime $\sth$ decomposes as a geodesic on $\Mtwo$ and a geodesic on $\mathbb{S}_2$. That is, if $I\subseteq\mathbb{R}$ is an interval and
\be c: I \to \sth: s\mapsto x^\alpha(s)=(x^A(s),x^a(s)) \label{geo-m4},\ee
is a geodesic on $\sth$, then 
\begin{eqnarray*} 
c_1 &:& I \to \Mtwo: s\mapsto x^A(s),\nonumber\\
c_2 &:& I \to \mathbb{S}_2: s\mapsto x^a(s),
\end{eqnarray*}
are geodesics on $\Mtwo$ and $\mathbb{S}_2$ respectively (the converse statement also holds). We refer to $c_1$ as the projection of the geodesic $c$ onto $\Mtwo$; there is a unique $c_1$ for a given geodesic $c$ of $\sth$. Now consider any pair of points $x^\alpha=(x^A,x^a),x^{\alpha'}=(x^{A'},x^{a'})$ of $\sth$. As proven in \cite{PhysRevD.92.104030}, there is a unique geodesic of $\Mtwo$ connecting $x^A$ and $x^{A'}$. When $x^a$ and $x^{a'}$ are neither antipodal points nor the same point (for which we would have $\gamma= \pi$ and $\gamma= 0$, respectively), there is a countably infinite family of geodesics of $\mathbb{S}_2$ connecting $x^a$ and $x^{a'}$, corresponding to multiple circuits of the appropriate great circle of the sphere. By lifting these geodesics from $\Mtwo$ and $\mathbb{S}_2$, we see that any pair of points of $\sth$ are connected by a countably infinite family of geodesics. Each of these geodesics projects to the same geodesic of $\Mtwo$. This resolves the question of the existence and multiplicity of geodesics on $\sth$. 

 Then, by conformal invariance of null geodesics, a null geodesic connects $x$ and $x'$ in Schwarzschild spacetime if and only if a null geodesic connects the corresponding points of the conformal Schwarzschild spacetime. This holds if and only if ${\hat{\sigma}}_k=0$, for some $k\in \mathbb{Z}$, where
\be {\hat{\sigma}}_k \equiv \sigma + \frac12\left(\gamma+ 2k\pi\right)^2. \label{sig-k-def}\ee

In this formula, $(\gamma+2\pi k)^2$ is the square of the total proper distance elapsed along the projection of the geodesic onto the 2-sphere. For $k\geq0$, the geodesic crosses $2k=2|k|$ caustics at $\gamma=0$ and $\gamma=\pi$. For $k<0$, the geodesic has crossed $2|k|-1$ caustics. Thus the number of caustics crossed is given by 
\be C_k \equiv \left\{ \begin{array}{ll} 2k, & k\geq0; \\ 2|k|-1, & k<0. \end{array}\right. \label{eq:caustic-counter} \ee 
We note that when the separation on $\Mtwo$ is timelike or null, we may write (\ref{sig-k-def}) as 
\be {\hat{\sigma}}_k = -\frac12\eta^2 + \frac12\left(\gamma+ 2k\pi\right)^2,\quad k\in \mathbb{Z}, \label{sig-k-eta}\ee
where $\eta$ 
is the geodesic distance (along causal geodesics) on $\Mtwo$ (so that $\sigma=-\frac12\eta^2$ and, particular,
it is
is proper time in $\Mtwo$ for timelike separations and is zero for null separations).
This `globalizes' the world function $\hat{\sigma}$ on the $4$-D conformal Schwarzschild spacetime:
$\hat{\sigma}_k$ is one-half of the square of the geodesic distance  between any two causally-related
 points in conformal Schwarzschild  spacetime along a timelike or null geodesic which has 
 passed through $C_k$ points with $\gamma=0$ or $\gamma=\pi$ (both of which are caustics in the case of a null geodesic).

Returning to the Green functions, we separate the angle variables in the usual way via a multipolar decomposition and write
\be \hat{G}_R(x,x') = \frac{1}{4\pi}\sum_{\ell=0}^\infty (2\ell+1)\Gl(x^A,x^{A'})P_\ell(\cos\gamma).\label{con-gret-mode}\ee
Here, $P_\ell$ are Legendre polynomials and $\Gl$ satisfy the partial differential equation (PDE) for a Green function on the $2$-D conformal space:
\be \square_2\Gl-V_\ell(\rs)\Gl = -4\pi\frac{r^2}{f}\delta_2(x^A-x^{A'}),\label{gl-pde}\ee
where $\square_2$ is the d'Alembertian operator of the $2$-D conformal space and the potential is 
\be V_\ell(\rs) \equiv \ell(\ell+1)+\frac{2M}{r}.\label{vl-def}\ee

The boundary conditions that the Green function $\Gl$  obey must be such that when they are introduced in Eq.~(\ref{con-gret-mode}), and use
is made of   Eq.~(\ref{con-gf}),
  the resulting Green function  $\GFret$ is the {\it retarded} Green function of Schwarzschild spacetime.
In fact,
$\Gl(x^A,x^{A'})$ must be equal to
 the retarded Green function of Eq.~(\ref{gl-pde}) since\footnote{We remind the reader that the  properties described in the text define the {\it unique}
 retarded Green function -- see Corollary 6.3.1 in~\cite{friedlander}.}
  it  satisfies the same PDE (Eq.~(\ref{gl-pde}))
 and it obeys the defining boundary conditions of a retarded Green function, as we now show. 
 From Eq.~(\ref{con-gret-mode}) and the orthogonality properties of the Legendre polynomials it follows that
 \begin{equation} \label{eq:G_l conf Schw}
\Gl(x^A,x^{A'})=2\pi\int_{-1}^{+1}d(\cos\gamma)P_{\ell}(\cos\gamma)\hat{G}_R(x,x').
\end{equation}

Now, from Eq.~(\ref{sig-k-def}), $\sigma>0$ implies $\hat{\sigma}_k>0$, $\forall k\in\mathbb{Z}$.
Since the retarded Green function $\hat{G}_R(x,x')$ of conformal Schwarzschild is zero if $x'\notin J^+(x)$,
it follows that $\Gl$ is  zero
if either $\sigma>0$ or $\Delta t\equiv t-t'<0$, which is the defining boundary condition of the retarded Green function in $\Mtwo$.
In order to illustrate the causality properties of the    $\Gl$, in Appendix \ref{sec:App} we calculate these modes
for two simple 4-D spacetimes: flat and Nariai spacetimes.

In coordinates $(t,\rs)$, Eq.~\eqref{gl-pde} takes the familiar form
\be -\pdd{\Gl}{t}+\pdd{\Gl}{\rs}-\frac{f}{r^2}V_\ell\Gl = -4\pi\delta_2(x^A-x^{A'}).\label{gl-pde-coords}\ee
Thus the structure of $\Gl$, and hence $\hat{G}_R$ and $\GFret$ can be probed using a variety of PDE techniques. In particular, there is a large body of work that exploits the amenability of (\ref{gl-pde-coords}) to a Fourier transform in the time coordinate. We will take an alternative approach that remains in the 2-dimensional setting and that applies (in particular) a result of Zauderer (\cite{zauderer1971modification}; see also Theorem 6.4.2 of \cite{friedlander}). This result  -- which we shall give explicitly below -- provides an expansion of $\Gl$ in terms of Bessel functions and of Hadamard coefficients for a `background' wave equation. The Green function $G$ for the background equation satisfies 
\be PG = -4\pi\frac{r^2}{f}\delta_2(x^A-x^{A'})\label{eq:background wave eq}\ee
where 
\be P \equiv \square_2 +\frac{1}{4}\left(1-\frac{8M}{r}\right),\label{pop-def}\ee
so that (\ref{gl-pde}) can be written in the perturbative (but exact) form
\be (P-\lam^2)\Gl=-4\pi\frac{r^2}{f}\delta_2(x^A-x^{A'}),\label{cgl-eqn}\ee
where $\lam\equiv \ell+1/2$.

The Green function of the  background wave equation (\ref{eq:background wave eq}) obeying retarded boundary conditions can be written in the 2-dimensional Hadamard form:
\be G(x^A,x^{A'})=2\pi\theta(t-t')\theta(-\sTCS)U(x^A,x^{A'}),\label{eq:Had form 2d} \ee
where the biscalar $U(x^A,x^{A'})$ is the solution of the homogeneous equation
$PU=0$
satisfying  boundary conditions derived from Eqs.~(\ref{eq:Had series 2d}) and (\ref{U0-def}) below
($U$ is the so-called \textit{Riemann function} of the operator $P$ on $\Mtwo$ -- see \cite{Garabedian} and Sec.~6.2 of \cite{friedlander}). This biscalar exists and is uniquely defined on $\Mtwo\times\Mtwo$, and for any point $p=x^{A}\in\Mtwo$, there is a neighbourhood $\Omega_p\subset \Mtwo$ such that  $\forall x^{A'}\in \Omega_p$ 
we can write  (we note that we are using the conventions of \cite{friedlander}):
\be U(x^A,x^{A'}) = \sum_{k=0}^{\infty}U_k(x^A,x^{A'}) \frac{(-2\sTCS)^k}{k!}, \label{eq:Had series 2d}\ee
where the series is uniformly convergent on $\Omega_p$  (see Theorem 6.2.1\cite{friedlander}): this is known as the {\it Hadamard series}.
{The numerical evidence in Sec.~\ref{sec:numerical} provides a strong indication that
 $\Omega_p$ extends well beyond the region corresponding to the {maximal} normal neighbourhood of a point 
of the 4-dimensional spacetime that projects to $p$.} Furthermore, in Appendix B, we consider convergence of the series in the limiting case of zero mass ($M=0$).

Using \eqref{eq:Had form 2d}, \eqref{eq:Had series 2d} and \eqref{sig-k-eta}, we have 
\be G(x^A,x^{A'})=2\pi\theta(t-t')\theta(-\sTCS) \sum_{k=0}^{\infty}U_k(x^A,x^{A'}) \frac{\eta^{2k}}{k!}.
\label{eq:Had form 2d eta}
\ee

The Hadamard coefficients $U_k$,  $k\geq0$,   in  $\Mtwo$    satisfy the following recurrence relations in the form of transport equations along the unique geodesic from $x^A$ to $x^{A'}$~\cite{DeWitt:1960,friedlander,Decanini:Folacci:2008,Ottewill:2009uj}. 
In two dimensions, $U_0$ is the square root of the so-called van Vleck determinant $\Delta_{2d}=\Delta_{2d}(x^A,x^{A'})$:
\be U_0=\Delta_{2d}^{1/2}\quad \Leftrightarrow \quad 2\sigma^A\nabla_AU_0=(2-\square_2\sigma)U_0,\quad \lim_{x'\to x}\ U_0=1,
\label{U0-def}\ee
and the $U_k, k\geq1$, are determined by solving the transport equations
\footnote{We thank David Q. Aruquipa for identifying a sign error in an earlier version of Eq.~\eqref{uk-transport}.}
\be 2\sigma^A\nabla_AU_k+\left(\square_2\sigma+2(k-1)\right)U_k=\frac12 PU_{k-1},\quad k\geq 1.\label{uk-transport}\ee
Regularity at $\sigma=0$ fixes constants of integration, so that the $U_k$ are uniquely determined.
Here, the operator $P$ denotes the operator of Eq.~\eqref{eq:background wave eq} satisfied by $G$. We note that $\sigma$ and $U_0$ depend only on the spacetime geometry, but the coefficients $U_k, k\geq 1$ depend on the details of the wave operator $P$. 
We also note that we have calculated 
$\sTCS$ and  $\Delta_{2d}$ in $\Mtwo$ numerically in~\cite{PhysRevD.92.104030}. 
The $2$-D Hadamard coefficients $U_k$ play an important part in the analysis of the GF in $4$-D Schwarzschild spacetime that we shall carry out in Sec.~\ref{sec:Sum Hads}.

We wish to derive a globally valid form for $\Gret$ on Schwarzschild spacetime and its full singularity structure.
 As seen in  Eqs.~(\ref{con-gf}) and (\ref{con-gret-mode}), this can be written as a sum of 2-dimensional Green functions:
\be G_R(x,x') = \frac{1}{2\pi r\cdot r'}\sum_{\ell=0}^\infty \left(\ell+\frac12\right)\Gl\left(x^A,x^{A'}\right)P_\ell(\cos\gamma),\label{sch-gret-mode}\ee
where the $\Gl$ satisfy Eq.~(\ref{cgl-eqn}).

In Sec.~\ref{sec:Sum Hads} below, we show how we can use a large$-\ell$ expansion to (i) determine the full singularity structure of $G_R$ and (ii) express the non-smooth contribution to $G_R$ as a `sum of Hadamard forms'. The resulting expressions are valid on regions of the form 
\begin{equation} \hat{\Omega}_P=\{Q\in \Mtwo\times\mathbb{S}_2: q\in\Omega_p \hbox{ and } \gamma(x^a,x^{a'})\in[0,\gamma_0)\},\label{eq:OmegaP} 
\end{equation}
where $P=(p,x^a)\in\Mtwo\times\mathbb{S}_2$  and $Q=(q,x^{a'})\in\Mtwo\times\mathbb{S}_2$. This arises through the application of a theorem due to Zauderer \cite{zauderer1971modification}, which is cited (with an alternative proof) as Theorem 6.4.2 of \cite{friedlander}. This result gives the following form for the retarded Green function $\Gl$ on $\Mtwo$ of the operator $P-L^2$ of (\ref{cgl-eqn}):
\be 
\Gl= 2\pi\theta(-\sigma)\theta(\Delta t)\cgl\left(x^A,x^{A'}\right),\quad
\cgl(x^A,x^{A'}) \equiv  \sum_{k=0}^\infty U_k\left(\frac{2\eta}{\lam}\right)^kJ_k(\lam\eta),
\label{bessel-exp-cgl}\ee
where the $J_k$ are Bessel functions,  $\cgl$ 
 is the Riemann function for the $2$-D wave equation \eqref{cgl-eqn} and the coefficients $U_k=U_k(x^A,x^{A'})$ are the 
$2$-D Hadamard coefficients of Eq.~(\ref{eq:Had series 2d}).
 These coefficients and the series Eq.~(\ref{bessel-exp-cgl}) are defined globally on $\Mtwo$.
  This result in \eqref{bessel-exp-cgl} is \textit{not} perturbative: it holds for all $\lam\in\mathbb{C}\setminus\{0\}$, and yields \eqref{eq:Had form 2d eta} in the limit $L\to 0$.  
According to Zauderer's theorem, the series in Eq.~(\ref{bessel-exp-cgl})
converges uniformly in the region 
$\Omega_p$ of any $p\in\Mtwo$ on which the Hadamard series Eq.~(\ref{eq:Had series 2d}) converges uniformly.

In the next sections we 
proceed to exploit, in different ways, Eq.~(\ref{sch-gret-mode}) with Eq.~\eqref{bessel-exp-cgl} in order to determine properties of the GF in Schwarzschild spacetime.


%


\section{Green Functions in Schwarzschild and in Pleba{\'n}ski-Hacyan spacetime $\mathbb{M}_2\times \mathbb{S}_2$}\label{sec:PH}


\subsection{General form}\label{sec:PH,gral}

We note that the GF in Schwarzschild spacetime given by \eqref{sch-gret-mode} and \eqref{bessel-exp-cgl} bears an interesting and useful relationship to the GF
$G_{R}^{PH}$
of Pleba{\'n}ski-Hacyan (PH) spacetime $\mathbb{M}_2\times \mathbb{S}_2$ (where $\mathbb{M}_2$ is two-dimensional Minkowski spacetime) for a scalar field with $m^2+2\xi=1/4$, where $m$ is the mass of the field. See Eq.~(134) in Ref.~\cite{Casals:2012px}\footnote{We note that in the last expression in Eq.~(134) in~\cite{Casals:2012px} there is a  missing factor $\theta(-\sigma_{\mathbb{M}_2})$.} for an expression for the latter GF:
\begin{eqnarray}
G_{R}^{PH}(\eta_{PH},\gamma)&=&
\theta(-\sigma_{\mathbb{M}_2})\theta(\Delta t)\sum_{\ell=0}^\infty \left(\ell+\frac12\right)P_\ell(\cos\gamma) J_0(\lam\, \eta_{PH})=
\theta(\eta_{PH})\sum_{\ell=0}^\infty \left(\ell+\frac12\right)P_\ell(\cos\gamma) J_0(\lam\, \eta_{PH}),
\nonumber \\
&& \label{eq:GR PH} \end{eqnarray}
where
the world function of the $\mathbb{M}_2$ factor of PH spacetime is given by $\sigma_{\mathbb{M}_2}=-\eta_{PH}^2/2$ and $\eta_{PH}$ is the geodesic distance in $\mathbb{M}_2$.
(In the last equality in \eqref{eq:GR PH} - and throughout the paper - we adopt the convention that proper time along future-directed timelike geodesics and the time coordinate $t$ are cosynchronous).  
We see that this expression for the GF in PH is obtained from (\ref{sch-gret-mode}) by removing the conformal factor $1/(r\cdot r')$, setting $U_0=1$ and $U_k=0, k\geq1$ in \eqref{bessel-exp-cgl}, and replacing $\eta$ by $\eta_{PH}$.
We will use this fact in Sec.~\ref{sec:caustics} to provide some basic checks on our caustic results  in Schwarzschild spacetime. 

We can in fact take the connection between the GF in Schwarzschild spacetime and the GF in PH further
and we next write the former as a sum of $\eta$-integrals of the latter.
By inserting Eq.~(\ref{bessel-exp-cgl}) into Eq.~(\ref{sch-gret-mode})  and swapping the order of the $\ell$- and $k$-summations, 
we obtain
\be\label{eq:GR-GkPH}
G_R(x,x') = 
\frac{\theta(\eta)}
{r\cdot r'}
\sum_{k=0}^\infty (2\eta)^k\, U_k(x^A,x^{A'})\,
G_{k}^{PH}(\eta,\gamma),
\ee
where we have defined
\be \label{eq:G_k^PH}
G_{k}^{PH}(\eta,\gamma)\equiv
\sum_{\ell=0}^\infty \left(\ell+\frac12\right)P_\ell(\cos\gamma) \frac{J_k(\lam\eta)}{\lam^k},
\ee 
and
the prefactor $\theta(-\sigma)\theta(\Delta t)=\theta(\eta)$ has been
taken outside as a common factor in \eqref{eq:GR-GkPH}.
We note that the GF in PH  is 
equal to $G_{R}^{PH}(\eta_{PH},\gamma)=\theta(-\sigma_{\mathbb{M}_2})\theta(\Delta t)G_{0}^{PH}(\eta_{PH},\gamma)$
(this is simply the identification mentioned above between GF's in Schwarzschild and in PH under $U_0=1$ and $U_k=0, k\geq1$).
In order to relate $G_{k}^{PH}$ for all $k\in \mathbb{Z}_{\ge 0}$ to the  GF in PH, we use Eq.~(10.6.6) in~\cite{NIST:DLMF} to first obtain:
\begin{align}
&
\frac{1}{\lam^{k}}
\left(\frac{1}{\eta}\frac{d}{d\eta}\right)^k\left(\eta^kJ_k(\lam\eta)\right)=J_0(\lam\eta),
\\ &
\frac{1}{\eta}\frac{d}{d\eta}\left(\eta^kJ_k(\lam\eta)\right)=L\ \eta^{k-1}J_{k-1}(\lam\eta).
\end{align} 
From these, it follows that
\be
\frac{J_k(L\eta)}{L^k}=\eta^{-k}\int_0^\eta d\eta'\,\left(\eta'\right)^k\frac{J_{k-1}(L\eta')}{L^{k-1}}=
\eta^{-k}\left(\int d\eta\,\eta\right)^kJ_{0}(L\eta),
\ee
where we have used the shorthand notation for the iterated integral 
\be \label{eq:iterated int}
\left(\int d\eta\, \eta\right)^k F(\eta) \equiv
\int_0^\eta d\eta_1\, \eta_1
\left(\int_0^{\eta_1} d\eta_2\,\eta_2 \left(\cdots\left(\int_0^{\eta_{k-1}} d\eta_{k}\,\eta_{k}F(\eta_{k})\right)\cdots\right)\right),
\ee
for an arbitrary function $F(\eta)$ for which the integrals exist.
Since the only $\eta$-dependence of $G_{k}^{PH}$ in \eqref{eq:G_k^PH} is via the $J_k(L\eta)$, it readily follows that
\be
G_{k}^{PH}(\eta,\gamma)=\eta^{-k}\int d\eta\ \eta^k\, G_{k-1}^{PH}(\eta,\gamma)=
\eta^{-k}\left(\int d\eta\, \eta\right)^kG_{0}^{PH}(\eta,\gamma).
\ee
 
We then have:
\be\label{eq:GF Schw from PH}
G_R(x,x') = 
\frac{\theta(\eta)}
{r\cdot r'}
\sum_{k=0}^\infty 2^k\, U_k(x^A,x^{A'})\,
\left(\int d\eta\, \eta\right)^kG_{0}^{PH}(\eta,\gamma).
\ee
 Finally, we note that the GF in PH  is $G_{R}^{PH}(\eta_{PH},\gamma)=
 \theta(\eta_{PH})G_{0}^{PH}(\eta_{PH},\gamma)$.
It is clear from the limits of integration in \eqref{eq:iterated int} that $\eta>0$ is equivalent to $\eta_k>0$. Thus, we can  write the GF in Schwarzschild in terms of the GF in PH   as
 \be\label{eq:GF Schw from GF PH}
G_R(x,x') = \frac{1}{r\cdot r'}
\sum_{k=0}^\infty 2^k\, U_k(x^A,x^{A'})\,
\left(\int d\eta\, \eta\right)^kG_{R}^{PH}(\eta,\gamma).
\ee
We have thus expressed the GF in Schwarzschild as a suggestive
sum of integrals of the GF in $\mathbb{M}_2\times \mathbb{S}_2$, weighted by the 2-dimensional Hadamard biscalars $U_k$.

In the next two subsections, we will use two different results for the GF in PH 
which we obtained in~\cite{Casals:2012px} in order to make some further progress with Schwarzschild's $G_R$.
Specifically, we will use the Hadamard form for $G_{R}^{PH}$ in the first subsection  and the globally-valid singularity structure of $G_{R}^{PH}$ (obtained using \eqref{eq:G_k^PH}) in the second subsection.


\subsection{Hadamard form}\label{sec:GF from PH}

The GF  in PH also admits, of course, its own Hadamard form, valid within normal neighbourhoods of points in PH:
\be\label{eq:Had PH}
G_{R}^{PH}(\eta_{PH},\gamma)=\theta(\Delta t)\left(\Delta_{PH}^{1/2}\delta(\sigma_{PH})+V_{PH}\theta(-\sigma_{PH})\right),
\ee
where  $\sigma_{PH}=(-\eta_{PH}^2+\gamma^2)/2$, $\Delta_{PH}$ and $V_{PH}$ are, respectively, the world function, the van Vleck determinant and 
the Hadamard tail biscalar in PH.
In~\cite{Casals:2012px} we managed to calculate, in closed form, $\Delta_{PH}$ (which is actually equal to the van Vleck determinant in the two-sphere $\mathbb{S}_2$) as well as the first two terms in the Hadamard series for $V_{PH}$, while the latter biscalar was calculated numerically in \cite{AruquipaCasals} for {\it any} pair of points.
Specifically, we found that 
\be \label{udef}
\Delta_{PH}=
\frac{\gamma}{\sin\gamma},
\quad
V_{PH}(\eta_{PH},\gamma)=\sum_{n=0}^\infty \nu_n(\gamma)\sigma_{PH}^n,
\ee
with (in the case $m^2+2\xi=1/4$)
\begin{align}
\nu_0(\gamma)=\frac18 \Delta_{PH}^{1/2}\left(
\frac{1}{\gamma^2}-\frac{\cot\gamma}{\gamma}\right)\label{v0def}
\end{align}
and  
\begin{align} \label{eq:hat V1 M2xS2} 
\nu_1=
\Delta_{PH}^{1/2}\frac{2\gamma^2-3 \csc ^2(\gamma ) \left[6 \gamma ^2+2 \gamma \sin (2 \gamma )+5 \cos (2 \gamma )-5\right]}{256 \gamma ^4}.
\end{align}
The higher orders $\nu_n$, $n>0$, can in principle be obtained from $\nu_{n-1}$ via a recurrence relation. These coefficients are closely related to the Hadamard coefficients of a certain wave equation in two-dimensional anti-de Sitter spacetime AdS$_2$ (see Appendix B), and to the coefficients that arise in an expansion of the Legendre polynomials in terms of Bessel functions. This expansion is due to Szego \cite{vonSzego01011934}; see (\ref{legendre-bessel}) below. 
We next introduce the Hadamard form \eqref{eq:Had PH} and \eqref{udef} for $G_{R}^{PH}$ in \eqref{eq:GF Schw from GF PH}, after replacing $\eta_{PH}$ by $\eta$, and so $\sigma_{PH}=(-\eta_{PH}^2+\gamma^2)/2$ by $\hat\sigma=(-\eta^2+\gamma^2)/2$.
Using the integral results, valid for all $k>0$,
\be
\left(\int d\eta\, \eta\right)^k \theta(\Delta t)\delta(\hat \sigma)=
\left(\int d\eta\, \eta\right)^k \theta(\eta)\delta(\hat \sigma)=
\frac{(-1)^{k+1}\theta(\eta-\gamma)}{(k-1)!}\hat\sigma^{k-1},
\ee
and
\be
\left(\int d\eta\, \eta\right)^k \theta(\Delta t)\theta(-\hat\sigma)\hat\sigma^{n}=
\left(\int d\eta\, \eta\right)^k \theta(\eta)\hat\sigma^{n}=
\frac{(-1)^{k}\theta(\eta-\gamma)n!}{(k+n)!}\hat\sigma^{k+n},
\ee
it readily follows that, separating out the $k=0$ term,
\begin{align}\label{eq:GF Schw Had PH}
&
G_R(x,x') = 
\frac{1}{r\cdot r'}
\Bigg\{
\theta(\Delta t)U_0(x^A,x^{A'})\left(\Delta_{PH}^{1/2}\delta(\hat\sigma)+
V_{PH}(\eta,\gamma)\theta(-\hat\sigma)\right)+
\\&\left.
\theta(\eta-\gamma)
\sum_{k=1}^\infty 
(-1)^k
2^k\, U_k(x^A,x^{A'})\hat\sigma^{k-1}\,
\left(-
\frac{\Delta_{PH}^{1/2}}{(k-1)!}+
\sum_{n=0}^{\infty}
\frac{n!}{(k+n)!}
\nu_n(\gamma)\hat\sigma^{n+1}
\right)
\right\},\nonumber
\end{align}
where we note that $\theta(\eta-\gamma)=\theta(-\hat\sigma)\theta(\Delta t)$
and $V_{PH}(\eta,\gamma)=\sum_{n=0}^\infty \nu_n(\gamma)\hat\sigma^n$.
The first term $\theta(\Delta t)U_0\Delta_{PH}^{1/2}\delta(\hat\sigma)/(r\cdot r')$ in \eqref{eq:GF Schw Had PH} is, of course, just equal (see Eqs.~(8) and (17) in~\cite{CNOW:2019}) to the direct part $\U\theta(\Delta t)\delta(\s)$ in the Hadamard form for the GF in Schwarzschild.
We have thus effectively written a Hadamard form for Schwarzschild's $G_R$ where the Hadamard tail $\V$ is manifestly written as a  series in $\hat\sigma$, with coefficients that 
depend on quantities defined in {\it two}-dimensional manifolds: $\Mtwo$ in the case of $U_k$ and
$\mathbb{S}_2$ 
in the case of $\nu_n(\gamma)$.
This is a manifest advantage over the standard Hadamard series of $\V$ in $\s$ because one can thus apply available machinery for calculating Hadamard coefficients in {\it four}-dimensional Schwarzschild to these easier two-dimensional cases.
As for the calculational machinery, Hadamard coefficients may be calculated, for example, by solving  transport equations~\cite{Ottewill:2009uj}.
Alternatively, one might prefer to obtain the coefficients analytically in terms of small coordinate-distance expansions (see Ref.~\cite{CDOWb} in Schwarzschild and Ref.~\cite{CNOW:2019} for explicit expansions of $U_0$ and $\eta$ in $\Mtwo$).


\subsection{Singularity structure}\label{sec:sings from PH}

Let us here look at the singularity structure that  expression \eqref{eq:GF Schw from GF PH} for $G_R$ yields. 
Specifically, we shall calculate $G_R^{disc}$, defined as the contribution to $G_R$ for which the coefficient of $\theta(\eta)$ in \eqref{eq:GF Schw from PH} does not take the form of a continuous function on the spacetime.
We obtained the equivalent discontinuous contribution to $G_{R}^{PH}(\eta_{PH},\gamma)$ in Eq.~156~\cite{Casals:2012px}.
By carrying out the $\eta$-integrations of $G_{R}^{PH}(\eta,\gamma)$ as required in \eqref{eq:GF Schw from GF PH}, the
integrals of the terms in Eq.~156~\cite{Casals:2012px} containing
$\theta$'s and $\log$'s of the world function yield continuous terms. Therefore, the
only other discontinuity in \eqref{eq:GF Schw from GF PH} 
arises from the one integral ($k=1$) of the 
terms in Eq.~(156) in Ref.~\cite{Casals:2012px} containing $\delta$'s and PV's of the world function:
\begin{align}
\label{eq:int PV}
&
\theta(\eta)
\int_0^{\eta} d\eta'\, \eta'\, \delta\left(\hat{\sigma}_n(\eta',\gamma)\right)=\theta\left(\eta-\gamma-2\pi n\right),
\quad \forall n\in \mathbb{Z}_{\ge 0},
\\ &
\theta(\eta)\int_0^{\eta} d\eta'\, \eta'\, \text{PV}\left(\hat{\sigma}_{-n}(\eta',\gamma)\right)=
\theta(\eta)\left(
-\ln\left(\hat{\sigma}_{-n}(\eta,\gamma)\right)+\ln\left(\frac{(2\pi n-\gamma)^2}{2}\right)\right),
\quad \forall n\in \mathbb{Z}_{\ge 1},
\nonumber
\end{align}
where
$\hat{\sigma}_n=\hat{\sigma}_n(\eta,\gamma)$ is given in \eqref{sig-k-eta}
\footnote{For $n\in \mathbb{Z}_{\ge 0}$, it is $\hat{\sigma}_n=\sigma_n^{\rm{even}}$ and $\hat{\sigma}_{-n}=\sigma_n^{\rm{odd}}$, where $\sigma_n^{\rm{even/odd}}$ are defined in Eqs.~(157) and (158) of Ref.~\cite{Casals:2012px}.}.
That is, the discontinuity $G_R^{disc}$ arising from \eqref{eq:GF Schw from GF PH} is given by its $k=0$ summand as per Eq.~(156) of Ref.~\cite{Casals:2012px} together with the 
discontinuous terms in \eqref{eq:int PV} (and so not including the $2\ln\left(2\pi-\gamma\right)$ within the  $k=1$ summand); in the next Sec.~\ref{sec:remainder} we prove that the $k$-sum does not bring in any new discontinuities.
After some calculations,
we find:
\begin{align}\label{eq:GF Schw from PH discont}
&
r\cdot r'\cdot  G_R^{disc}=
\theta(\eta)\left[
\frac{U_0}{\sqrt{\eta\sin\gamma}}\sum_{n=0}^{\infty}(-1)^n\delta\left(\eta-(\gamma+2\pi n)\right)
+
\frac{U_0}{\pi\sqrt{\sin\gamma}}\sum_{n=1}^{\infty}\frac{(-1)^n}{\sqrt{2\pi n-\gamma}}
\text{PV}\left(\frac{1}{\eta+(\gamma-2\pi n)}\right)
+
\right.
\nonumber \\ &
\frac{1}{8\pi\sqrt{\sin\gamma}}\sum_{n=1}^{\infty}\frac{(-1)^n}{\sqrt{2\pi n-\gamma}}\left(
U_0\left(\frac{1}{2\pi n-\gamma}+\cot\gamma\right)+16(2\pi n-\gamma)U_1
\right)\ln\left((2\pi n-\gamma)-\eta\right)+
\nonumber \\ &
\left.
\frac{1}{8\sqrt{\sin\gamma}}
\sum_{n=0}^{\infty}\frac{(-1)^n}{\sqrt{2\pi n+\gamma}}
\left(
U_0\left(\frac{1}{2\pi n+\gamma}-\cot\gamma\right)+
16(\gamma+2\pi n)U_1
\right)
\theta(\eta-(\gamma+2\pi n))
\right].
\end{align}
This expression for the discontinuities in $G_R$ should be compared against the terms in (\ref{grl-k-zero2}) and (\ref{grl-k-one1}) below.
By focusing on the behaviour at the discontinuities  and so allowing  the evaluation at the discontinuities themselves of coefficients of the discontinuities (i.e., as indicated below Eq.~(154) in Ref.~\cite{Casals:2012px}),
it can be easily checked that the asymptotics at the discontinuities given by \eqref{eq:GF Schw from PH discont} agree with those given by  (\ref{grl-k-zero2}) and (\ref{grl-k-one1}).




\section{Global analysis of the Schwarzschild Green function  using a large-$\ell$ expansion}\label{sec:Sum Hads}

In this section, we will use an asymptotic expansion for Bessel functions to obtain a representation for the retarded Green function of Schwarzschild spacetime given in (\ref{sch-gret-mode}). Bessel functions enter via Zauderer's theorem as expressed in (\ref{bessel-exp-cgl}), and via a representation of the Legendre polynomials as sums of Bessel functions (see Eq.(\ref{legendre-bessel}) below). We then sum over the mode index $\ell$ to identify the global singularity structure of $G_R$, and to determine the non-smooth contribution to $G_R$ as a sum over Hadamard forms. Here, the distinction between smooth and non-smooth is made modulo the presence of an overall factor of $\theta(-\sigma)\theta(\Delta t)$ in $G_R$: that is, the distinction between smooth, non-smooth and singular behaviour is made with respect to the coefficient of this term. The results of this section are contingent on convergence of the background Hadamard series (\ref{eq:Had series 2d}), and so apply on the region $\Omega_p$ of $\Mtwo$ on which convergence holds.

\subsection{The large$-\ell$ expansion}\label{sec:remainder}

The large$-\ell$ expansion of Eq.~(\ref{sch-gret-mode}) is underpinned by the following large-argument expansion for Bessel functions (Eq.~(8.451) of Ref.~\cite{GradRyz}):
\be J_k(z) = {\sqrt{\frac{2}{\pi z}}}\left[\sum_{m=0}^N E_m\left(z-\frac{\pi}{2}k-\frac{\pi}{4}\right)\frac{a_{k,m}}{(2z)^m} + R_{k,N}(z)\right],
\quad |\arg(z)|<\pi,
\label{bessel-large-ell}\ee 
where $N\in \mathbb{N}$ is odd,
and
\be
E_m(x) \equiv \frac{e^{im\pi/2}}{2}\left(e^{ix}+(-1)^me^{-ix}\right) = \cos\left(x+\frac{\pi}{2}m\right),\quad a_{k,m} \equiv \frac{\Gamma\left(k+m+\frac12\right)}{m!\Gamma\left(k-m+\frac12\right)},\label{ek-akm-def}
\ee
and the remainder term has the form 
\be 
R_{k,N}(z) = \cos\left(z-\frac{\pi}{2}k-\frac{\pi}{4}\right)R_1-\sin\left(z-\frac{\pi}{2}k-\frac{\pi}{4}\right) R_2, \label{RN-def}
\ee
with the bounds
\begin{eqnarray}
|R_1| & < & \left|\frac{\Gamma\left(k+N+\frac32\right)}{(2z)^{N+1}(N+1)!\Gamma\left(k-N-\frac12\right)}\right|,\quad N>k-\frac32,\label{R1-bound} \\
|R_2| & < & \left| \frac{\Gamma\left(k+N+\frac52\right)}{(2z)^{N+2}(N+2)!\Gamma\left(k-2N-\frac32\right)}\right|,\quad N\geq k-\frac52.\label{R2-bound} 
\end{eqnarray}
A key feature of the expansion \eqref{bessel-large-ell} is that it does not apply to all orders: for each positive integer choice of $N$ in \eqref{bessel-large-ell}, we must cut off the sum in \eqref{bessel-exp-cgl} 
at the appropriate value of $k$ as determined by Eqs.~\eqref{R1-bound} and \eqref{R2-bound}.
We note that, in Eq.~(\ref{bessel-exp-cgl}), the argument is $z=L\eta$, and so large $\ell=L-\frac12$ corresponds to large $z$ (given a fixed $\eta>0$). 

Using Eq.~(\ref{bessel-large-ell}), we can write down a corresponding equation for the $2$-D Green function, based on Eq.~(\ref{bessel-exp-cgl}). This equation holds for any odd $N\geq 1$, and we extract features of the Green function by taking a limit $N\to\infty$. We write
\be \cgl(x^A,x^{A'}) = {\cgl}^{(N)}+\check{R}_{\ell,N}+{\cgl}^{(N,\infty)}, \label{eq:Ul-decompose}\ee
where
\begin{eqnarray}
{\cgl}^{(N)} &\equiv &  \sum_{j=0}^{N} \frac{(2\eta)^jU_j}{L^j}\sqrt{\frac{2}{\pi\lam\eta}}\left(\sum_{m=0}^N E_m\left(L\eta-\frac{\pi}{2}j-\frac{\pi}{4}\right)\frac{a_{j,m}}{(2L\eta)^m}\right), \label{def:finite-sum} \\
\check{R}_{\ell,N}
&\equiv& \sum_{j=0}^{N}\frac{(2\eta)^jU_j}{L^j}\sqrt{\frac{2}{\pi\lam\eta}}R_{j,N}(L\eta),\label{def:remainder}\\
{\cgl}^{(N,\infty)}&\equiv& \sum_{j=N+1}^\infty \frac{(2\eta)^jU_j}{L^j}J_j(L\eta).
\label{def:infinite-tail}
\end{eqnarray}
In the discussion below, we will refer to ${\cgl}^{(N)}$ as the \textit{finite sum}, to 
$\check{R}_{\ell,N}$ as the \textit{remainder} and to ${\cgl}^{(N,\infty)}$ as the \textit{infinite tail}.

Our next step is to rewrite these three terms as sums of ascending powers of $L^{-1}$, and to then sum over $\ell$ to determine the contribution of each term to Eq.~(\ref{sch-gret-mode}).  Our key conclusion is that we can take the limit $N\to\infty$ to obtain the following results: in the limit, only the finite sum ${\cgl}^{(N)}$ contributes to the singular and non-smooth parts of $G_R(x,x')$, the remainder term $\check{R}_{\ell,N}$ contributes only to the smooth part and the infinite tail ${\cgl}^{(N,\infty)}$ does not contribute to the GF.

First, we consider the infinite tail contribution to (\ref{sch-gret-mode}). For any finite $N$, this term is continuous. We apply the following large order asymptotic relation for Bessel functions~\cite{NIST:DLMF}:
\be J_k(z) \sim \frac{1}{k!}\left(\frac{z}{2}\right)^k,\quad k\to\infty. \label{bessel-large-order} \ee
Then for large $N$, the infinite tail has the following asymptotic behaviour:
\be
\cgl^{(N,\infty)}\equiv\sum_{j=N+1}^\infty \frac{(2\eta)^jU_j}{L^j}J_j(L\eta) \sim {\cal{G}}^{(N,\infty)},\quad N\to \infty,
\label{eq:infinite-tail-asymptotics}\ee
where 
\be 
{\cal{G}}^{(N,\infty)} \equiv \sum_{j=N+1}^\infty U_j\frac{\eta^{2j}}{j!}.
\label{infinite-tail-limit}
\ee
From (\ref{eq:Had form 2d eta}), we recognise this as the tail of the infinite series which generates the background Green function $G$: note that the approximation is independent of $L$. 
It follows that, in the region where the Hadamard series Eq.~(\ref{eq:Had form 2d eta}) converges, the infinite tail satisfies
\be
\cgl^{(N,\infty)}(x^A,x^{A'}) \to 0, \quad N\to\infty, 
\label{infinite-tail-limit}
\ee
and that for each $x^A$, the convergence is uniform in $L$ and $x^{A'}$.  
Summing over $\ell$ yields the contribution of the infinite tail to $G_R$ in Eq.~(\ref{sch-gret-mode}). This corresponds to 
\begin{eqnarray}
\sum_{\ell=0}^\infty \left(\ell+\frac12\right)\cgl^{(N,\infty)}P_\ell(\cos\gam) &\sim & \sum_{\ell=0}^\infty \left(\ell+\frac12\right){\cal{G}}^{(N,\infty)}P_\ell(\cos\gam),\quad N\to\infty, \label{eq:infinite-sum-in-GR}
\end{eqnarray}
and we have 
\be 
 \sum_{\ell=0}^\infty \left(\ell+\frac12\right){\cal{G}}^{(N,\infty)}P_\ell(\cos\gam)
= {\cal{G}}^{(N,\infty)}\sum_{\ell=0}^\infty \left(\ell+\frac12\right)P_\ell(\cos\gam) =
 {\cal{G}}^{(N,\infty)}\sum_{k\in\mathbb{Z}} \delta\left(\gam+2k\pi\right),
\label{infinite-tail-contrib}
\ee
where we have used the fact that ${\cal{G}}^{(N,\infty)}$ is independent of $\ell$. This contribution is a large-$N$ error term which, by Eq.~(\ref{infinite-tail-limit}), vanishes in the limit as $N\to\infty$. \textit{This term does not contribute to the physical Green function.}

Next, we consider the contribution of the remainder term (\ref{def:remainder}) to (\ref{sch-gret-mode}).

The quantities $R_{k,N}$ introduced in \eqref{RN-def} are subject to the bounds of Eqs.~(\ref{R1-bound}) and (\ref{R2-bound}), and so it follows that, for a fixed $\eta>0$, 
\be \check{R}_{\ell,N} = O\left(\ell^{-N-3/2}\right), \quad \ell\to\infty.
\label{rem-ell-bound}
\ee
The corresponding contribution to (\ref{sch-gret-mode}) then satisfies
\be \left| \sum_{\ell=0}^\infty \left(\ell+\frac12\right)\check{R}_{\ell,N} P_\ell\left(\cos\gam\right)\right| \leq \sum_{\ell=0}^\infty O\left(\ell^{-N-1}\right),
\label{rem-contrib}
\ee
where we have used $P_\ell(\cos\gamma)=O(\ell^{-1/2}), \ell\to+\infty, \gamma\in(0,\pi)$ (recall that $N\geq1$). 
Since the series in \eqref{rem-contrib} comprises a series of continuous functions, the upper bound stated allows us to apply the Weierstrass $M-$test and deduce that this contribution must be both finite and continuous. {Furthermore, by differentiating the summands on the left-hand side of (\ref{def:remainder}), applying recursion formulae for Bessel functions and Legendre polynomials along with the relations  (\ref{bessel-large-ell}), (\ref{RN-def}) and (\ref{def:remainder}), we obtain a bound equivalent to (\ref{rem-contrib}) for any first order partial derivative. (Recall that the Hadamard coefficients $U_k(x^A,x^{A'})$ are smooth functions of their arguments.) There is a loss of one power of $\ell$ on the right-hand side, so that, in general, $\ell^{-N-1}$ must be replaced by $\ell^{-N}$. This process can be repeated iteratively, and it shows that we can always choose $N$ large enough so that derivatives of the remainder terms of arbitrarily high order converge to a continuous function. Hence: \textit{the remainder term does not contribute to the non-smooth part of the retarded Green function\footnote{This is true for a fixed $\eta>0$, which, for causal separation, merely excludes the coincidence limit (i.e., $\eta, \gamma \to 0$) which is covered by the Hadamard form Eq.~\eqref{eqhada}.}.}} 

Next, we consider the contribution of the finite sum (\ref{def:finite-sum}) to the Green function (\ref{sch-gret-mode}). Our aim is to show that only this term contributes non-smooth (including singular) terms to the Green function. To see this, it is necessary to include again all contributions to $\mathcal{U}_\ell$: the finite sum, the remainder, and the infinite tail. We can do this by rewriting the asymptotic expansion of the Bessel functions in the following way: for any (odd) $N\geq 1$, we have
\be
J_k(z) = \sqrt{\frac{2}{\pi z}}\sum_{m=0}^{N+2} E_m\left(z-\frac{\pi}{2}k-\frac\pi4\right)\frac{\rma_{k,m}}{(2z)^m},\label{take2:belle-k-expansion}\ee
where $E_m$ is defined as above, $\rma_{k,m}=a_{k,m}$ for $0\leq m\leq N$, and 
\begin{eqnarray}
\rma_{k,N+1}&=& b_{k,N+1}(z)\frac{\Gamma(k+N+\frac32)}{(N+1)!\Gamma(k-N-\frac12)},\label{take2:remainder1} \\
\rma_{k,N+2}&=& b_{k,N+2}(z)\frac{\Gamma(k+N+\frac52)}{(N+2)!\Gamma(k-N-\frac32)},\label{take2:remainder2}
\end{eqnarray}
where  the quantities $b_{k,N+1}(z)$ and $b_{k,N+2}(z)$ are, for each $k$ and $N$, bounded functions of $z$ - in fact, bounded by unity (this is equivalent to the bounds on $R_1$ and $R_2$ given in (\ref{R1-bound}) and (\ref{R2-bound}) above). Then for each odd $N\geq1$, we can write
\begin{eqnarray} \mathcal{U}_\ell &=& \sqrt{\frac{2}{\pi\eta L}}\sum_{k=0}^N \left(\frac{(2\eta)^kU_k}{L^k}\sum_{m=0}^{N+2} E_m\left(L\eta-\frac\pi2 k-\frac\pi4\right)\frac{\rma_{k,m}}{(2\eta)^m}L^{-m}\right) + \mathcal{U}_{\ell}^{(N,\infty)} \nonumber \\
&=& \sqrt{\frac{2}{\pi\eta L}}\left(\sum_{k=0}^{N} E_k\left(L\eta-\frac{\pi}{4}\right)\hat{\alpha}_k(x^A,x^{A'})L^{-k} + \sum_{k=N+1}^{2N+2}\bar{\alpha}_k(x^A,x^{A'};L)L^{-k}\right) + \mathcal{U}_{\ell}^{(N,\infty)}
\label{take2:ul-sum-tail}
\end{eqnarray}
where the coefficients ${\hat{\alpha}}_k(x^A,x^{A'})$, $k=0,\dots,N$  and $\bar{\alpha}_k(x^A,x^{A'};L), k=N+1,\dots,2N+2$ arise by formally collecting like powers of $L$, treating $b_{k,N+1}(L\eta)$ and $b_{k,N+2}(L\eta)$ as though they were independent of $L$. This procedure is well-defined, and yields coefficients $\bar{\alpha}_k(x^A,x^{A'};L)$ that are ``weakly" dependent on $L$: these coefficients have the form of linear combinations of functions of $\eta$ multiplied by the zero order (i.e.\ $O(L^0)=O(1)$) terms $b_{k,N+1}, b_{k,N+2}$. In particular, these cannot contribute positive integer powers of $L$  that would ``contaminate" the coefficients ${\hat{\alpha}}_k$ for $k\leq N$: that is, these coefficients are completely determined by $\rma_{m,k}=a_{m,k}$:
\be \hat{\alpha}_k\left(x^A,x^{A'}\right)=\sum_{m=0}^k (-1)^ma_{m,k-m}\frac{U_m(x^A,x^{A'})}{(2\eta)^{k-2m}}.\label{alpha-hat-def1}
\ee
Note that these coefficients are independent of $L$. Recall that the functions $E_k$ are trigonometric, and so are $O(1)$. As above, the term $\mathcal{U}_{\ell}^{(N,\infty)}$ in (\ref{take2:ul-sum-tail}) is the infinite tail contribution (\ref{def:infinite-tail}).

We can derive a corresponding expansion for the Legendre polynomials, based on the following expansion in terms of Bessel functions (see~\cite{vonSzego01011934} and volume 2, page 58, Eq.~(15) of Ref.~\cite{Erdelyi:1953}):
\be P_\ell(\cos\gamma) =\left(\frac{\gamma}{\sin\gamma}\right)^{1/2} \sum_{j=0}^\infty V_j(\gamma) \frac{J_j(\rlam\gamma)}{\rlam^j},\label{legendre-bessel}\ee
where the coefficients $V_j(\gamma)$ are elementary functions which are regular for $\gamma\in[0,\pi)$. See Appendix \ref{sec:M=0} for more details. 
In particular,
\be V_0(\gamma)=1,\qquad V_1(\gamma)=\frac18\left(\cot\gamma-\frac{1}{\gamma}\right).\label{al-0-1}\ee 
The series (\ref{legendre-bessel}) is uniformly convergent in any interval of the form $[0,\gamma_0-\epsilon]$ with $\epsilon>0$ and $\gamma_0=2(\sqrt{2}-1)\pi\simeq 0.828\pi$ \cite{vonSzego01011934}. Note, in particular, that the expansion is not valid at $\gamma=\pi$. An alternative approach to the calculation of the coefficients $V_k$ is given in Appendix \ref{sec:M=0}. 

Using the representation (\ref{take2:belle-k-expansion}) for $J_k(\gamma)$, we can write 
\be P_\ell(\cos\gamma) =\left( \frac{\gamma}{\sin\gamma}\right)^{1/2}\sqrt{\frac{2}{\pi L\gamma}}\left(\sum_{k=0}^{N} E_k\left(L\gamma-\frac{\pi}{4}\right)\hat{\beta}_k(\gamma)L^{-k} + \sum_{k=N+1}^{2N+2} \bar{\beta}_k(\gamma;L)L^{-k}\right) + P_{\ell}^{(N,\infty)} 
\label{take2:pl-sum-tail}
\ee
where the $\bar{\beta}_k(\gamma;L)$ are weakly dependent on $L$ in the sense defined above and 
\be {\hat{\beta}}_j(\gamma)=\sum_{m=0}^j (-1)^ma_{m,j-m}\frac{V_m(\gamma)}{(2\gamma)^{j-m}}.\label{be-hat-def1}\ee
The infinite tail term $P_{\ell}^{(N,\infty)}$ is defined analogously to (\ref{def:infinite-tail}):
\be P_{\ell}^{(N,\infty)} \equiv 
\left(\frac{\gamma}{\sin\gamma}\right)^{1/2} \sum_{j=N+1}^\infty V_j(\gamma) \frac{J_j(\rlam\gamma)}{\rlam^j}.\label{legendre-infinite-tail}\ee

Note that both $\mathcal{U}_{\ell}^{(N,\infty)}$ and $P_{\ell}^{(N,\infty)}$ are zero in the limit as $N\to \infty$: these terms correspond to tails of infinite series that converge in the domain under consideration. That is,
\begin{eqnarray}
\lim_{N\to\infty} \sqrt{\frac{2}{\pi\eta L}}\left(\sum_{k=0}^{N} E_k\left(L\eta-\frac{\pi}{4}\right){{\hat{\alpha}}}_k(x^A,x^{A'})L^{-k} + \sum_{k=N+1}^{2N+2}\bar{\alpha}_k(x^A,x^{A'};L)L^{-k}\right)&=& \mathcal{U}_\ell,\label{take2:ulim} \\
\lim_{N\to\infty} \left( \frac{\gamma}{\sin\gamma}\right)^{1/2} \sqrt{\frac{2}{\pi L\gamma}}\left(\sum_{k=0}^{N} E_k\left(L\gamma-\frac{\pi}{4}\right)\hat{\beta}_k(\gamma)L^{-k} + \sum_{k=N+1}^{2N+2} \bar{\beta}_k(\gamma;L)L^{-k}\right)&=& {P}_\ell.\label{take2:plim} 
\end{eqnarray}

Multiplying and collecting inverse powers of $L$ (again, in a well-defined manner), we can write 
\begin{eqnarray}
\left(\ell+\frac12\right)\mathcal{U}_\ell P_\ell
&=&
\frac{1}{\pi\sqrt{\eta\sin\gamma}}
\times \nonumber \\
&& \left[
\sum_{k=0}^{2N} 
\left( 
\sum_{j=\rm{max}\{0,k-N\}}^{\rm{min}\{k,N\}} \left(E_k\left(L\eta+L\gamma-\frac{\pi}{2}\right)+(-1)^{j-k}E_k\left(L\eta-L\gamma\right)\right)\hat{\alpha}_j\hat{\beta}_{k-j}\right)L^{-k}
+ O(L^{-N-1})\right]
\nonumber \\
&=&
\frac{1}{\pi\sqrt{\eta\sin\gamma}}
\times \nonumber \\
&& \left[
\sum_{k=0}^{N} 
\left( 
\sum_{j=0}^{k} \left(E_k\left(L\eta+L\gamma-\frac{\pi}{2}\right)+(-1)^{j-k}E_k\left(L\eta-L\gamma\right)\right)\hat{\alpha}_j\hat{\beta}_{k-j}\right)L^{-k}
+ O(L^{-N-1})\right], \label{take2:UlPlV2}
\end{eqnarray}
where (we recall) $N$ is an odd integer with $N\geq 1$. 
The functional dependence of the sinusoidal functions $E_k$ on terms proportional to $\eta\pm\gamma$ in \eqref{take2:UlPlV2} is worth noting. These combinations {in the arguments of the $E_k$}  are the seeds that will later give rise to the functional dependence of the GF on {the global world function} $\hat\sigma_k$ {(see \eqref{sig-k-eta})} in the expressions for the singularity structure and the expression as a sum of Hadamard forms (see Eq.~\eqref{eq:sum Hads}  below).

We \textit{cannot} at this point take the limit $N\to\infty$  and neglect the $O(L^{-N-1})$ term in \eqref{take2:UlPlV2}. However, as $N$ can be arbitrarily large, when we sum {the $O(L^{-N-1})$ term} over $\ell$, as we will see below, we obtain a smooth (i.e.\ $C^\infty$) contribution to the full Green function. The sum to $N$ terms in (\ref{take2:UlPlV2}) yields the full non-smooth contribution to the Green function, including the singular contribution. 
We proceed to calculate this non-smooth contribution (and show that the contribution from the remainder term in (\ref{take2:UlPlV2}) must be smooth (see the paragraph below following Eq. \eqref{vk-minus-def}). 

{As already mentioned,} the function $\cgl(x^A,x^{A'})$ in Eq.~(\ref{bessel-exp-cgl})
 is the Riemann function for the $2$-D wave equation \eqref{cgl-eqn}. This is a smooth function on $\Mtwo$, and so the anticipated non-smoothness in Eq.~(\ref{sch-gret-mode}) is due to the infinite sum: any cut-off at a finite $\ell=\ell_{\rm{max}}$ would yield a smooth quantity. Thus the non-smooth nature of $G_R(x,x')$ arises from the large$-\ell$ behaviour of Eq.~(\ref{sch-gret-mode}), and so we used a large$-\ell$ expansion above to determine the non-smooth contributions. 

The large-$\ell$ expansion requires that we first subtract the $\ell=0$ contribution to $G_R$, and so we define
\be \grl(x,x') \equiv \frac{1}{r\cdot r'}\sum_{\ell=1}^\infty \left(\ell+\frac12\right)\cgl(x^A,x^{A'})P_\ell(\cos\gamma),\label{sch-grl-mode}\ee
so that
\be G_R =\theta(-\sigma)\theta(\Delta t)\left( \frac{\cgo}{2 r\cdot r'}+\grl\right). 
\label{eq:Gl=0+Gl>0}
\ee
From the comments above, the only discontinuity in the first term on the right is due to the presence of the Heaviside distributions
 $\theta(-\sigma)$
 and $\theta(\Delta t)$. So our aim now is to expand the summand in Eq.\eqref{sch-grl-mode} for $\grl$ in inverse powers of $\ell$ and to identify the non-smooth contributions. 
 We note that presence of the causal terms $\theta(-\sigma)\theta(\Delta t)$ in (\ref{eq:Gl=0+Gl>0}) implies that $G_R$ is zero for $\eta(x^A,x^{A'})<0$. Thus in the remainder of the paper we take $\eta\geq0$, and so we set  $\theta(-\sigma)\theta(\Delta t)=\theta(\eta)=1$.
 
 Next, we define the contribution to $\grl(x,x')$ that arises by including the finite sum to $N$ terms in (\ref{take2:UlPlV2}), and then take the limit as $N\to\infty$. This results in the following expression (with the subscript ``$NS$" for non-smooth):

\be \grlns = \frac{2}{\pi r\cdot r'}\frac{1}{\sqrt{\eta\sin\gamma}}\sum_{\ell=1}^\infty \left( \sum_{k=0}^\infty \frac{\hat{\nu}_k}{\ell^k}\right),
\quad \gamma\in (0,\gamma_0),
\label{grl-sum1}\ee
where
\begin{eqnarray}
\hat{\nu}_k &\equiv& \sum_{j=0}^k\frac{(-1)^{k-j}}{2^{k-j}}\left(\begin{array}{c}k-1\\k-j\end{array}\right)\nu_j,\label{nu-hat-def}\\
\nu_j &\equiv& E_j\left(\gamma\rlam+\eta\rlam-\frac{\pi}{2}\right)\nu_j^{(+)}(x^A,x^{A'},\gamma)+E_j\left(\gamma\rlam-\eta\rlam\right)\nu_j^{(-)}(x^A,x^{A'},\gamma),\label{nu-def}\\
\nu_j^{(+)}&\equiv&\frac12\sum_{k=0}^j{\hat{\alpha}}_{j-k}(x^A,x^{A'}){{\hat{\beta}}}_k(\gamma),\label{nu-plus-def}\\
\nu_j^{(-)}&\equiv&\frac12\sum_{k=0}^j(-1)^k{\hat{\alpha}}_{j-k}(x^A,x^{A'}){{\hat{\beta}}}_k(\gamma),\label{nu-minus-def}
\end{eqnarray}
and where $\hat{\alpha}_k$ and $\hat{\beta}_k$ are defined in (\ref{alpha-hat-def1}) and (\ref{be-hat-def1}) respectively. We note that in (\ref{nu-hat-def}), we use the binomial coefficients
\be \left(\begin{array}{c}-1\\0\end{array}\right)=\left(\begin{array}{c}0\\0\end{array}\right)=1,\quad
\left(\begin{array}{c}j\\k\end{array}\right)=0 \quad \hbox{for all} \quad k>j\geq0.\label{binom-alt}\ee

In (\ref{grl-sum1}), the summation index $\ell$ appears only as an inverse power and in the phase functions $E_j$. This leads to the next step, which involves collecting like phases and then summing over $\ell$:
\be \grlns = \frac{2}{\pi r\cdot r'}\frac{1}{\sqrt{\eta\sin\gamma}}\sum_{k=0}^\infty\left[\ca_k(\gamma+\eta)V_k^{(+)}+\bar{\ca}_k(\gamma+\eta){\bar{V}}_k^{(+)}+\ca_k(\gamma-\eta)V_k^{(-)}+\bar{\ca}_k(\gamma-\eta){\bar{V}}_k^{(-)}
\right],\label{grl-sum2}
\ee
where
\begin{eqnarray} \ca_k(x) &\equiv& \sum_{\ell=1}^\infty \frac{e^{i\ell x}}{\ell^k},\label{ca-def}\\
 V_k^{(+)} &\equiv& e^{i(\gamma+\eta)/2} \sum_{m=0}^k \frac{e^{i\pi(2k-m-1)/2}}{2^{k-m+1}}\left(\begin{array}{c}k-1\\k-m\end{array}\right)\nu_m^{(+)},\label{vk-plus-def}\\
V_k^{(-)} &\equiv& e^{i(\gamma-\eta)/2} \sum_{m=0}^k \frac{e^{i\pi(2k-m)/2}}{2^{k-m+1}}\left(\begin{array}{c}k-1\\k-m\end{array}\right)\nu_m^{(-)}.\label{vk-minus-def}
\end{eqnarray}
 We note that the $V_k^{(\pm)}$ are smooth functions on the conformal Schwarzschild spacetime which are independent of $\ell$. 

The distributions $\ca_k, k\geq 0$ play a key role in our description of (\ref{grl-sum2}) as a sum of Hadamard forms. The basis of this structure is the observation that the $\ca_k$ are periodic in their argument. From the spacetime point of view, this periodicity is reflected in structural similarities in the form of (\ref{grl-sum2}) that repeat periodically as null geodesics emerging from the base point $x$ execute multiple orbits around the black hole. The forms of the different $\ca_k, k\geq 0$ are also crucial in identifying the singular and non-smooth contribution to $G_R$. We note that inclusion of the remainder term from (\ref{take2:UlPlV2}) yields terms with coefficients of the form $\mathcal{A}_N$, with $N$ arbitrarily large. The sum over $\ell$ produces smooth functions - essentially trigonometric functions with arguments $\eta\pm\gamma$. This establishes the fact that $\grlns$ includes all non-smooth contributions to the retarded Green function, and that the remainder term (\ref{def:remainder}) ultimately contributes a smooth term to $G_R$.

\subsection{Structure of the $\ca_k$.}\label{sec:ca_k}
We write
\be \ca_k(x) = \cc_k(x) + i\cs_k(x),\label{A-trig-split}\ee
where
\be \cc_k(x) \equiv \sum_{\ell=1}^\infty \frac{\cos\left(\ell x\right)}{\ell^k},\quad k\geq 0\label{ck-def}\ee
and
\be \cs_k(x) \equiv\sum_{\ell=1}^\infty \frac{\sin\left(\ell x\right)}{\ell^k},\quad k\geq 0.\label{sk-def}\ee
We note the distributional result
\be \label{eq:rec rln ca} \ca_k'(x) = i\ca_{k-1}(x),\quad k\geq 1, \ee
and correspondingly
\be \ca_k(x) = \zeta(k) + i\int_0^x\ca_{k-1}(y)dy,\quad k\geq 2,\ee
where $\zeta(k)=\ca_k(0)$ is the Riemann zeta function, and so we have the regularity results
\be \ca_k \in C^{k-2}(\mathbb{R}),\quad \ca_k^{(k-1)}\in L^1_{\rm{loc}}(\mathbb{R}),\quad k\geq2.\label{ca-reg}\ee

For $k\geq 1$, we have 
\be\label{eq:Csk} \cc_{2k}(x) = \check{\cb}_{2k}\left(\frac{x}{2\pi}\right), \quad 0\leq x\leq 2\pi,
\ee
and
\be \cs_{2k+1}(x) =\hat{\cb}_{2k+1}\left(\frac{x}{2\pi}\right),\quad 0\leq x\leq 2\pi,
\ee
where
\be
 \check{\cb}_{2k}(x)\equiv
\frac{(-1)^{k-1}}{2}\frac{(2\pi)^{2k}}{(2k)!}\cb_{2k}(x),\quad x\in \mathbb{R},\label{ck-even}
\ee
\be
\hat{\cb}_{2k+1}(x)\equiv \frac{(-1)^{k-1}}{2}\frac{(2\pi)^{2k+1}}{(2k+1)!}\cb_{2k+1}\left(x\right),\quad x\in \mathbb{R},\label{sk-odd}\ee
and where $\cb_n(x), n\geq 1$ is the $n^{th}$ Bernoulli polynomial~\cite{NIST:DLMF}. It follows from Fourier theory that for $k\geq 2$ even, $\cc_k(x), x\in\mathbb{R}$ is the periodic continuation to the real line of the corresponding Bernoulli polynomial, and likewise for $\cs_k(x), x\in\mathbb{R}$ with $k\geq 3$ odd. To see this, we consider the following theorem (see e.g.\ Theorem 14.29 of~\cite{wadeintroduction}).

\begin{theorem}
If $f:\mathbb{R}\to\mathbb{R}$ is periodic on $\mathbb{R}$ with period $2\pi$, has bounded variation on $[0,2\pi]$ and is continuous on the closed interval $I$, then the Fourier series of $f$ converges uniformly to $f$ on $I$. 
\end{theorem} 

Applying Theorem 1 then allows us to write $\cc_{2k}$ and $\cs_{2k+1}$ as the periodic continuation of the polynomials appearing in (\ref{ck-even}) and (\ref{sk-odd}) respectively:
\be \cc_{2k}(x)=
\sum_{\ell=1}^\infty \frac{\cos\left(\ell x\right)}{\ell^{2k}}=
 \sum_{n\in\mathbb{Z}} \theta\left(x-2n\pi)\theta(2(n+1)\pi-x\right)\check{\cb}_{2k}\left(\frac{x-2n\pi}{2\pi}\right),\quad k\geq 1, x\in\mathbb{R},\label{cos-even-had-sum}\ee
and
\be \cs_{2k+1}(x)=
\sum_{\ell=1}^\infty \frac{\sin\left(\ell x\right)}{\ell^{2k+1}}=
 \sum_{n\in\mathbb{Z}} \theta(x-2n\pi)\theta\left(2(n+1)\pi-x\right)\hat{\cb}_{2k+1}\left(\frac{x-2n\pi}{2\pi}\right),\quad k\geq 1, x\in\mathbb{R}.\label{sin-odd-had-sum}\ee

This accounts for approximately half of the terms in (\ref{grl-sum2}). To deal with the terms $\cc_{2k+1}$ and $\cs_{2k}, k\geq 1$, we must first recall some facts established in \cite{Casals:2012px} regarding $\ca_0$ and $\ca_1$.

In \cite{Casals:2012px}, we showed that (using the notation of (\ref{A-trig-split}) and with $\mathbb{Z}_0=\mathbb{Z}\setminus\{0\}$)
\begin{eqnarray} \cc_1(x) &=& -\ln|x|- \sum_{n\in\mathbb{Z}_0}^\infty\ln\left|1-\frac{x}{2n\pi}\right|,\label{c1-form}\\
\cs_1(x) &=& -\frac{x}{2}+\pi\sum_{n\in\mathbb{Z}} \left[\theta(x-2n\pi)-\frac12\right],\label{s1-form}
\end{eqnarray}
with the distributional derivatives
\begin{eqnarray}
\cc_0(x) &=& \cs_1'(x) = -\frac12+\pi\sum_{n\in\mathbb{Z}}\delta(x-2n\pi) \label{c0-form}\\
\cs_0(x) &=& -\cc_1'(x) = \sum_{n\in\mathbb{Z}}\PV\left(\frac{1}{x-2n\pi}\right). \label{s0-form}
\end{eqnarray}

We define 
\be \cw_k(x) \equiv \left\{ \begin{array}{ll} i\cc_k(x),& k \hbox{ odd}; \\ \cs_k(x), & k \hbox{ even}, \end{array}\right. \quad k\geq 1, x\in\mathbb{R}. \label{wk-def}\ee
Then
\be \cw_k'(x) = -i\cw_{k-1}(x),\quad k\geq 2, x\in\mathbb{R}, \label{wk-deriv-recur}\ee
and so 
\be 
\cw_k(x) = \cw_k(0) -i\int_0^x \cw_{k-1}(t) dt,\quad k\geq 2, x\in\mathbb{R}.\label{wk-int-recur}\ee
We note the values
\be
\cw_k(0) = \left\{ \begin{array}{ll} i\zeta(k),& k \hbox{ odd}; \\ 0, & k \hbox{ even}. \end{array}\right. \label{w0}
\ee
Integrating (\ref{wk-deriv-recur}) with $k=2$ using (\ref{c1-form}) then yields
\be \cw_2(x) = x(1-\ln|x|)+\sum_{n\in\zo}\left[x+(2n\pi-x)\ln\left|1-\frac{x}{2n\pi}\right|\right]\in C^0(\mathbb{R}),\label{w2-form}\ee
where we use $\zo=\{n\in \mathbb{Z}: n\neq 0\}$. Integrating repeatedly yields the form below for $\cw_k, k\geq 2$. We state the result as a proposition: this is proven by showing that the given functions satisfy the sequence of initial value problems (IVPs)
\be \cw_k'(x) = -i\cw_{k-1}(x),\quad x\in\mathbb{R},\quad \cw_k(0) = \left\{ \begin{array}{ll} i\zeta(k),& k \hbox{ odd}; \\ 0, & k \hbox{ even}. \end{array}\right.\quad k\geq 3. \label{w-ivp}
\ee
It follows by an inductive argument that each IVP in this sequence has a unique solution (essentially by virtue of the continuity of the sequence of right hand sides that emerges): this solution - which takes the form stated in Proposition 1 - must be the required function $\cw_k$. 

\begin{proposition}
Let $\cw_k(x), k\geq 2$ be as defined in (\ref{wk-def}). Then 
\begin{eqnarray}
\cw_k(x) &=& \underbrace{{}_{0\!}Q_k(x) +A_kx^{k-1}\ln|x|}_{n=0} +\sum_{n\in\zo}\left[ {}_{n\!}Q_k(x) + A_k(x-2n\pi)^{k-1}\ln\left|1-\frac{x}{2n\pi}\right|\right],
\label{wk-log-sum}
\end{eqnarray}
where $A_k$ are constants and 
\begin{eqnarray} {}_{0\!}Q_k(x) &=& b_{k,0}+b_{k,1}x+\cdots+b_{k,k-1}x^{k-1}, \label{q-poly-def}\\
{}_{n\!}Q_k(x) &=& {}_{n\!}B_{k,0}+{}_{n\!}B_{k,1}(x-2n\pi)+\cdots+{}_{n\!}B_{k,k-1}(x-2n\pi)^{k-1},\quad n\in\zo \label{Q-poly-def}\end{eqnarray} 
are polynomials of degree $k-1$. The constant and the polynomial coefficients for the $n=0$ contribution are determined by the relations (valid for $k\geq 2$)
\begin{eqnarray} 
A_k &=& \frac{(-i)^k}{(k-1)!},\nonumber\\
b_{k,j} &=& \frac{(-i)^j}{j!}\cw_{k-j}(0),\quad 0\leq j \leq k-2, \nonumber \\
b_{k,k-1} &=& - \frac{(-i)^k}{(k-1)!}(\gamma+\psi(k)), \label{n-zero-coeffts}
\end{eqnarray}
where $\gamma$ is the Euler-Mascheroni constant and $\psi(k)=\Gamma'(k)/\Gamma(k)$ is the digamma function.
The  polynomial coefficients for the $n\in\zo$ contributions are determined by 
\begin{eqnarray}
{}_{n\!}B_{k,k-1} &=& - \frac{(-i)^k}{(k-1)!}(\gamma+\psi(k)), \quad k\geq 3, \nonumber \\
{}_{n\!}B_{k,j} &=& \frac{(-i)^j}{j!}{}_{n\!}B_{k-j,0},\quad 0<j\leq k-2,\nonumber \\
{}_{n\!}B_{k,0}&=& - \sum_{j=1}^{k-1} (-2n\pi)^j{}_{n\!}B_{k,j}, \quad k\geq 3,
\end{eqnarray} 
with the initial values (read off from (\ref{w2-form}))
\be {}_{n\!}B_{2,0} = 2n\pi,\quad {}_{n\!}B_{2,1}=1. \ee
\end{proposition}

Note that, as anticipated, the functions $\cw_k(x), k\geq 2$ are smooth (i.e.\ infinitely differentiable) everywhere except at points of the form $x=2n\pi, n\in\mathbb{Z}$. The term labelled ``$n=0$" is non-smooth at $x=0$. 

By way of summary, we note that the formulae (\ref{cos-even-had-sum})-(\ref{s0-form}) and (\ref{wk-log-sum}) provide, via \eqref{A-trig-split}, the complete description of the functions/distributions $\ca_k$ that is required to determine (a) singularity structure of the GF and (b) the sum-of-Hadamard forms representation of GF. We turn now to the first of these.  

\subsection{The singularity of the GF}\label{Singular GF}
The singular contribution to $\grl$ arises from the $k=0$ and $k=1$ terms of $\grlns$ in (\ref{grl-sum2}). This follows from the regularity conditions (\ref{ca-reg}), and from the argument in Sec.~\ref{sec:remainder} above that the remainder term (\ref{def:remainder}) and the infinite tail (\ref{def:infinite-tail}) do not contribute to the singularities of $G_R$. The result in this subsection is valid everywhere that both the Hadamard series (\ref{eq:Had series 2d}) and Szego's series (\ref{legendre-bessel}) converge.

Collecting the relevant terms, we see that the `most' singular contribution, the $k=0$ term, is
\begin{eqnarray} \left.G_R^{\ell\geq1}\right|_{k=0} &=& \frac{1}{\pi r\cdot r'}\frac{U_0(x^A,x^{A'})}{\sqrt{\eta\sin\gamma}}\left[
\sin\left(\frac{\gamma+\eta}{2}\right)\cc_0(\gamma+\eta)+\cos\left(\frac{\gamma+\eta}{2}\right)\cs_0(\gamma+\eta)\right.\nonumber\\&& \left. + \cos\left(\frac{\gamma-\eta}{2}\right)\cc_0(\gamma-\eta) + \sin\left(\frac{\gamma-\eta}{2}\right)\cs_0(\gamma-\eta)\right].\label{grl-k-zero1}\end{eqnarray}

Some simplification is possible here using the distributional identity $f(x)\delta(x-a)=f(a)\delta(x-a)$. This yields
\begin{eqnarray} \sin\left(\frac{\gamma+\eta}{2}\right)\cc_0(\gamma+\eta) &=& \sin\left(\frac{\gamma+\eta}{2}\right)\left[-\frac12+\pi\sum_{n\in\mathbb{Z}}\delta(\gamma+\eta-2n\pi)\right]\nonumber\\
&=& -\frac12\sin\left(\frac{\gamma+\eta}{2}\right),
\end{eqnarray}
and
\begin{eqnarray}
\cos\left(\frac{\gamma-\eta}{2}\right)\cc_0(\gamma-\eta) &=& \cos\left(\frac{\gamma-\eta}{2}\right)\left[-\frac12+\pi\sum_{n\in\mathbb{Z}}\delta(\gamma-\eta-2n\pi)\right]\nonumber\\
&=& -\frac12\cos\left(\frac{\gamma-\eta}{2}\right)+\pi\sum_{n\in\mathbb{Z}}(-1)^n\delta(\gamma-\eta-2n\pi).
\end{eqnarray}

Thus
\begin{eqnarray} \left.G_R^{\ell\geq1}\right|_{k=0} &=& \frac{1}{\pi r\cdot r'}\frac{U_0(x^A,x^{A'})}{\sqrt{\eta\sin\gamma}}\left\{
-\frac12\sin\left(\frac{\gamma+\eta}{2}\right)-\frac12\cos\left(\frac{\gamma-\eta}{2}\right)\right.
\nonumber\\&& \left.+\sum_{n\in\mathbb{Z}}\left[(-1)^n\pi\delta(\gamma-\eta-2n\pi)+ \cos\left(\frac{\gamma+\eta}{2}\right)PV\left(\frac{1}{\gamma+\eta-2n\pi}\right)+ \sin\left(\frac{\gamma-\eta}{2}\right)PV\left(\frac{1}{\gamma-\eta-2n\pi}\right)\right] \right\}.\nonumber
\\ \label{grl-k-zero2}
\end{eqnarray}
Corresponding to this equation for $k=1$ we have
\begin{eqnarray}
\left.G_R^{\ell\geq1}\right|_{k=1} &=& -\frac{1}{8\pi r\cdot r'}\frac{1}{\sqrt{\eta\sin\gamma}}
{\Bigg{\{}}\nonumber\\
&&
\VHads(\gamma)
\left[\cos\left(\frac{\gamma+\eta}{2}\right)\cc_1(\gamma+\eta)-\sin\left(\frac{\gamma+\eta}{2}\right)\cs_1(\gamma+\eta)\right]\nonumber\\
&& +
\VHads(-\gamma)
\left[\sin\left(\frac{\gamma-\eta}{2}\right)\cc_1(\gamma-\eta)-\cos\left(\frac{\gamma-\eta}{2}\right)\cs_1(\gamma-\eta)\right]{\Bigg{\}}} \\
&=& -\frac{1}{8\pi r\cdot r'}\frac{1}{\sqrt{\eta\sin\gamma}}
{\Bigg{\{}}\nonumber\\
&& 
\VHads(\gamma)
\left[-\cos\left(\frac{\gamma+\eta}{2}\right)\left(\log|\gamma+\eta|+\sum_{n\in\zo}\log\left|1-\frac{\gamma+\eta}{2n\pi}\right|\right) \right. \nonumber \\
&& \left. +\sin\left(\frac{\gamma+\eta}{2}\right)\left(\frac{\gamma+\eta}{2}-\pi\sum_{n\in\mathbb{Z}}\left[\theta(\gamma+\eta-2n\pi)-\frac12\right]\right) \right] \nonumber \\
&& 
+
\VHads(-\gamma)
\left[-\sin\left(\frac{\gamma-\eta}{2}\right)\left(\log|\gamma-\eta|+\sum_{n\in\zo}\log\left|1-\frac{\gamma-\eta}{2n\pi}\right|\right) \right. \nonumber \\
&& \left. +\cos\left(\frac{\gamma-\eta}{2}\right)\left(\frac{\gamma-\eta}{2}-\pi\sum_{n\in\mathbb{Z}}\left[\theta(\gamma-\eta-2n\pi)-\frac12\right]\right) \right] {\Bigg{\}}},
\label{grl-k-one1}
\end{eqnarray}
where
\be \label{eq:VHads}
\VHads(\gamma) \equiv \left(\frac{1}{\eta}+\cot\gamma\right)U_0+16\eta U_1.
\ee
(For ease of notation, we omit the explicit dependence of $\VHads$ on $x^A$ and $x^{A'}$: $\VHads(\gamma)=\VHads(x^A,x^{A'},\gamma)$.)
Thus, all the discontinuities of $G_R$ arise
from the terms in $\left.G_R^{\ell\geq1}\right|_{k=0}$, given by \eqref{grl-k-zero2}, and in  $\left.G_R^{\ell\geq1}\right|_{k=1}$, given by \eqref{grl-k-one1}, which are  discontinuous at values $\eta>0$. As mentioned in Sec.~\ref{sec:sings from PH}, it is easy to check that the sum of these discontinuities in   $G_R$ agrees with \eqref{eq:GF Schw from PH discont}.

We note that the $n=0$ contributions in (\ref{grl-k-zero2}) and (\ref{grl-k-one1}) correspond to the discontinuity terms of the Hadamard form of the GF in the normal neighbourhood of the base point $x$ - the familiar ``$\delta+\theta$" terms of Eq.~(\ref{eqhada}). 
For $n\geq 1$, we note the presence of the four-fold singularity structure mentioned in the introduction (see Eqs.~(\ref{4-fold}) and (\ref{4-fold,sublead})).  
Equations \eqref{grl-k-zero2} and \eqref{grl-k-one1} together provide the full singularity structure of the GF
at any points in Schwarzschild spacetime
(subject to the convergence condition mentioned at the beginning of this subsection). This excludes caustic points (where $\gamma=0,\pi$): we obtain the singularity structure at caustics in Sec.\ref{sec:caustics}.

\subsection{The sum of Hadamard forms}\label{sec:sum Hads}
To obtain the complete `sum of Hadamard forms' representation of the non-smooth contribution to the retarded Green function, we now focus our attention on the $k\geq2$ terms in (\ref{grl-sum2}). So we define
\be \left.\grlns\right|_{k\geq 2} \equiv \grlns-\left(\left.G_R^{\ell\geq1}\right|_{k=0}+\left.G_R^{\ell\geq1}\right|_{k=1}\right). \label{grl-k-2up}\ee
We can then immediately write down 
\be \left.\grlns\right|_{k\geq 2} = \frac{1}{\pi r\cdot r'}\frac{1}{\sqrt{\eta\sin\gamma}}\sum_{n\in\mathbb{Z}} G_{R,NS}^{(n)}(x^A,x^{A'},\gamma),\label{had-sum}\ee
where 
\begin{eqnarray} G_{R,NS}^{(n)} &\equiv& \sum_{k=1}^\infty\left[ \left( X_{2k}^{(+)}\check{\cb}_{2k}(\gamma+\eta-2n\pi)+Y_{2k+1}^{(+)}\hat{\cb}_{2k+1}(\gamma+\eta-2n\pi)\right)\theta(\gamma+\eta-2n\pi)\theta(2(n+1)\pi-\gamma-\eta)\right. \nonumber \\
&& + \left( X_{2k}^{(-)}\check{\cb}_{2k}(\gamma-\eta-2n\pi)+Y_{2k+1}^{(-)}\hat{\cb}_{2k+1}(\gamma-\eta-2n\pi)\right)\theta(\gamma-\eta-2n\pi)\theta(2(n+1)\pi-\gamma+\eta) \nonumber \\
&&+ Y_{2k}^{(+)}\left({}_{n\!}q_{2k}(\gamma+\eta)+a_{2k}(\gamma+\eta-2n\pi)^{2k-1}\log\left|1-\frac{\gamma+\eta}{2n\pi}\right|\right) \nonumber\\
&&+ X_{2k+1}^{(+)}\left({}_{n\!}q_{2k+1}(\gamma+\eta)+a_{2k+1}(\gamma+\eta-2n\pi)^{2k}\log\left|1-\frac{\gamma+\eta}{2n\pi}\right|\right) \nonumber\\
&&+ Y_{2k}^{(-)}\left({}_{n\!}q_{2k}(\gamma-\eta)+a_{2k}(\gamma-\eta-2n\pi)^{2k-1}\log\left|1-\frac{\gamma-\eta}{2n\pi}\right|\right) \nonumber\\
&& \left. + X_{2k+1}^{(-)}\left({}_{n\!}q_{2k+1}(\gamma-\eta)+a_{2k+1}(\gamma-\eta-2n\pi)^{2k}\log\left|1-\frac{\gamma-\eta}{2n\pi}\right|\right) \right],\quad n\in\zo, \label{had-sum-terms}
\end{eqnarray}
where 
\be 
X_k^{(\pm)}=X_k^{(\pm)}(x^A,x^{A'},\gamma)\equiv V_k^{(\pm)} + \bar{V}_k^{(\pm)},\quad 
Y_k^{(\pm)}=Y_k^{(\pm)}(x^A,x^{A'},\gamma)\equiv  i(V_k^{(\pm)}-\bar{V}_k^{(\pm)}),
\label{xy-def}
\ee
and
\be {}_{n\!}q_{k} \equiv \left\{ \begin{array}{rl} -i{}_{n\!}Q_{k}, & k \hbox{ odd}; \\ {}_{n\!}Q_{k}, & k \hbox{ even}; \end{array} \right.\quad a_{k} \equiv \left\{ \begin{array}{rl} -iA_{k}, & k \hbox{ odd}; \\ A_{k}, & k \hbox{ even}. \end{array} \right. \label{a-q-real} 
\ee
(The latter definitions are required to remove the imaginary unit in the definition (\ref{wk-def}) of $\cw_k$.) The $n=0$ contribution, $G_{R,NS}^{(0)}$, has the same form as (\ref{had-sum-terms}) but with the replacements 
\be \log\left|1-\frac{\gamma\pm\eta}{2n\pi}\right| \to \log\left|\gamma\pm\eta\right|. \label{grn-zero}\ee

It is a straightforward algebraic manipulation to put together Eqs.~(\ref{eq:Gl=0+Gl>0}), (\ref{grl-k-zero2}), (\ref{grl-k-one1}) and (\ref{had-sum})
and regroup them in the following form:
\begin{align}\label{eq:sum Hads}
&
G_{R,NS} =\frac{\theta(-\sigma)\theta(\Delta t)}{\pi r\cdot r' \sqrt{\eta \sin\gamma}}
\left\{
\sum_{n\in\mathbb{Z}} 
\bigg[
U_0
\left(\pi (-1)^n\delta(\xm{n})+\cos\left(\frac{\xp{0}}{2}\right)\PV\left(\frac{1}{\xp{n}}\right)+
\sin\left(\frac{\xm{0}}{2}\right)\PV\left(\frac{1}{\xm{n}}\right)
\right)
+
\right.\nonumber  \\ &
\Vnp{n} \theta\left(\xp{n}\right)\theta\left(-\xp{n+1}\right)
+
\Vnm{n} \theta\left(\xm{n}\right)\theta\left(-\xm{n+1}\right)+
\Vnpb{n}\ln\left|\xpt{n}\right|+
\Vnmb{n}\ln\left|\xmt{n}\right|
\bigg]
+\Wn{n}
\Bigg\},
\end{align}
where 
\begin{equation}\label{eq:xpm}
\xpm{n}\equiv \gamma\pm \eta-2n\pi
\end{equation}
and
\be 
\xpmt{n}\equiv 
\begin{cases}
\displaystyle  \frac{\xpm{n}}{2n\pi}, &n\in\mathbb{Z}_0, \\
\displaystyle  \xpm{0},  &  n=0.
\end{cases}
\ee
We note that the $\hat{\sigma}_k$ in Eq.~(\ref{sig-k-eta}), which is the `globalization' of the world function $\hat{\sigma}$ on the $4$-D conformal Schwarzschild spacetime, can be expressed as
\be \label{eq:sigma_k - x+/-}
\hat{\sigma}_k=\frac 12\xp{-k}\cdot \xm{-k},\quad k\in\mathbb{Z}.
\ee
This means that $\xpm{n}=0$, $n\in \mathbb{Z}$, corresponds to null geodesics  in the $4$-D conformal Schwarzschild spacetime and, therefore, also
in Schwarzschild spacetime.
The coefficients in Eq.~(\ref{eq:sum Hads}) are obtained as follows.
The coefficients of the Heaviside distributions are:
\be
\Vnpm{n}\equiv \sum_{k=1}^{\infty}\left(X_{2k}^{(\pm)}\check{\cb}_{2k}\left(\xpm{n}\right)
+Y_{2k+1}^{(\pm)}\hat{\cb}_{2k+1}\left(\xpm{n}\right)\right).
\ee
The coefficients of the logarithms are:
\begin{align}\label{eq:Vnb}
&
\Vnpb{n}\equiv 
\frac{\VHads(\gamma)}{8}\cos\left(\frac{\xp{0}}{2}\right)+
\sum_{k=1}^{\infty}\left(X_{2k+1}^{(+)}a_{2k+1}\xp{n}^{2k}+
Y_{2k}^{(+)}a_{2k}\xp{n}^{2k-1}\right),
\nonumber \\ &
\Vnmb{n}\equiv 
\frac{\VHads(-\gamma)}{8}\sin\left(\frac{\xm{0}}{2}\right)+
\sum_{k=1}^{\infty}\left(X_{2k+1}^{(-)}a_{2k+1}\xm{n}^{2k}+
Y_{2k}^{(-)}a_{2k}\xm{n}^{2k-1}\right).
\end{align}
Finally, the coefficients in Eq.~(\ref{eq:sum Hads}) which are not multiplying any distribution are:
\begin{align}\label{eq:Wn}
&
\Wn{n}\equiv \hat{\mathcal{W}}_n,\quad  \forall n\in\mathbb{Z}_0,
\\&
\Wn{0}\equiv \hat{\mathcal{W}}_0+
\pi \sqrt{\eta\sin\gamma}\ \cgo
-\frac{1}{2}
\left[
\left(U_0+\xp{0}\frac{V(\gamma)}{8}\right)\sin\left(\frac{\xp{0}}{2}\right)+
\left(U_0+\xm{0}\frac{V(-\gamma)}{8}\right)\cos\left(\frac{\xm{0}}{2}\right)
\right].
\nonumber
\end{align}
where
\begin{align}
&
 \hat{\mathcal{W}}_n
 \equiv
\sum_{k=1}^{\infty}\left(
Y_{2k}^{(+)}{}_nq_{2k}(\gamma+\eta)+X_{2k+1}^{(+)}{}_nq_{2k+1}(\gamma+\eta)+
Y_{2k}^{(-)}{}_nq_{2k}(\gamma-\eta)+X_{2k+1}^{(-)}{}_nq_{2k+1}(\gamma-\eta)\right)+
\nonumber  \\ &
\frac{\pi}{8}\VHads(\gamma)\sin\left(\frac{\xp{0}}{2}\right)\left(\theta(\xp{n})-\frac{1}{2}\right)+
\frac{\pi}{8}\VHads(-\gamma)\cos\left(\frac{\xm{0}}{2}\right)\left(\theta(\xm{n})-\frac{1}{2}\right),
\quad \forall n\in\mathbb{Z}.
\end{align}
We remind the reader that:
\begin{itemize}
\item[$\bullet$]
$\check{\cb}_{2k}\left(\xpm{n}\right)$ and $\hat{\cb}_{2k+1}\left(\xpm{n}\right)$, as given by Eqs.~(\ref{ck-even}) and (\ref{sk-odd}),
are polynomials (of order $2k$ and $2k+1$ respectively) in $\xpm{n}$;
\item
${}_nq_{k}(\gamma\pm\eta)$, given by Eqs.~(\ref{Q-poly-def}) and (\ref{a-q-real}),  can be written as a polynomial
(of order $k-1$) in $\xpm{n}$;
\item
$Y_{2k}^{(\pm)}$ and $X_{2k+1}^{(\pm)}$, 
given by Eqs.~(\ref{xy-def}), (\ref{vk-plus-def}), (\ref{vk-minus-def}), (\ref{nu-plus-def}) and (\ref{nu-minus-def}),
depend on the coordinates $\gamma$, $x^A$ and $x^{A'}$;
\item
the $a_k$, given by Eqs.~(\ref{a-q-real}) and (\ref{n-zero-coeffts}) are numbers.
\end{itemize}
It then follows that:
\begin{itemize}
\item[$\bullet$]
the $\Vnpm{n}$ can be formally expressed as an expansion in powers of $\xpm{n}$;
\item
the $k$-sums in Eq.~(\ref{eq:Vnb}) for $\Vnpmb{n}$ can be formally expressed as an expansion in powers of $\xpm{n}$;
\item
the $k$-sum of the terms containing $Y_{2k}^{(\pm)}$ and $X_{2k+1}^{(\pm)}$
in Eq.~(\ref{eq:Wn}) for $\Wn{n}$  can be formally expressed as an expansion in powers of $\xp{n}$ plus another expansion in powers of $\xm{n}$.
\end{itemize}
Here, an ``expansion in powers of $\xpm{n}$" means a series expansion in non-negative integer powers of $\xpm{n}$, with coefficients that may depend on $(x^A,x^{A'},\gamma)$, but that are independent of $n$. We note that the coefficients in these expansions depend crucially on the $2$-D Hadamard coefficients $U_k$ of Eq.~(\ref{eq:Had series 2d}).
Because of these various properties, together with the fact that Eq.~(\ref{eq:sum Hads}) makes explicit the full singularity structure of the GF globally, we refer to  the expression in Eq.~(\ref{eq:sum Hads})
 as a `sum of Hadamard forms'.
We note that Eqs.~(\ref{grl-k-zero2}), (\ref{grl-k-one1}) and (\ref{eq:sum Hads}) are  not valid when  $\eta=0$,
$\gamma=0$ or $\gamma\in[\gamma_0,\pi]$ (where $\gamma_0=2(\sqrt{2}-1)\pi\simeq 0.828\pi$) -- we deal with the
 points $\gamma=0$ and $\pi$ in the next section.


\section{Singularity at caustics}\label{sec:caustics}

In this section we investigate the singularity of the GF at  caustic points of Schwarzschild spacetime, which are points where null geodesics focus.
Because of this property and the fact that the GF diverges when the two points in its argument are connected via a null geodesic, it is not surprising that the type of the singularity  of the GF
at caustics  is different  from that we have derived above (Eqs.~(\ref{grl-k-zero2}) and (\ref{grl-k-one1})), which is valid away from  $\gamma=0$ and $\pi$, i.e., the values of $\gamma$ for caustics in Schwarzschild spacetime.

Let us first deal with the caustics with $\gamma=0$.
In this case, we  insert $P_\ell(1)=1$, $\forall \ell$, in
Eq.~\eqref{sch-grl-mode} with \eqref{bessel-exp-cgl}.
We then just proceed as in Sec.~\ref{sec:Sum Hads} and use Eq.~(\ref{bessel-large-ell}) in order to first gather like powers of $\lam$ (where 
we are now of course spared the use of Eq.~(\ref{legendre-bessel})) and obtain:
\be \label{grl gamma=0} 
\left.\grl\right|_{\gamma=0} =
\frac{1}
{rr'\sqrt{2\pi\eta}}
\sum_{m=0}^{\infty}\left\{
e^{-i\pi/4}
\tilde V_m
\sum_{\ell=1}^{\infty}\frac{e^{i\lam\eta}}{\lam^{m-1/2}}
+c.c.
\right\},
\ee
where  `$c.c.$' denotes the complex conjugation of the term that is preceding it and where
\be
\tilde V_m\equiv 
\sum_{j=0}^{m}
e^{i\pi(j+m/2)}(2\eta)^{2j-m}U_{j}a_{j,m-j}.
\ee
We note that in this case we have half-integer powers of $\lam$, as opposed to Sec.~\ref{sec:Sum Hads} where we had integer powers.
The highest power of $\lam$ here is $1/2$, as opposed to $0$ in Sec.~\ref{sec:Sum Hads}, anticipating a `stronger' divergence of the GF at caustic points. 
As we are interested here only in the divergent terms,
we introduce the notation $A \doteq B$ to mean that $A-B$ is a continuous function.
We then have
\be\label{eq:G_R, gamma=0}
\left.\grl\right|_{\gamma=0} 
\doteq 
\frac{1}
{rr'\sqrt{2\pi\eta}}
 \left[ \left.\tilde G_R^0\right|_{\gamma=0} +\left.\tilde G_R^1\right|_{\gamma=0}  \right],  
 \quad
\ee
where
\be \label{eq:G_R^0,1 gamma=0}
 \left.\tilde G_R^0\right|_{\gamma=0}
\equiv
e^{-i\pi/4} e^{i\eta/2}\ca_{-1/2}(\eta) \tilde V_0 +c.c.,
\quad
\left.\tilde G_R^1\right|_{\gamma=0}
\equiv e^{-i\pi/4}e^{i\eta/2}\ca_{1/2}(\eta) \left(\frac14\tilde V_0 +\tilde V_1 \right)+c.c.,
\ee
and we have 
\be \tilde V_0 = U_0, \qquad \tilde V_1 = -\frac{i}{4}\frac{\left(U_0+16\eta^2 U_1\right)}{2\eta}\equiv -\frac{i}{4}\tilde U_1. \label{eq:Vt0-Vt1} \ee

We already dealt with the distributions $\ca_{k}$ for $k\in\mathbb{Z}$ in Sec.~\ref{sec:ca_k} and, in the cases $\ca_{0}$ and $\ca_{1}$, originally in~\cite{Casals:2012px}. Here we deal similarly with the distributions
$\ca_{\pm 1/2}$.
We have~\cite{NIST:DLMF},
\be \label{eq:ca_3/2}
\ca_{3/2}(\eta)=\Gamma\left(-\tfrac{1}{2}\right)(-i\eta)^{1/2}+\sum_{j=0}^{\infty}\zeta\left(\tfrac{3}{2}-j\right)\frac{(i\eta)^j}{j!},\quad |\eta|<2\pi,
\ee
where $s\mapsto\zeta(s)$ is the Riemann zeta function. 
It is straightforward to show that the infinite series in (\ref{eq:ca_3/2}) is uniformly convergent and hence analytic for $|\eta|<2\pi$. It follows that the series and its derivatives are smooth on $|\eta|<2\pi$. Also, the first expression on the right-hand side of Eq.~(\ref{eq:ca_3/2}) involves the principal branch of the square root. Then we can write 
\be \ca_{3/2}(\eta) =s_{3/2}(\eta) + r_{3/2}(\eta),\quad |\eta|<2\pi, \label{eq:A32-sing-reg} \ee
where 
\be
s_{3/2}(\eta) \equiv  -2\sqrt{\pi}(-i\eta)^{1/2}=-2\sqrt{\pi}|\eta|^{1/2}e^{i\pi (1-2\theta(\eta))/4} \label{eq:s32-def} \ee
is the singular (non-smooth) part of $\ca_{3/2}$ on $|\eta|<2\pi$, and $r_{3/2}$ is the regular part. By analyticity of $r_{3/2}$, it follows that the singular and regular parts of $\ca_{1/2}$ and $\ca_{-1/2}$ are obtained by applying Eqs.~(\ref{eq:rec rln ca}) to the singular and regular parts of $\ca_{3/2}$ separately. Thus we define the singular parts of $\ca_{\pm1/2}$ (for $|\eta|<2\pi$) by 
\begin{eqnarray}
s_{1/2} &=&-is_{3/2}',\label{eq:s12-def} \\
s_{-1/2} &=&-is_{1/2}',\label{eq:s-12-def} 
\end{eqnarray}
where the derivative is taken in the sense of distributions.
Calculating the first of these derivatives yields
\be s_{1/2}(\eta) = \delta_1(\eta)  + i \delta_2(\eta) \label{eq:s12-form}\ee
where 
\begin{eqnarray}
\delta_1(\eta) &\equiv& \sqrt{\frac{\pi}{2}}|\eta|^{-1/2}, \label{eq:delta1-def} \\
\delta_2(\eta) &\equiv& \sqrt{\frac{\pi}{2}}|\eta|^{-1/2}(2\theta(\eta)-1). \label{eq:delta2-def} 
\end{eqnarray}
Then 
\be s_{-1/2}(\eta) = \delta_2'(\eta)-i\delta_1'(\eta). \label{eq:s-12--form} \ee
We note that $s_{1/2}$ has the form of a locally integrable function, but the distribution $s_{-1/2}$ does not. The derivatives $\delta_i'$ are calculated (and applied) by using the distributional rule 
\be <\delta_i',u> = - <\delta_i,u'>,\quad i=1,2,\label{eq:dist-deriv} \ee
for all test functions $u$ and where $<f,g>=\int_\mathbb{R}f\, g\, dx$.

Collecting terms yields 
\begin{eqnarray} 
\left.\tilde G_R^0 \right|_{\gamma=0}
&\doteq& 
\sqrt{\pi}\left(\cos\frac{\eta}{2}(\delta_2'-\delta_1')+\sin\frac{\eta}{2}(\delta_2'+\delta_1')\right)U_0,\quad |\eta|<2\pi, \label{eq:Grt0-v1} 
\end{eqnarray}
and
\begin{eqnarray}
\left.\tilde G_R^1 \right|_{\gamma=0} &\doteq&
\frac{1}{2\sqrt{2}}\left(\left(\cos\frac{\eta}{2}+\sin\frac{\eta}{2}\right)U_0\delta_1+\left(\cos\frac{\eta}{2}-\sin\frac{\eta}{2}\right)U_0\delta_2 \right. \nonumber \\
&& \left. + \left(\cos\frac{\eta}{2}+\sin\frac{\eta}{2}\right)\tilde U_1\delta_2 - \left(\cos\frac{\eta}{2}-\sin\frac{\eta}{2}\right)\tilde U_1\delta_1 \right), \quad |\eta|<2\pi.
\label{eq:Grt1-v1} 
\end{eqnarray}
It is straightforward to establish that 
\begin{eqnarray}
\eta^p\delta_i(\eta) &\doteq& 0, \quad i=1,2,\quad p\geq 1, \label{eq:eta-p-delta} \\
\eta^q\delta_i'(\eta) &\doteq& 0, \quad i=1,2,\quad q\geq 2. \label{eq:eta-p-delta-prime} 
\end{eqnarray}
Then in a Taylor expansion of the trigonometric coefficients in (\ref{eq:Grt0-v1}) and (\ref{eq:Grt1-v1}), only $O(\eta^p), p\leq 1$ and $O(1)$ terms contribute to the right hand sides of (\ref{eq:Grt0-v1}) and (\ref{eq:Grt1-v1}) respectively. Thus we can write

\begin{eqnarray} 
\left.\tilde G_R^0 \right|_{\gamma=0}
&\doteq& 
\sqrt{\pi}\left(-\left(1-\frac{\eta}{2}\right)U_0\delta_1'(\eta)+\left(1+\frac{\eta}{2}\right)U_0\delta_2'(\eta)\right),\quad |\eta|<2\pi, \label{eq:Grt0-v2} 
\end{eqnarray}
and
\begin{eqnarray}
\left.\tilde G_R^1 \right|_{\gamma=0} &\doteq&
\frac{1}{2\sqrt{2}}\left((U_0-\tilde U_1)\delta_1(\eta)+(U_0+\tilde U_1)\delta_2(\eta)\right), \quad |\eta|<2\pi.
\label{eq:Grt1-v2} 
\end{eqnarray}

To determine the corresponding globally valid expressions, we note that the distributions $\ca_k(\eta)$ are periodic with period $2\pi$. This implies that there will also be discontinuous contributions to $\left.\tilde G_R^0 \right|_{\gamma=0}$ and $\left.\tilde G_R^1 \right|_{\gamma=0}$ when $\eta=2n\pi, n\in\mathbb{Z}$. These contributions are determined by repeating the calculation above, using 
\be \label{eq:ca_3/2-extend}
\ca_{3/2}(\eta)=\ca_{3/2}(\etane)=\Gamma\left(-\tfrac{1}{2}\right)(-i\etane)^{1/2}+\sum_{j=0}^{\infty}\zeta\left(\tfrac{3}{2}-j\right)\frac{(i\etane)^j}{j!},\quad |\etane|<2\pi,
\ee
where 
\be \etane = \eta +2n\pi,\quad n\in\mathbb{Z}. \label{eq:etane-def} \ee
Note that 
\be e^{i\etane/2} = (-1)^ne^{i\eta/2}. \label{eq:exo-etane} \ee
This results in 

\begin{eqnarray} 
\left.\tilde G_R^0 \right|_{\gamma=0}
&\doteq& 
(-1)^n\sqrt{\pi}\left(-\left(1-\frac{\etane}{2}\right)U_0\delta_1'(\etane)+\left(1+\frac{\etane}{2}\right)U_0\delta_2'(\etane)\right),\quad |\etane|<2\pi, \label{eq:Grt0-vn} 
\end{eqnarray}
and
\begin{eqnarray}
\left.\tilde G_R^1 \right|_{\gamma=0} &\doteq&
\frac{(-1)^n}{2\sqrt{2}}\left((U_0-\tilde U_1)\delta_1(\etane)+(U_0+\tilde U_1)\delta_2(\etane)\right), \quad |\etane|<2\pi.
\label{eq:Grt1-vn} 
\end{eqnarray}

We then collect terms and use the distributional identities
\be \eta \delta_i'(\eta) = -\frac12\delta_i(\eta),\quad i=1,2 \label{eq:dist-deriv-id} \ee
to obtain (reintroducing the $\theta$'s as per \eqref{eq:Gl=0+Gl>0})
\begin{eqnarray}
\left.G_R\right|_{\gamma=0}
 &\doteq & 
\theta(\Delta t)\theta(-\sigma) \frac{\sqrt{2}}{rr'\sqrt{\eta}} \times \nonumber \\
&& \sum_{n\in\mathbb{Z}} (-1)^n\left(\frac{U_0+16\eta^2 U_1}{8\eta} H_{-\!\frac12}(-\etane) + U_0(\eta)H_{-\!\frac12}'(-\etane)\right),
\label{eq:Grl-caustic-0}
\end{eqnarray}
where 
\be H_{-\!\frac12}(x) \equiv |x|^{-\frac12}\theta(x),\qquad x\in\mathbb{R}. \label{eq:Hk-def} \ee
This equation
shows the singular behaviour of the GF at caustics with $\gamma=0$\footnote{As mentioned in Sec.~\ref{sec:remainder}, the coincidence limit (i.e., $\eta, \gamma \to 0$) is excluded from the region of validity of this result.}.

In order to deal with the caustics with $\gamma=\pi$, we note that $P_{\ell}(-1)=(-1)^{\ell}=e^{\pm i\pi \ell}$.
In this case, Eq.~(\ref{eq:G_R, gamma=0}) and (\ref{eq:G_R^0,1 gamma=0})
become respectively
\be\label{eq:G_R, gamma=pi}
\left.\grl\right|_{\gamma=\pi} 
\doteq 
\frac{1}
{rr'\sqrt{2\pi\eta}}
 \left[ \left.\tilde G_R^0\right|_{\gamma=\pi} +\left.\tilde G_R^1\right|_{\gamma=\pi}  \right],  
 \quad
\ee
and 
\be \label{eq:G_R^0,1 gamma=pi}
 \left.\tilde G_R^0\right|_{\gamma=\pi}
\equiv
e^{-i\pi/4} e^{i\eta/2}\ca_{-1/2}(\eta+\pi) \tilde V_0 +c.c.,
\quad
\left.\tilde G_R^1\right|_{\gamma=\pi}
\equiv e^{-i\pi/4}e^{i\eta/2}\ca_{1/2}(\eta+\pi) \left(\frac14\tilde V_0 +\tilde V_1 \right)+c.c.
\ee

 Otherwise, the calculation proceeds as above, and we find

\begin{eqnarray}
\left.G_R\right|_{\gamma=\pi} 
&\doteq & 
\theta(\Delta t)\theta(-\sigma) \frac{\sqrt{2}}{rr'\sqrt{\eta}} \times \nonumber \\
&& \sum_{n\in\mathbb{Z}} (-1)^{n+1}\left(\frac{U_0+16\eta^2 U_1}{8\eta} H_{-\!\frac12}(\etano) + U_0(\eta)H_{-\!\frac12}'(\etano)\right),
\label{eq:Grl-caustic-pi}
\end{eqnarray}
where 
\be \etano = \eta + (2n+1)\pi,\quad n\in\mathbb{Z}. \label{eq:etano-def} \ee

This equation shows the singular behaviour of the GF at caustics with $\gamma=\pi$.

These results show that the singularity of the GF at caustic points is `stronger'  than at points which are not caustics and that
its structure is two-fold instead of the observed four-fold structure
at points which are not caustics.

As stated in Sec.~\ref{sec:PH,gral}, our results in this paper for the GF in Schwarzschild spacetime should reproduce those in $\mathbb{M}_2\times \mathbb{S}_2$, times the conformal factor $1/(rr')$, when setting $U_0=1$ and $U_k=0$, $\forall k>0$.
In particular that should be true for 
Eq.~\eqref{eq:Grl-caustic-pi}
above when compared with Eq.~(132) of Ref.~\cite{Casals:2012px} and it is easy to check that that is indeed the case.
Furthermore, in Schwarzschild spacetime, Fig.~13~\cite{Zenginoglu:2012xe} shows a numerical approximation to the GF near a caustic point both for the cases $\gamma=0$ and $\gamma=\pi$.
This figure seems to suggest that if the form, at least to leading order, of the GF about $\eta_e=0$ when $\gamma=0$ is, say, `$F(\eta_e)$', then its form about $\eta_o=0$ when $\gamma=\pi$ is `$-F(-\eta_o)$'. 
This symmetry is manifest to leading order in 
Eqs.~\eqref{eq:Grl-caustic-0} and \eqref{eq:Grl-caustic-pi}
for $G_R$ (note, however, that this symmetry is not satisfied by the terms $\tilde G_R^0$ and $\tilde G_R^1$ separately), thus serving as further support for our results at caustic points in Schwarzschild spacetime. In the next section we also present numerical result for our results away from caustics.


\section{Numerical evidence} \label{sec:numerical}

In this section we provide  numerical evidence for the  analytical  leading-order divergences  that we have obtained:
(i) in the Bessel expansion Eq.~(\ref{bessel-exp-cgl}), and
(ii) in the sum of Hadamard forms in Eq.~\eqref{eq:sum Hads}.
We provide the evidence for (i) and (ii) in, respectively, the first subsection and
 the second (and third) subsection(s).
By ``leading-order divergences" here we mean the divergences whose coefficients involve 
$U_0$ but no $U_k$ with $k>0$.
We compare both of our analytical results to a semi-analytic/numerical calculation of the full GF.
In particular, we shall see that the comparison is good for 
points which are quite far apart, thus indicating that the region $\Omega_p$ of convergence of the 
 series in Eq.~(\ref{bessel-exp-cgl}) is either equal  to  the whole of   $\Mtwo$ or, at least, that it covers a large enough region of physical interest.
 {We further note that~\cite{JACMK2019,caribe2023lensing} also numerically corroborated the singularity structure that we have here obtained beyond the maximal normal neighbourhood, and applied it to the setting of quantum field theory}. 
 In the third subsection, which is the last one, we do not present a new comparison but we briefly discuss
 the use of the ``direct" divergence in Eq.~\eqref{eq:sum Hads} which has been made in~\cite{JACMK2019, CNOW:2019,OToole:2021} for a practical
 calculation of the GF, thus corroborating the validity of this expression of ours for this direct divergence.
 
 The comparison in $G_R(x,x')$ will be made specifically
  for points $x$ and $x'$
 on a timelike circular geodesic at $r=6M$  in Schwarzschild spacetime.
One can think of the field point $x$ as being fixed 
  and the base point $x'$ varying
to  the past 
   along 
  the timelike
  geodesic. Then $x$ and all points $x'$ are connected via this timelike geodesic, and some of the points $x'$ might also be connected to $x$ via a null geodesic that has orbited around the black hole a certain number of times (a different number for each of these such points $x'$).
We refer to these points $x'$ which are connected to $x$ via a null geodesic (as well as via the timelike geodesic) as {\it light-crossings};
the first light-crossing signals the end of the normal neighbourhood of $x$.
As we know, the GF  diverges at the light-crossings.
For an illustration of the light-crossings and the corresponding null geodesics in the case of the timelike circular geodesic at $r=6M$, see Fig.~1(a) in~\cite{CDOW13}.
  For this specific timelike geodesic, 
  it is easy to check that the light-crossings occur at the following times: $\Delta t /M \approx 27.62, 51.84, 58.05, 75.96, 100.09, 108.55, 124.21,\dots$.

Before we proceed to the comparison, we give a brief description of the semi-analytic/numerical solution.
This solution is also obtained  by doing an $\ell$-mode decomposition but, in this method, each $\ell$-mode is calculated as
 the semi-analytic sum of its quasinormal mode and branch cut contributions -- see~\cite{CDOW13} for details.
 This  solution is very accurate but it is not exact, at least for three reasons: (i) its $\ell$-modes have been calculated up to a certain numerical accuracy;
 (ii) the infinite sum has been truncated at the large but finite vaue of $\ell=100$;
(iii) a  smoothing  factor $e^{-\ell^2/(2 \ell_{cut}^2)}$ with $\ell_{cut}=25$ has been included
 in the $\ell$-sum in order to avert spurious oscillations (see~\cite{CDOW13,Hardy}).
This means that the fine features of the exact GF  close to the singularities, as well as its diverging behaviour at the singularities themselves,  are not  captured
exactly by the semi-analytic/numerical solution (in particular, the singularities are smeared out).
Finally, we note that, in a region ``close enough" to $x$,  the $\ell$-sum as just described does not offer a practical way of calculating the GF.
That region is the so-called  `quasi-local'  region, which is within a normal neighbourhood of $x$ and so, if the points are timelike-separated, the GF
is just equal to the $4$-D Hadamard biscalar $V(x,x')$ of Eq.~(\ref{eqhada}). Therefore, for calculating the GF very accurately in a quasi-local region, instead of using the
mentioned $\ell$-sum, we directly calculate  $V(x,x')$ using the techniques described in~\cite{CDOWb,CDOW13}.

We next proceed to the comparison between the semi-analytic/numerical solution and our mentioned leading-order analytical results.


\subsection{Leading term in the Bessel expansion}\label{eq:num Bessel}

In this subsection we use the values of $\sTCS$ and $U_0$ that we numerically calculated in~\cite{PhysRevD.92.104030}\footnote{In Fig.~6(a) in~\cite{PhysRevD.92.104030} we said that we plotted the  final values of $\sTCS$ (which on the vertical axis there is labelled $\sigma_2$), i.e., the values upon return to the original radius $r=6M$ after bouncing off at the turning point. However, we actually plotted the values of $\sTCS$ only when reaching the turning point. Equivalently,
Fig.~6(a) in~\cite{PhysRevD.92.104030} is really a plot of  $\sTCS/4$  (not of   $\sTCS$ as wrongly indicated there) as a function of the total coordinate time intervals $\Delta t/M$.
[All the other plots in~\cite{PhysRevD.92.104030} are correct as they are, including Fig.~6(b), which is indeed a plot of the final values of $\Delta^{1/2}$ as a function of  $\Delta t/M$, as indicated there.]}
in order to evaluate the leading ($k=0$) term in the Bessel expansion Eq.~(\ref{bessel-exp-cgl}).
This leading term provides an approximation for $\cgl$, which we
 then insert  into Eq.~(\ref{sch-gret-mode}), in order to obtain an approximation for the GF in Schwarzschild. 
As we  demonstrated in  Sec.~\ref{sec:Sum Hads}, the leading $k=0$ term is sufficient to determine the `leading order' of the singularity of the GF 
along null geodesics;
for the next-to-leading order singularity and the regular part of the GF, other $k>0$ terms contribute as well as the $k=0$ term.
We truncate the $\ell$-sum in  Eq.~(\ref{sch-gret-mode})  at $\ell=200$  and, as in the semi-analytic/numerical GF,
we include
a  smoothing  factor
$e^{-\ell^2/(2 \ell_{cut}^2)}$ with $\ell_{cut}=25$ -- as a consequence, the singularities are smeared out.

\begin{figure}[h!]
\begin{center}
\includegraphics[width=13cm]{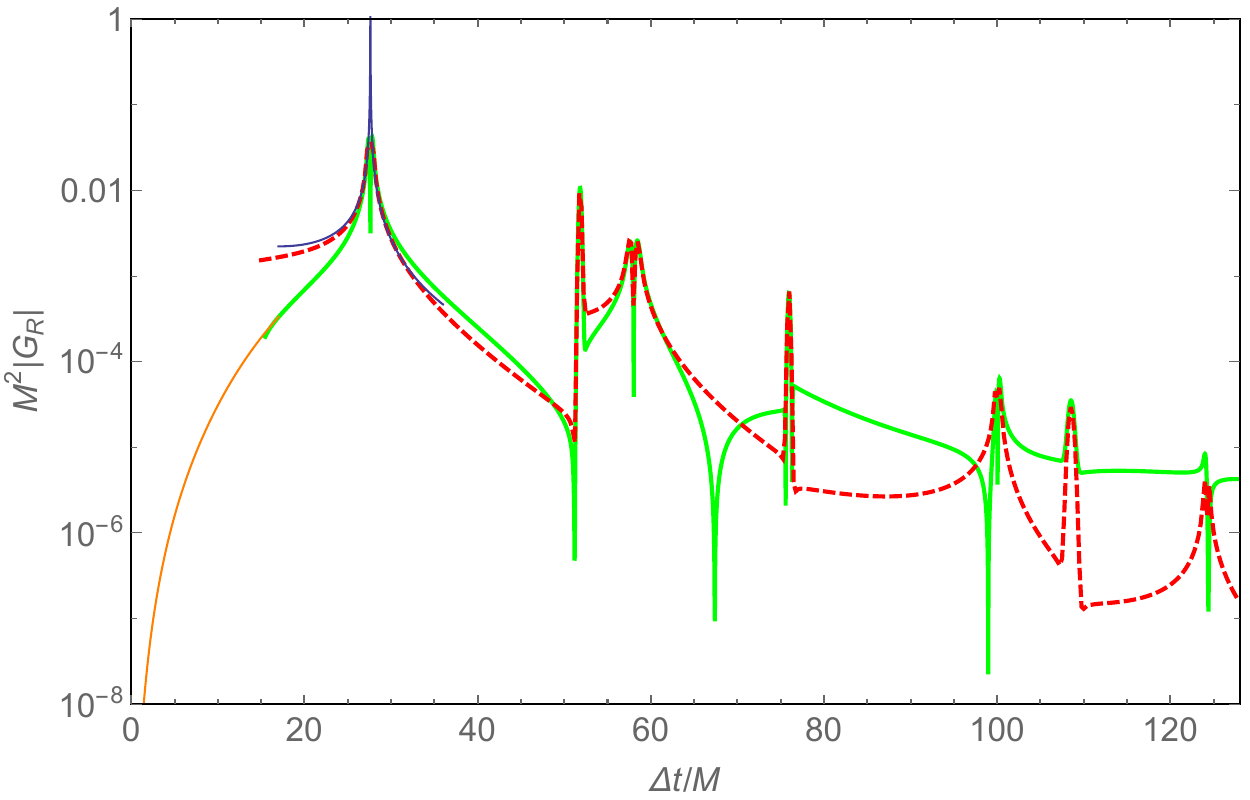}
  \end{center}
\caption{
Log-plots of
the retarded Green function and approximations to it
on Schwarzschild spacetime
 as functions of coordinate time  interval $\Delta t$ for points on a timelike circular geodesic at $r=6M$. 
The `exact' GF is plotted in the solid orange curve (up to $\Delta t\approx 17M$, which is within a  `quasi-local'  region, 
where it is  equal to $V(x,x')$ of Eq.~(\ref{eqhada}))
matched on to the solid green curve  (from $\Delta t\approx 17M$, i.e.,  outside a `quasi-local'  region, where it is
calculated using a sum of quasinormal mode and branch cut contributions).
The dashed red curve is the Bessel expansion Eq.~(\ref{sch-gret-mode}) with (\ref{bessel-exp-cgl}), 
where we only use the leading $k=0$ term and we sum up to a finite value of $\ell=200$ including a smoothing factor $e^{-\ell^2/(2 \ell_{cut}^2)}$ with $\ell_{cut}=25$ (see \cite{CDOW13}).
The solid blue curve, which is concentrated around the first light-crossing only, is the large-$\ell$ asymptotics of the quasinormal mode contribution to the GF as given by Eq.~(55) in Ref.~\cite{Dolan:2011fh}.
}
\label{fig:Bessel exp}
\end{figure}

We plot 
the mentioned leading $k=0$ approximation as a function of time
in 
 Fig.~\ref{fig:Bessel exp},
which is
   to be compared with Figs.~1(b), 10(a), 11(a) in \cite{CDOW13}.
  In the figure, we compare our 
  approximation
  to the semi-analytic/numerical calculation of the GF mentioned above.
  We also compare it to another approximation near the first light-crossing
  as given by Eq.~55~\cite{Dolan:2011fh}.
  We note a couple of features about the plot.
  Firstly, as expected, the leading $k=0$ approximation captures well the light-crossing times given above as well as the behaviour of the (smeared-out) GF near the singularities; this is
  true up to the fourth, if not even seventh, light-crossing.
  Secondly, the leading $k=0$ term does not approximate well the GF {\it before} very near the first light-crossing \footnote{We note that the  behaviour of the semi-analytic/numerical calculation (solid green curve) around the 1st light-crossing  manifests a Dawsonian shape (see Eq.~\eqref{eq:Dawsonian}), corresponding to the $\ell$-smoothing that has been applied. Although similar smoothing has also been applied to the calculation using the Bessel expansion Eq.~(\ref{sch-gret-mode}) with (\ref{bessel-exp-cgl}) (dashed red curve),  it  manifests no similar Dawsonian shape  around the 1st light-crossing (although the peak at the light-crossing is manifest) just due to the
  large stepsize in time used in the calculation (so it is just a `numerical artifact');
  the Dawsonian shape in this curve, however, is indeed manifest around some later light-crossings. In its turn, the calculation using the large-$\ell$ asymptotics of the quasinormal mode contribution (solid blue curve) exhibits a peak at the 1st light-crossing and no Dawsonian shape because, in this case, no $\ell$-smoothing was actually  applied.}. This is not surprising, since the time regimes in-between singularities
  essentially correspond to a backscattering (due to the gravitational potential) `tail' whereas we are only including the first $2$-D Hadamard coefficient $U_0$
  (the square root of the van Vleck determinant $\Delta_{2d}$ in the 2-D spacetime $\Mtwo$).
  Considering this,
  the agreement between the leading $k=0$ term and the
  semi-analytic/numerical
  GF is rather remarkable 
in the time regimes between the first, second, third and fourth light-crossings.

\subsection{Leading term in the global Hadamard form}\label{eq:num global}

In this subsection we provide  numerical evidence for our result in Eq.~\eqref{eq:sum Hads} by 
comparing the semi-analytic/numerical calculation of the GF with 
the result of subtracting from it (a smeared version of)
some of the divergences in that equation.
Let us next describe exactly what we subtracted from the semi-analytic/numerical calculation.
We first define the following quantities, which contain
the Dirac-$\delta$-singularities in Eq.~\eqref{eq:sum Hads}:
\begin{align}\label{eq:divs delta}
&
G_{\delta,n}\equiv 
(-1)^n
\left.\frac{U_0}{ r\cdot r' \sqrt{\eta \sin\gamma}}\right|_{\xm{n}=0} \delta(\xm{n}).
\end{align}
We now define  part of the PV- and logarithmic-singularities in Eq.~\eqref{eq:sum Hads} 
with the coefficients evaluated at the singularities
as:
\begin{align}\label{eq:divs PV+log}
&
G_{PVL,n}\equiv
\frac{(-1)^n\theta(-\sigma)\theta(\Delta t)}{\pi}
\left.\frac{U_0}{ r\cdot r' \sqrt{\eta \sin\gamma}}\right|_{\xp{n}=0}
\bigg(
\PV\left(\frac{1}{\xp{n}}\right)
+
\frac{1}{8}\left(\frac{1}{\eta}+\cot\gamma\right)_{\xp{n}=0}\ln\left|
\xp{n}
\right|
\bigg).
\end{align}
We note that, here, we have
only included the $U_0$ term\footnote{Including the $U_1$ term  would involve an extra, more difficult numerical calculation.} in $\VHads$ in Eq.~\eqref{eq:VHads}
and we have evaluated $\cos\left(\frac{\xp{0}}{2}\right)$ at $x=\xp{n}$, and so we replaced it by $(-1)^n$.

We would use the quantities $G_{\delta,n}$ and $G_{PVL,n}$ if we were to subtract quantities from the exact GF.
However, as mentioned, our semi-analytic/numerical GF includes a finite upper limit in the $\ell$-sum and  a smoothing factor $e^{-\ell^2/(2 \ell_{cut}^2)}$, thus smearing out the divergences.
Therefore, instead of using  $G_{\delta,n}$ and $G_{PVL,n}$, we shall use versions where their divergences have been smeared-out in a similar manner.
Remembering that the leading-order divergences come from $\ca_0(\xpm{n})= \sum_{\ell=1}^\infty e^{i\ell \xpm{n}}$ (Eq.~\eqref{ca-def})
 and
taking cognizance of Eq.~(50) of Ref.~\cite{Dolan:2011fh} but also including in the integrand the smoothing factor $e^{-\ell^2/(2 \ell_{cut}^2)}$, 
the  Dirac-$\delta$-singularity is smeared out as the following Gaussian-like distribution:
\begin{align} \label{eq:Gaussian}
G_{Gaus}\left(\xm{n}\right)
\equiv &
\frac{1}{\pi}
\text{Re}\left(\int_0^{\infty}d\ell \ e^{i(\ell+1/2)\xm{n}-\ell^2/(2 \ell_{cut}^2)}\right)
=
\nonumber \\ &
\frac{\ell_{cut}}{\sqrt{2\pi}}
\, e^{-\ell_{cut}^2\xm{n}^2/2}\left(-\sin\left(\frac{\xm{n}}{2}\right)\text{Erfi}\left(\frac{\ell_{cut}\xm{n}}{\sqrt 2}\right)+\cos\left(\frac{\xm{n}}{ 2}\right)\right),
\end{align}
where Erfi is the imaginary error function.
Similarly, but taking the imaginary part instead of the real part, 
the PV-singularity  is smeared out as the following Dawson-like distribution (``Dawsonian") (see Eq.10~\cite{Zenginoglu:2012xe}):
\begin{align}\label{eq:Dawsonian}
G_{Daws}\left(\xp{n}\right)
\equiv &
\text{Im}\left(\int_0^{\infty}d\ell \ e^{i(\ell+1/2)\xp{n}-\ell^2/(2 \ell_{cut}^2)}\right)
=
\nonumber \\ &
\ell_{cut}\, \sqrt{\frac{\pi}{2}}\, e^{-\ell_{cut}^2\xp{n}^2/2}\left(\cos\left(\frac{\xp{n}}{2}\right)\text{Erfi}\left(\frac{\ell_{cut}\xp{n}}{\sqrt 2}\right)+\sin\left(\frac{\xp{n}}{ 2}\right)\right).
\end{align}
We now denote by $G_{G,n}$ the result of replacing  $\delta(\xm{n})$ by $G_{Gaus}\left(\xm{n}\right)$ in Eq.~\eqref{eq:divs delta} for $G_{\delta,n}$,
and we denote  by $G_{DL,n}$ the result of replacing  $\PV\left(\frac{1}{\xp{n}}\right)$ by $G_{Daws}\left(\xp{n}\right)$ in Eq.~\eqref{eq:divs PV+log} for $G_{PVL,n}$.
The quantities that we  subtracted from our semi-analytic/numerical GF were, separately in different plots:  
 $G_{PVL,+1}$, $G_{PVL,+2}$ and $G_{\delta,-1}$,
i.e.,
the PV- and logarithmic-singularities both at the first light-crossing (i.e., at $\hat{\sigma}_1=0$, i.e., $\eta=2\pi-\gamma$, i.e., $x_{+,+1}=\gamma+\eta-2\pi=0$)
and at the third light-crossing (i.e., at $\hat{\sigma}_3=0$, i.e., $\eta=4\pi-\gamma$, i.e., $x_{+,+2}=\gamma+\eta-4\pi=0$), and the Dirac-$\delta$-singularity at the second light-crossing (i.e., at $\hat{\sigma}_2=0$, i.e., $\eta=2\pi+\gamma$, i.e., $x_{-,-1}=\gamma-\eta+2\pi=0$).

\begin{figure}[h!]
\begin{center}
                      \includegraphics[width=.45\textwidth]{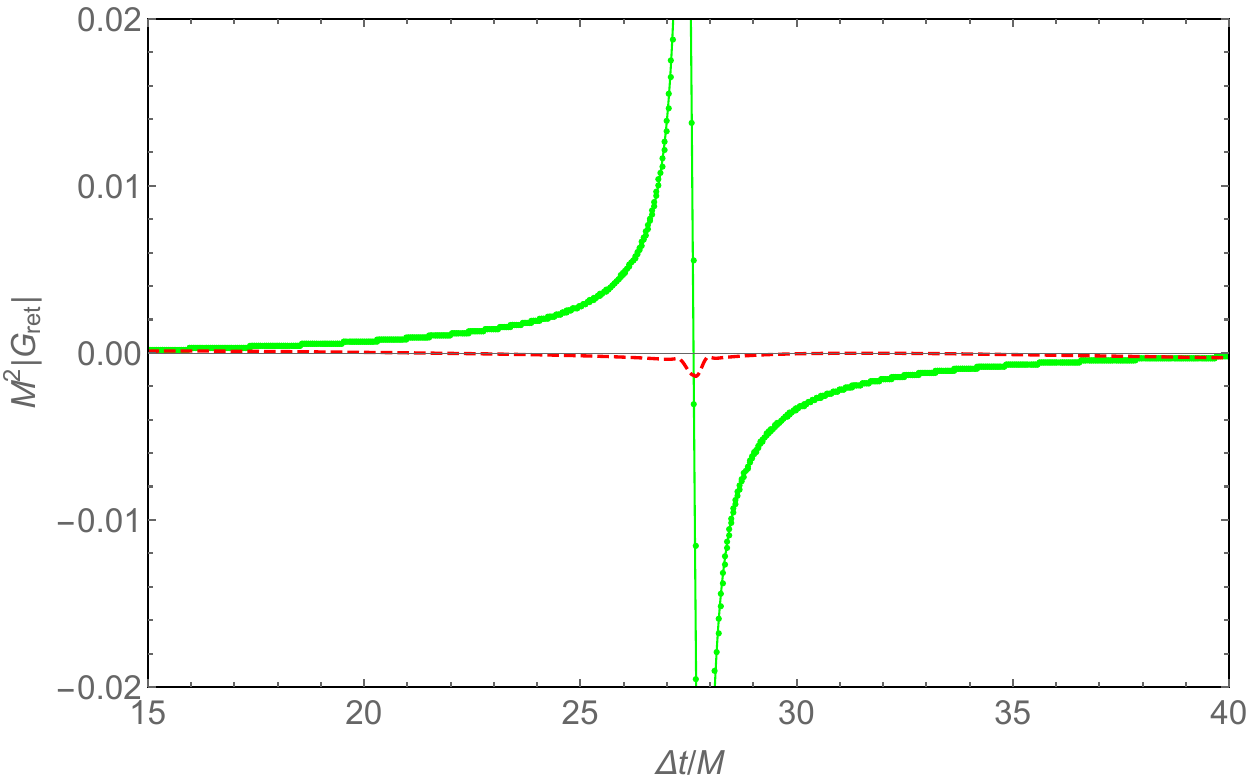}      
                       \includegraphics[width=.45\textwidth]{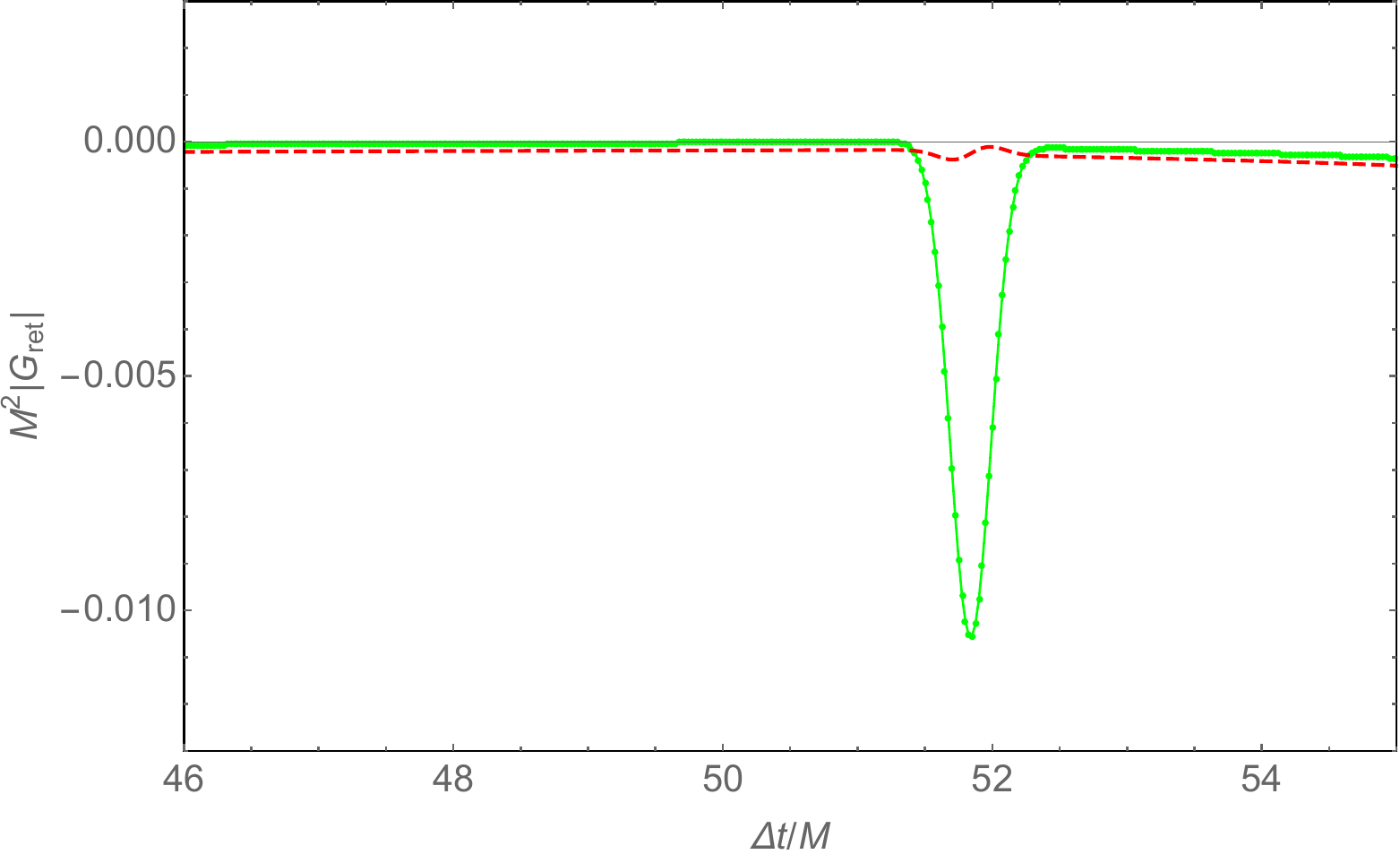}      
                       \includegraphics[width=.45\textwidth]{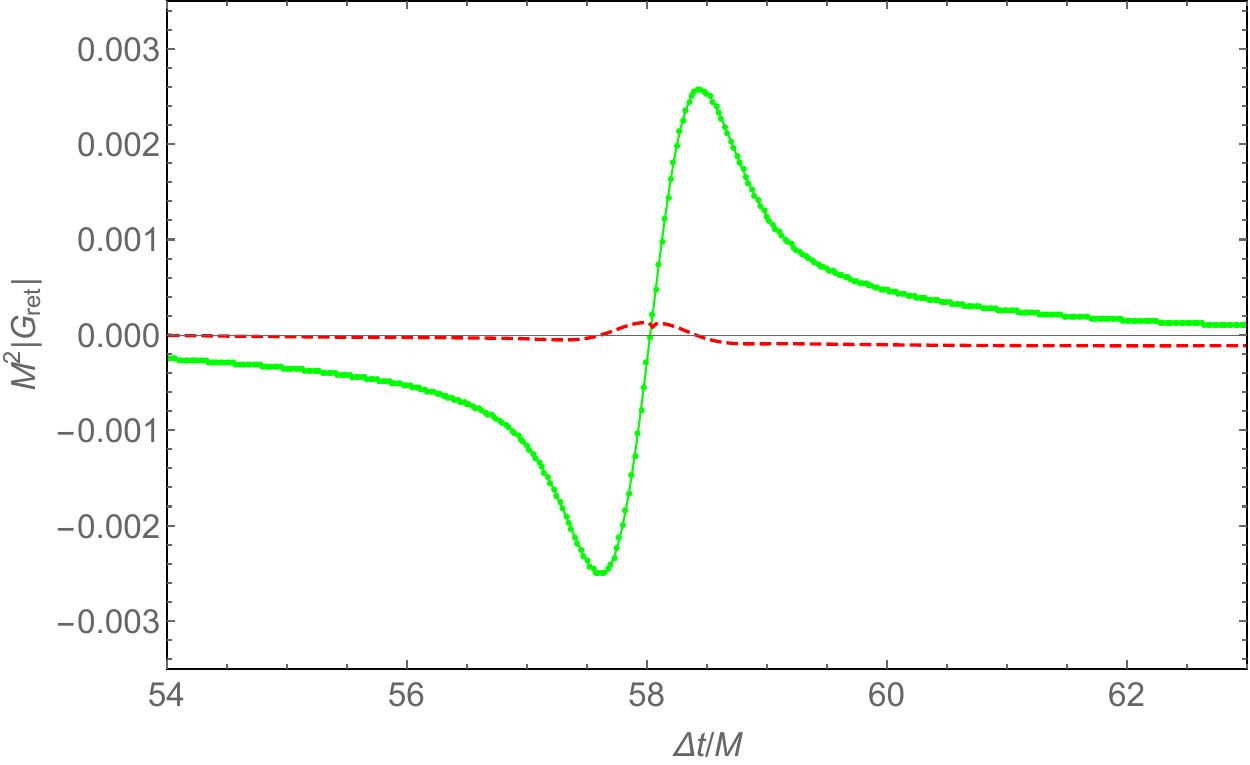}      
   \end{center}
\caption{
Comparison of the  semi-analytic/numerical GF and its divergent behaviour derived in Eq.~\eqref{eq:sum Hads} as a function of $\Delta t$ for points on a timelike circular geodesic at $r=6M$. 
The solid green curve is the semi-analytic/numerical GF and the
dashed red curve is the semi-analytic/numerical GF minus one of the following diverging terms, depending on the plot:
(a) (top left) $G_{PVL,+1}$, plotted near the first light-crossing;
(b) (top right) $G_{\delta,-1}$, plotted near the second light-crossing;
(c) (bottom) $G_{PVL,+2}$, plotted near the third light-crossing.
That is, the solid green curve  is the GF minus a smeared version of the  leading-PV and subleading-log divergences at the first and third light-crossings, and minus the  leading-Dirac-$\delta$ divergence at the second light-crossing; see Eqs.~\eqref{eq:Gaussian} and \eqref{eq:Dawsonian}.}
 \label{fig:GF subt divs extra}
\end{figure}

In 
Fig.~\ref{fig:GF subt divs extra}
we show three plots which contain both the full semi-analytic/numerical GF (solid green curve; same curve as in Fig.~\ref{fig:Bessel exp}) as well as  the result of subtracting 
(a)  $G_{PVL,+1}$, (b) $G_{\delta,-1}$ and (c)  $G_{PVL,+2}$ from it (dashed red curves).
We do it for the same setting as in Fig.~\ref{fig:Bessel exp}, i.e., as a function of $\Delta t$ for points on a timelike circular geodesic at $r=6M$. 
In Fig.~\ref{fig:GF subt divs extra}
 we can see
that the singularities  in Eq.~\eqref{eq:sum Hads} smeared as per $G_{G,n}$ and $G_{DL,n}$ capture well the behaviour of the semi-analytic/numerical GF.
This is true up to at least the third light-crossing (after that, the overall magnitude has decreased too much to allow for such a precise check).
The removal of $G_{PVL,n}$ and $G_{\delta,n}$, which only contain the $U_0$-dependent coefficients in Eq.~\eqref{eq:sum Hads}, ``flattens" out the GF in a striking way, as can be seen
in Fig.~\ref{fig:GF subt divs extra}; these terms thus seem to form the ``backbone" of the GF.
The ``flattening" of the divergences is not perfect since, after all, we are just carrying out approximations to the GF: 
there is an $\ell$-mode truncation and a smoothing-factor in the semi-analytic/numerical solution, which we somewhat mirror in $G_{PVL,n}$ and $G_{\delta,n}$; 
in Eq.~\eqref{eq:divs PV+log} we only include the term that contains $U_0$
in the coefficient of the logarithmic divergence, not the full coefficient; the semi-analytic/numerical solution has some numerical error; etc.

\subsection{Direct term in the global Hadamard form}\label{eq:num global}

A separate numerical corroboration of, specifically, our expression for the ``direct" divergence, namely, the $\delta(\xm{n=0})$ term in Eq.~\eqref{eq:sum Hads}, i.e.,  $G_{\delta,n=0}$ in Eq.~\eqref{eq:divs delta},
  is provided in~\cite{JACMK2019, CNOW:2019,OToole:2021}. In these works, our
expression for such a term proves to be very useful for a practical calculation of the GF.
As mentioned above, in a quasi-local region, the semi-analytic/numerical $\ell$-sum is not practical for calculating the GF and so one typically needs to supplement it with a calculation
of $V(x,x')$ inside the quasi-local region. 
The reason is essentially due to the fact that truncation of the $\ell$-sum and inclusion of a smoothing factor cause the direct $\delta(\s)$-singularity in Eq.~\eqref{eqhada} to
smear out in a way that it 
``contaminates" the calculation at field points $x'$ near the base point $x$.
In~\cite{CNOW:2019} it is shown that by obtaining the $\ell$-modes of $G_{\delta,n=0}$ and subtracting them from the $\ell$-modes of the numerical/semi-analytic $\ell$-sum, the resulting sum converges at points $x'$ much closer to $x$ than without such subtraction.
This means that a calculation of $V(x,x')$ might potentially not be needed or that, even if it is needed, it will be at a smaller region (and so  if it
 is calculated via, e.g., an expansion near $x$, then many less terms are needed in the expansion).
 The application of this subtraction trick in~\cite{JACMK2019}, enabled a practical calculation of the GF in a much larger region of pairs of points in Schwarzschild spacetime,  
which  is there applied to obtaining the  signal strength between two quantum particle detectors in this spacetime.

We finish by noting a connection for the leading term in the exact  Bessel expansion Eq.~\eqref{bessel-exp-cgl}, i.e., the term considered in Sec.~\ref{eq:num Bessel}.
By using Eq.~\eqref{legendre-bessel}, it can be seen\footnote{The statements in this sentence and the one that follows it are true modulo a factor of $4\pi$, since the definitions of $\Gl$ here and of $G_{\ell}^{\text{ret}}$ in~\cite{CNOW:2019} differ by such a factor.} that Eq.~20 in~\cite{CNOW:2019} for $\Gld$, which are the $\ell$-modes of $\theta(\Delta t)G_{\delta,n=0}$,
  reduces, for large-$\ell$, to the $k=0$ term in Eq.~\eqref{bessel-exp-cgl}.
  That is, it reduces to the term $\theta(-\sigma)\theta(\Delta t)U_0\cdot J_0\left(\left(\ell+1/2\right)\eta\right)$, except 
  for a factor $\theta(\pi-\eta)$ instead of $\theta(-\sigma)$, which is acceptable since, while
  $\Gld$ in~\cite{CNOW:2019} is a clean-cut expression for the $\ell$-modes of the direct part,  
the $k=0$ term in Eq.~\eqref{bessel-exp-cgl} does not correspond {\it just} to the direct part.


\section{Conclusions} \label{sec:conclusions}

In this paper we have derived representations for the retarded
 Green function of the wave equation for scalar field perturbations
of Schwarzschild spacetime. Our starting point is the \textit{globally valid} expression 
\be
G_R(x.x')=\frac{1}{r\cdot r'}\theta(-\sigma)\theta(\Delta t)\sum_{\ell=0}^\infty \left(\ell+\frac12\right)\mathcal{U}_\ell(x^A,x^{A'})P_\ell(\cos\theta). \label{eq:conclusions-GR} \ee
See (\ref{sch-gret-mode}) and (\ref{bessel-exp-cgl}). This expression  arises naturally by re-expressing the Schwarzschild metric as conformal to the direct product of the two-sphere and a 2-dimensional
spacetime, $\Mtwo$.
Global validity stems from the global existence (and uniqueness) of the 2-dimensional Riemann functions $\mathcal{U}_\ell$, defined throughout $\Mtwo$. A necessary ingredient of this statement is the fact that $\Mtwo$ is a causal domain: the normal neighbourhood of any event of $\Mtwo$ is the whole spacetime. This was established in \cite{PhysRevD.92.104030}. We then express $\mathcal{U}_\ell$ as a sum of Bessel functions with coefficients given by the Hadamard coefficients of the Riemann function $U$. This sum converges whenever the Hadamard series of $U$ converges: numerical evidence (see Sec.~\ref{sec:numerical} {as well as~\cite{JACMK2019,caribe2023lensing}}) indicates that this holds on regions $\Omega_p$ of $\Mtwo$ that yield results extending well beyond the normal neighbourhood of (the 4-dimensional) Schwarzschild spacetime. 

We then took two different approaches to obtain complementary details of $G_R$ on Schwarzschild spacetime. The first (see Sec.~\ref{sec:PH}) exploited a connection with the retarded Green function of Pleba{\'n}ski-Hacyan spacetime, whereas the second (see Sec.~\ref{sec:Sum Hads}) employed asymptotic expansions of Bessel functions. In each case, we obtained results describing the singularity structure of $G_R$  on Schwarzschild and its representation as, using the former approach, integrals of the Pleba{\'n}ski-Hacyan  Green function  and, using latter approach, a `sum of Hadamard forms'.
These representations make explicit
 the {\it complete}  singularity structure of the   Schwarzschild Green function 
 and is given in terms of the Hadamard coefficients of the  `background' $(1+1)$-wave equation, Eq.~(\ref{eq:background wave eq}). 
 Thus, Eq.~(\ref{eq:sum Hads}) is effectively an  extension of the Hadamard form
 for the Schwarzschild Green function, which is only
valid in normal neighbourhoods.
 Specifically, our representation shows that,
away from caustics, the singularity structure is  fourfold:
the `direct' part changes as
$\delta(\s)$, $\PV(1/\s)$, $-\delta(\s)$, $-\PV(1/\s)$, $\delta(\s)$, \dots,
whereas the `tail' part changes as
$\theta(-\s)$, $-\ln\left|\s\right|$, $-\theta(-\s)$, $\ln\left|\s\right|$, $\theta(-\s)$,\dots.
Here, $\s$  represents a globally well-defined generalization of the world-function in Schwarzschild spacetime
once a  geodesic is specified (see Eqs.~(\ref{sig-k-def}) and (\ref{eq:xpm})). 
The changes in the character of the singularity take place as the null wavefront passes through caustics of the spacetime.
We have separately analyzed the singularity of the Green function  along the line of caustics in Schwarzschild spacetime.
We have shown that, in this case,
the singularity structure is instead twofold and is `stronger' than away from caustics -- the result is a little
more involved than in the non-caustic case and is given in 
Eq.~\eqref{eq:Grl-caustic-0},
for $\gamma=0$ and in 
\eqref{eq:Grl-caustic-pi}
for $\gamma=\pi$.
The type of singular behaviour encountered here is strongly tied to the fact that we are at a caustic where, in the present case, a 2-parameter family of null geodesics from $x'$ reconverges at $x$. This is distinct to the situation addressed elsewhere in the paper, where singular behaviour is encountered at light crossing, whereat a single causal geodesic from $x'$ intersects the future light cone of $x'$. This behaviour that we found at caustics occurs in Schwarzschild spacetime and in (some) other spherically symmetric spacetimes (such as in Pleba{\'n}ski-Hacyan spacetime, see~\cite{Casals:2012px}), but it is not clear how general this behaviour is.

In~\cite{CDOWa,Casals:2012px} the following heuristic explanation was put forward for the fourfold singularity structure of the `direct' part of the Green function.
The retarded Green function can be obtained from the Feynman Green function as $G_R(x,x')=2\textrm{Re}(G_F(x,x'))\theta_+(x,x')$~\cite{DeWitt:1960}.
The Hadamard form for the Feynman Green function in $(3+1)$-dimensions is:
\begin{equation}\label{eq:Feynman GF Had}
G_F(x,x')=\lim_{\epsilon \rightarrow 0^+}\frac{i}{2\pi}\left[\frac{\U(x,x')}{\s+i\epsilon}-\V(x,x')\ln\left(\s+i\epsilon\right)+W(x,x')\right],
\end{equation}
where $W(x,x')$, like $\U$(x,x') and $\V$(x,x'), is a regular and real-valued biscalar in a normal neighbourhood of $x'$.
Since $\lim_{\epsilon \rightarrow 0^+}1/(\s + i \epsilon) =\text{P.V.}\left({\frac{1}{\s}} \right)-i\pi \delta(\s)$
and  $\lim_{\epsilon \rightarrow 0^+} \ln\left(\s+i\epsilon\right)=\ln|\s|+i\pi\theta(-\s)$, 
the Hadamard form Eq.~(\ref{eqhada}) for the retarded Green function readily follows from that of the Feynman Green function, Eq.~(\ref{eq:Feynman GF Had}).

Now, in $(3+1)$-dimensions, the biscalar $\U(x,x')$
 is related to the van Vleck determinant $\Delta(x,x')$ {in that $(3+1)$-dimensional spacetime} as $\U(x,x')=\Delta(x,x')^{1/2}$.
It can be argued~\cite{CDOWa} that, in a spherically symmetric spacetime, the van Vleck determinant picks up a phase of `$-\pi$' as the geodesic along which it is evaluated crosses a caustic point.
That is, $\Delta^{1/2}=e^{-i\pi/2}\left|\Delta\right|^{1/2}$ after the geodesic  has crossed  a first caustic.
If one then tentatively evaluated the retarded Green function from the form Eq.~(\ref{eq:Feynman GF Had}) for the Feynman Green function
 after the geodesic has crossed  a first caustic 
point (this is, of course, not rigorously justified since the Hadamard form is only valid within a normal neighbourhood, which cannot contain caustic points; however,~\cite{BUSS2018168} provides numerical evidence that this works)
the singularity $\PV(1/\s)$  -- instead of the $\delta(\s)$ -- would be obtained for the `direct' part. As the geodesic crosses later caustics, $\Delta^{1/2}$ 
picks up a phase `$-\pi/2$' every time and the fourfold
 structure (considering only the distributions, not their coefficients) for the leading singularity of the retarded Green function (that we have  derived in this paper {-- see Eq.~\eqref{4-fold}}) would ensue.
As suggested in~\cite{Casals:2012px}, if the same phase `$-\pi/2$' were picked up by the biscalar $\V(x,x')$ at each caustic crossing (this is known to be true --at least for  {$\nu_0$}, the first term in the Hadamard series for $\V(x,x')$, see Eq.~\eqref{v0def}-- in the case of PH spacetime), then the fourfold
 structure for the sub-leading discontinuity of the retarded Green function that we have derived {(see Eq.~\eqref{4-fold,sublead})} would  follow similarly.
 {The corresponding leading and subleading singularity structures for the imaginary part of the Feynman Green function would then be, respectively: $\PV(1/\s) \to -\delta(\s) \to -\PV(1/\s) \to \delta(\s) \to \PV(1/\s) \dots$ (corroborated numerically in~\cite{BUSS2018168,caribe2023lensing})
 and $-\ln\left(\s\right)\to -\theta\left(-\s\right) \to \ln\left(\s\right)\to \theta\left(-\s\right)\to -\ln\left(\s\right)$.
 }

Although we have proved these singularity structures for the retarded Green function specifically for the case of a scalar field, we expect 
that the leading singularity structures ($\delta(\s) \to \PV(1/\s)\dots$) carry over to higher-spin field perturbations of Schwarzschild spacetime.
The reason is that
the leading singularity structure of the Green function is dictated by the coefficients of the second derivatives in the wave equation --
which may be any of the generalized Regge-Wheeler or Teukolsky equations.
The spin does not appear in the second order terms of these equations.

Apart from the intrinsic theoretical interest in having a `global Hadamard form' for the retarded Green function, 
such an expression may have various applications.
For example, in the calculation of the self-force acting on a particle moving in a curved background spacetime.
In~\cite{CDOWa,CDOW13,PhysRevD.89.084021} the self-force was calculated via a time integration of the GF along the past worldline of the particle.
These calculations illustrated how the self-force may be seen as arising from backscattering of the field perturbation and from trapping of null geodesics, both of which are
perfectly encapsulated within our `global Hadamard form'.
{In its turn, in the context of quantum field theory, the global singularity structure/Hadamard form has found applications within quantum communication~\cite{JACMK2019} as well as entanglement harvesting~\cite{caribe2023lensing}.}
In terms of specific practical use, it has been found in~\cite{JACMK2019, CNOW:2019,OToole:2021} 
that subtracting, mode-by-mode, an $\ell$-mode decomposition of an approximation of
the direct term $\U(x,x')\delta(\s)$  that we have found here (specifically,
the term containing $\delta$ in Eq.~(\ref{grl-k-zero2}) for $n=0$) from a numerical
calculation (such as the one carried out in~\cite{PhysRevD.89.084021}) of the $\ell$-modes of the retarded Green function
 significantly increases the accuracy of the resulting, numerical retarded Green function.

One other possible application of our results to the self-force  would be to combine one of the methods used in~\cite{CDOW13,PhysRevD.89.084021} (these methods are valid outside the quasi-local region; 
it can either be a numerical calculation of the retarded Green function or else a semi-analytic sum of its quasinormal mode and branch cut contributions) together with our Eq.~(\ref{eq:sum Hads})
in different time regimes. 
We hope to investigate these  possible applications in the future.


\begin{acknowledgments}
We are thankful to Chad Galley, Abraham Harte, Adrian Ottewill, Lorenzo Pisani, Peter Taylor and Barry Wardell for useful discussions.
M.C. acknowledges partial financial support by CNPq (Brazil), process numbers 308556/2014-3 and 310200/2017-2. 
\end{acknowledgments}


\appendix

\section{Examples of Multipolar Modes in Other Spacetimes}\label{sec:App}
\numberwithin{figure}{section}
\setcounter{figure}{0}

For illustration purposes, in this appendix we present the multipolar modes $\Gl$ of the retarded Green function Eq.~(\ref{sch-gret-mode})  in two simple $4$-dimensional  cases\footnote{The multipolar modes  of the  direct part of the retarded Green function in Schwarzschild spacetime have been obtained in~\cite{CNOW:2019}.}: flat spacetime
and a static region of Nariai spacetime (in this case we only do it for the direct part of the Hadamard form).
N.B.: the modes  $\Gl$  of
PH spacetime are readily inferrable from Eq.~\eqref{eq:G_k^PH}.


\subsection{Flat spacetime}\label{sec:flat Gl}

Consider 4-dimensional flat spacetime. The line element in spherical coordinates {can be obtained from that in Schwarzschild in \eqref{sch-lel} by setting the mass to zero ($M=0$):}
\be
ds^2=-dt^2+dr^2+r^2d\Omega_2^2=r^2\left(ds^2_2+d\Omega_2^2\right),
\ee
where 
\be\label{eq:ds 2D conf flat}
ds^2_2=
\frac{1}{r^2}\left(-dt^2+dr^2\right)
\ee
is the line element of the $2$-D conformal space $\Mtwo$ corresponding to this case, whose Synge world function we denote in the current section by $\sigma$.

\begin{figure}[h!]
\begin{center}
 \includegraphics[width=4cm]  {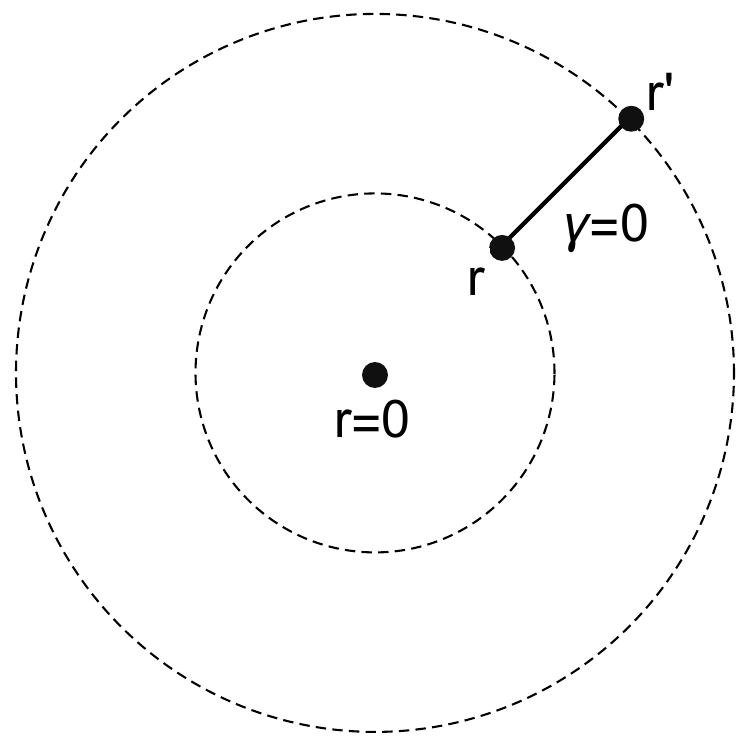}
 \hspace{2cm}
  \includegraphics[width=4cm]  {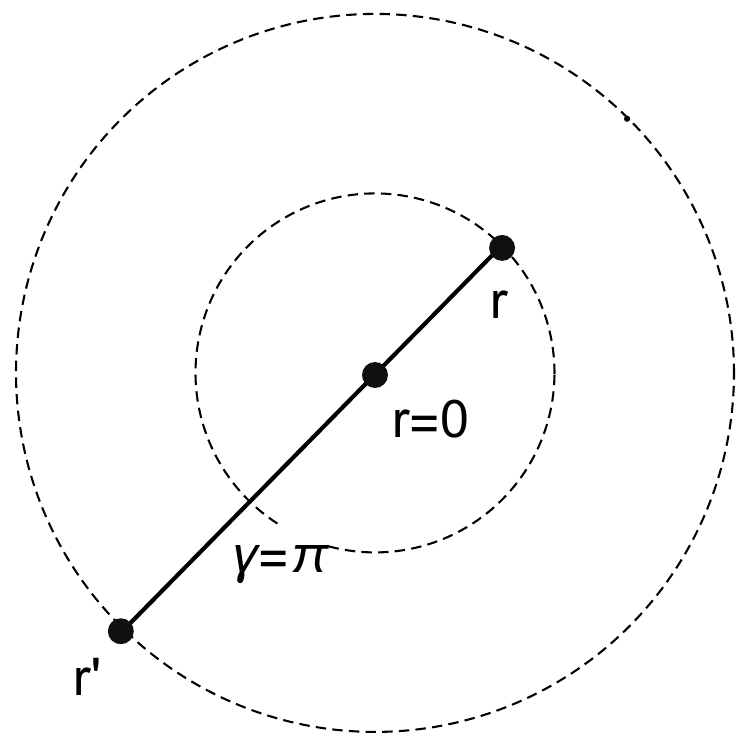}
 \end{center}
\label{fig:Circles Flat}
\caption{Illustration of two spatial points in flat spacetime at the shortest (a) and largest (b) possible spatial distances given fixed radii $r$ and $r'$.}
\end{figure}

The Hadamard form for the GF of the wave equation for a massless scalar field in flat spacetime is valid everywhere in the spacetime.
In spherical coordinates it is given by:
\begin{equation}\label{eq:GR flat}
\GFret=\theta(\Delta t)\delta(\s)=
\frac{\theta(\Delta t)}{r\cdot r'}
\delta\left(z-u\right),
\end{equation}
where $\gamma$ is the angular separation between the two points,
\be
z\equiv \cos\gamma,
\quad u\equiv \frac{-\Delta t^2+r^2+r'^2}{2r\cdot r'},
\ee
and
$\s=r\cdot r'(u-z)$ 
is Synge's world function in flat spacetime 
(after a trivial use of the law of cosines).
From Eq.~(\ref{sch-gret-mode}), and making use of the orthogonality of the Legendre polynomials, the multipolar modes $\Gl$ can  readily be  evaluated as:
\begin{equation} \label{eq:G_l flat}
\Gl=2\pi r\cdot r'\int_{-1}^{+1}
dz\, P_{\ell}(z)
\GFret=
2\pi\, 
\theta(\Delta t)\, \theta\left(-\s(\gamma=0)\right)\theta\left(\s(\gamma=\pi)\right)
P_\ell\left(u\right).
\end{equation}
The presence of the $\theta$-step distributions comes from ensuring that the argument of the second $\delta$-distribution in Eq.~(\ref{eq:GR flat})
 is within the limits of integration in the integral in Eq.~(\ref{eq:G_l flat}).
We note that $\theta(\Delta t)\theta\left(-\s(\gamma=0)\right)=\theta\left(\Delta t-|r-r'|\right)$ corresponds to the shortest possible spatial distance
between two points in flat spacetime given $\Delta t$ positive), $r$ and $r'$,
 whereas 
$\theta(\Delta t)\theta\left(\s(\gamma=\pi)\right)=\theta(\Delta t)\theta\left(r+r'-\Delta t\right)$ corresponds to the largest possible one -- see Fig.~\ref{fig:Circles Flat} for an illustration. 
This combination of $\theta$'s is, therefore, as constraining as possible if their arguments are to be $\ell$-independent: there are no two
points in flat spacetime  with
$\Delta t<|r-r'|$ or with $r+r'<\Delta t$ which may be joined by a null geodesic (and a massless field in flat spacetime propagates purely along null geodesics).

We also note that the Legendre function $P_\ell\left(u\right)$ in Eq.~(\ref{eq:G_l flat}) is precisely the Riemann function {$\mathcal{U}_{\ell}$} of the 
wave equation {Eq.~\eqref{gl-pde-coords}} on the $2$-D conformal space with $M=0$ (see also App.~\ref{sec:M=0}). This is because: (1) it is a homogeneous solution of this equation,
\be
\left(-\partial_t^2+\partial_r^2-\frac{\ell (\ell+1)}{r^2} \right)P_\ell\left(\frac{-\Delta t^2+r^2+r'^2}{2r\cdot r'}\right)=0,
\ee 
and (2) it satisfies the boundary condition: $P_\ell\left(u=1\right)=1$, as can be readily checked.
We note that $\theta\left(-\s(\gamma=0)\right)=\theta(-\sigma)$ and that, in normal neighbourhoods in $\Mtwo$ (as explained below Eq.~\eqref{con-sch-lel,M=0}, $\Mtwo$ is not a convex domain when $M=0$, so that, in this case, the maximal normal neighbourhood of a point is not the whole of $\Mtwo$), $\theta\left(\s(\gamma=\pi)\right)=1$.
Therefore, in  normal neighbourhoods in $\Mtwo$, the expression (\ref{eq:G_l flat}) for $\Gl$  reduces to the expression \eqref{bessel-exp-cgl}
for the $\ell$-modes of the GF, as expected.
Thus, for $M=0$, the exact \eqref{eq:G_l flat}
 extends globally the expression \eqref{bessel-exp-cgl} which is only valid in normal neighbourhoods in $\Mtwo$, rendering the $\ell$-modes exactly zero beyond these neighbourhoods due to the distributional factor $\theta\left(\s(\gamma=\pi)\right)$ (such a Heaviside distribution cannot appear in an expression
 for $\Gl$ in a general spherically-symmetric spacetime because, as opposed to the flat spacetime case here, the Hadamard tail is generally non-zero).

\subsection{Nariai spacetime}

In a static region of Nariai spacetime ($dS_2\times \mathbb{S}_2$) the line-element can be written as~\cite{CDOWa}
\begin{equation}
ds^2=-(1-\rho^2)dt^2+(1-\rho^2)^{-1}d\rho^2+d\Omega_2^2,
\end{equation}
where $\rho \in (-1,+1)$, $t\in(-\infty,+\infty$) and we have set the cosmological constant to be equal to unity.
 Synge's world function is given by $\s=\sigma_2+\gamma^2/2$ where
 $\sigma_2$ is the world function in two-dimensional de Sitter spacetime, $dS_2$, and is given by:
\be 
\cosh \sqrt{-2\sigma_2}=\rho\cdot \rho'+(1-\rho^2)^{1/2}(1-\rho'^2)^{1/2}\cosh\Delta t.
\ee
In this case, we were not able to find the multipolar decomposition  $\Gl$ of the full $G_R$. However, it is straight-forward to find  the multipolar decomposition of the direct
part of the Hadamard form for $G_R$.
This direct part is given by $G^{dir}(x,x')\equiv \U(x,x')\delta(\s)\theta(\Delta t)$ where~\cite{CDOWa,Casals:2012px}
\be
\U(x,x')=\left(\frac{\sqrt{-2\sigma_2}}{\sinh \sqrt{-2\sigma_2}}\right)^{1/2}\left(\frac{\gamma}{\sin \gamma}\right)^{1/2}
\ee
and 
\be
\delta(\s)=\frac{\delta(\gamma+\sqrt{-2\sigma_2})+\delta(\gamma-\sqrt{-2\sigma_2})}{|\sqrt{-2\sigma_2}|}.
\ee
We can now readily calculate the $\ell$-modes of the direct part,
\begin{equation}
\int_{-1}^{+1}d(\cos\gamma)P_{\ell}(\cos\gamma)G^{dir}=
\theta(\Delta t)\theta\left(-\s(\gamma=0)\right)\theta\left(\s(\gamma=\pi)\right)P_\ell\left(\cos\sqrt{-2\sigma_2}\right)\left(\frac{\sin\sqrt{-2\sigma_2}}{\sinh\sqrt{-2\sigma_2}}\right)^{1/2}.
\end{equation}
The causality structure is equivalent to that in flat spacetime, Eq.~(\ref{eq:G_l flat}).


\section{Case of mass $M=0$}\label{sec:M=0}

In this appendix, we consider the case  of vanishing mass, $M=0$. This allows us to investigate the convergence of the Hadamard series \eqref{eq:Had series 2d} in this limiting case, and to make connections between Green functions on different maximally symmetric manifolds, and direct products of such manifolds. As a by-product, we obtain an improved approach to the calculation of Hadamard coefficients in certain cases, via the result of Szego mentioned above \cite{vonSzego01011934}.


For $M=0$, the line-element \eqref{con-sch-lel} of conformal Schwarzschild spacetime becomes
\be d\hat{s}^2  = -\frac{1}{r^2}(dt^2-dr^2)+d\Omega_2^2.\label{con-sch-lel,M=0}\ee
This can readily be recognized as the line-element of Bertotti-Robinson spacetime, $AdS_2\times \mathbb{S}_2$~\cite{Bertotti,{robinson1959solution}}, i.e.,
$\Mtwo$ is (a patch of) two-dimensional anti-de Sitter spacetime $AdS_2$.
In this case we have an `enhanced' symmetry between the two factors of conformal Schwarzschild spacetime $\sth$, namely between the factor $\Mtwo=AdS_2$ and 
 the factor $\mathbb{S}_2$: both are maximally-symmetric manifolds and the corresponding Riemann functions for certain differential operators 
 are formally the same (as functions of their corresponding geodesic distances), as we shall {next} see. 
 Before that, though, we point out that the global structures of $AdS_2$ and the 2-dimensional conformal Schwarzschild spacetime for $M\neq 0$ are very different. The former is not geodesically convex: there are events $x$ in the causal future of any event $x'$ which are not linked by any future-directed causal geodesic from $x'$ to $x$. (We note that the proof of the geodesic convexity of the 2-dimensional conformal Schwarzschild spacetime of \cite{PhysRevD.92.104030} is not valid when $M=0$.)  We can link this to the fact that the `soft' potential barrier of $\Mtwo$ at $r=3M$ (see \cite{PhysRevD.92.104030}) is replaced by a repulsive barrier at $r=0$ which forms part of the boundary of the patch of $AdS_2$ under consideration: $\{r=0\}$ is accessible only to null geodesics of infinite affine length. However, this difference does not impact the analysis below which applies on normal neighbourhoods of $AdS_2$. In particular, the analysis is valid for regions of the form $\{x^A: \eta(x^A,x^{A'})\in[0,\pi)\}$ where $\eta(x^A,x^{A'})$ is the geodesic distance
 along 
 geodesics emanating from $x^A\in AdS_2$: geodesics remain unique for (at least) this amount of 
 geodesic distance.  So, in the particular case that $x^A$ and $x^{A'}$ are connected by a timelike geodesic, $\eta$ is the proper time along that geodesic.

As seen in Section \ref{sec:GF} above, the 4-dimensional retarded Green function $G_R$ can be constructred from a sequence of 2-dimensional retarded Green functions $G_\ell, \ell\geq 0$ - which in the present case we write as $G_{\ell,AdS_2}$. This takes the form {(see \eqref{bessel-exp-cgl})}\be G_{\ell,AdS_2} = 2\pi\mathcal{U}_\ell\theta(-\sigma_{AdS_2})\theta(\Delta t), \label {eq:Gl-AdS2}\ee
where{\footnote{In the generic-mass case throughout the main body of the paper, $\sigma_{AdS_2}$ is denoted by $\sigma$.}}
$\sigma_{AdS_2}=\epsilon\,\eta^2/2$ 
and  $\epsilon=-1,0,+1$  for, respectively, timelike, null, and spacelike
geodesics.
 The two-point function $\mathcal{U}_\ell$ is the Riemann function of the operator $P=\Box_2+\lambda$  in the case  $\lambda=-\ell(\ell+1)$ {(i.e.\ the operator in \eqref{cgl-eqn} with $M=0$)}. It is convenient to consider the Riemann function $U$  of $P$ more generally, which satisfies
\be (\Box_2+\lambda)U=0,\quad \left.U\right|_{\eta=0}  = 1. \label{eq:Riem-AdS2} \ee
In this maximally symmetric spacetime, $U$ depends only on the geodesic distance $\eta$ (see e.g.\ \cite{allen1986vector}). So $U=U(\eta)$, and we can then show that 
\be U''+\cot\eta U' -\lambda U=0,\quad U(0)=1,\label{eq:U-Leg} \ee
where the prime represents derivative with respect to argument. This problem has a unique solution given by the Legendre function $U=P_\nu(\cos\eta)$ with $\nu\equiv (-1+\sqrt{1-4\lambda})/2$. We note that the second linearly independent solution of the ODE (\ref{eq:U-Leg}) is singular at $\eta=0$). Specialising to the case $\lambda=-\ell(\ell+1)$, we see that {(as already noted in App.~\ref{sec:flat Gl})}
\be \mathcal{U}_\ell(x^A,x^{A'}) = P_\ell(\cos\eta),\quad \eta\in[0,\pi),\quad \ell\geq 0. \label{eq:Ul BR} \ee
This is the Riemann function of the operator $\Box_2-\ell(\ell+1) = \Box_2+\frac14-L^2$ where $L=\ell+1/2$. Thus in this case, the `background' wave operator is $\Box_2+\frac14$ (consistent with (\ref{pop-def}) in the case $M=0$), and the background Riemann function $U$ (cf. Eq. (\ref{eq:Had series 2d})) is given in this case by setting $\lambda=1/4$, which yields $\nu=-1/2$:
\be\label{eq:U M=0}
U(x^A,x^{A'})=P_{-1/2}(\cos\eta),
\quad \eta\in[0,\pi).
\ee

Alternatively, we can use the fact that, for $M=0$, Schwarzschild spacetime is just Minkowski spacetime and so its GF is just (see \eqref{eq:GR flat}\footnote{In normal neighbourhoods of $\Mtwo=AdS_2$ and for causal separations, $u$ in \eqref{eq:GR flat} is equal to $\cos\eta$ here.} but also written here in terms of van Vleck determinants as per, e.g.,~\cite{CNOW:2019}):
\be\label{eq:GR flat,VVs}
G_R(x,x')=\theta(\Delta t)\delta(\s)=\frac{\theta(\Delta t)}{r\cdot r'}
\Delta_{2d}^{1/2}\Delta_{\mathbb{S}_2}^{1/2}\delta(\hat\sigma)=
\frac{1}{r\cdot r'}
\frac{\delta(\eta-\gamma)}{\sin\eta},
\ee
where 
\be
\Delta_{2d}=\frac{\eta}{\sin\eta},\quad
\Delta_{\mathbb{S}_2}=\frac{\gamma}{\sin\gamma}
\ee
are the van Vleck determinants in $\Mtwo=AdS_2$ (already introduced in \eqref{U0-def}) and in $\mathbb{S}_2$, respectively. Using the standard distributional representation of  the Dirac delta function in terms of Legendre polynomials (e.g, Eq.~1.17.22~\cite{NIST:DLMF}), 
\be \label{eq:Delta-Trig}
\sum_{k=-\infty}^{\infty}
\frac{\delta(\eta+2\pi+\gamma)+\delta(\eta+2\pi-\gamma)}{\left|\sin\eta\right|}=
\delta(\cos\eta-\cos\gamma)=\sum_{\ell=0}^{\infty}\left(\ell+\frac12\right)P_\ell(\cos\eta)P_\ell(\cos\gamma).
\ee
Using this representation in \eqref{eq:GR flat,VVs}, yields
\be\label{eq:GR flat1}
G_R(x,x')=
\frac{1}{r\cdot r'}
\sum_{\ell=0}^{\infty}\left(\ell+\frac12\right)P_\ell(\cos\eta)P_\ell(\cos\gamma),
\quad \text{for}\ \eta\in(0,\pi).
\ee
Comparing this expression with Eqs.~\eqref{sch-gret-mode} and \eqref{bessel-exp-cgl}, we reproduce (\ref{eq:Ul BR}).

A third quite distinct approach is available to us. This is to use a Hadamard series representation for the Riemann functions $U$ and $\mathcal{U}_\ell$. As seen above, Zauderer's theorem allows us to relate the latter to the former. That is, writing {(see \eqref{eq:Had series 2d})}
\be U = \sum_{k=0}^\infty U_k\frac{\eta^{2k}}{k!} \label{eq:Had Series AdS2} \ee
yields {(see \eqref{bessel-exp-cgl})}
\be \mathcal{U}_\ell = \sum_{k=0}^\infty U_k \left(\frac{2\eta}{L}\right)^k J_k(L\eta),\quad L\geq 0. \label{eq:Had Series Zauder} \ee
The Hadamard coefficients $U_k=U_k(\eta)$, $k\geq 0$, obey a sequence of transport equations, and can, in theory, be calculated recursively. 

On the other hand, since $\mathcal{U}_\ell=P_\ell(\cos\eta)$, Szego's result expressing $P_\ell(\cos\eta)$ as a series of Bessel functions provides a very similar representation: we have 
\be P_\ell(\cos\eta)=
\Delta_{2d}^{1/2}
\sum_{k=0}^{\infty}
V_k(\eta)
\frac{J_k\left(L\eta\right)}{L^k}.
\label{eq:AdS2-Szego}
\ee

The coefficients $V_k$ may be obtained in the following manner as given in \cite{vonSzego01011934}:
\begin{equation} \label{eq:coeffs a}
V_0(\gamma)=1, \quad
V_j(\gamma)=(2j-1)!!\ \gamma^j\psi_j(\gamma),\ \forall j>0,
\end{equation}
where the $\psi_k$ are given via
\begin{equation}\label{eq:psi_k}
\left[1+\sum_{k=1}^{\infty}(\gamma^2-v^2)^k\phi_{k+1}(\gamma)\right]^{-1/2}=\sum_{k=0}^{\infty}(\gamma^2-v^2)^k\psi_k(\gamma),
\end{equation}
with $v\in[0,\gamma]$ and
\begin{equation}\label{eq:phi_k}
\phi_k(\gamma)\equiv \frac{(2\gamma)^{1-k}}{k!}\frac{J_{k-1/2}(\gamma)}{J_{1/2}(\gamma)},
\end{equation}

Crucially for the current discussion, Szego provides a convergence result for the series (\ref{eq:AdS2-Szego}): this is uniformly convergent on any interval $[0,\gamma_0-\epsilon]$ with $0<\epsilon<\gamma_0$ where  $\gamma_0=2(\sqrt{2}-1)\pi\simeq 0.828\pi$.

Thus we have two representations for the Riemann function $\cgl$ in \eqref{eq:Ul BR} in terms of Bessel functions: 
comparing (\ref{eq:Ul BR}), (\ref{eq:Had Series Zauder}) and (\ref{eq:AdS2-Szego}), we see that 
\be
\sum_{k=0}^{\infty}
\left(2\eta\right)^kU_k(\eta)
\frac{J_k\left(L\eta\right)}{L^k}=
P_\ell(\cos\eta)=
\Delta_{2d}^{1/2}
\sum_{k=0}^{\infty}
V_k(\eta)
\frac{J_k\left(L\eta\right)}{L^k},
\ee
and so
\be
\label{eq:exp U-V}
0=\sum_{k=0}^{\infty}
C_k
\frac{J_k\left(L\eta\right)}{L^k},\quad
C_k\equiv \left(2\eta\right)^kU_k(\eta)-\Delta_{2d}^{1/2}V_k(\eta).
\ee
Furthermore, we know from
\eqref{U0-def} that 
$U_0=\Delta_{2d}^{1/2}$ and from
\eqref{eq:coeffs a} that $V_0=1$, so that
$C_0=0$.
We now prove that
$C_k=0$ for any $k>0$.

Crucial to the proof are the following ``orthogonality" properties of the Bessel functions with respect to their argument (Eqs.~(10.22.62) and (10.22.63) in Ref.~\cite{NIST:DLMF}):
\be\label{eq:orth J}
\int_0^{\infty}dL\, L^{m-k+1} J_m\left(L\eta\right)J_k\left(L\eta\right)=
\begin{cases}
      0, & \text{if $k>m>-1$ and $m\neq k$ and $m\neq k-1$,}\\
      \frac{1}{2\eta}, & \text{if $m=k-1$.}
    \end{cases}   
\ee
So, multiplying \eqref{eq:exp U-V} by $L\,J_0\left(L\eta\right)$, integrating over $L:0\to \infty$  (the relevant expansions above being valid for all $L\geq 0$: see Eq.\ 15 on page 58 of \cite{Erdelyi:1953}, volume 2 and  Theorem 6.4.2 of \cite{friedlander}) and using \eqref{eq:orth J} together with $C_0=0$, it readily follows that $C_1=0$ (interchange of the integral and the sum is permitted by uniform convergence of the series).
Next, we multiply \eqref{eq:exp U-V} by $L^2J_1\left(L\eta\right)$, integrate over $L:0\to \infty$ and use \eqref{eq:orth J} together with $C_0=C_1=0$, from which it follows  that $C_2=0$.
Proceeding similarly for $k=3,4,\dots$, it follows that $C_k=0$, and so \be\label{eq:Uk-Vk,M=0}
U_k=\left(2\eta\right)^{-k}\Delta_{2d}^{1/2}V_k(\eta), \quad \forall k\geq 0.
\ee

We thus  have that, in this zero-mass case, the 
terms in the
Bessel series \eqref{bessel-exp-cgl} are the same as the terms in 
 Szego's expansion \eqref{legendre-bessel}, and therefore the former series converges where we already know that the latter does: namely, in the case $M=0$, \eqref{bessel-exp-cgl}
 converges uniformly in (at least)  $\eta\in [0,\gamma_0)$, with  $\gamma_0=2(\sqrt{2}-1)\pi$.

 Two further observations are worth making, the first in relation to the calculation of Szego's coefficients $V_k$, or equivalently, the Hadamard coefficients $U_k$. As noted, the latter coefficients are determined by a sequence of transport equations. These take the form 
 \be 2\eta U_k'+(2k-1+\eta\cot\eta)U_k = -\frac12\left(U_{k-1}''+\cot\eta U_{k-1}' -\frac14 U_{k-1}\right),\quad k\geq 1,\label{eq:Uk recur} \ee
 and where $U_0=(\eta/\sin\eta)^{1/2}$. The (exact) integration is non-trivial, but is amenable to calculation using a computer algebra system. Szego's formulas (\ref{eq:coeffs a})--(\ref{eq:phi_k}) provide a purely algebraic approach to the calculation of the coefficients $V_k$. However, the calculations involved are again non-trivial - but are simplified via the following approach. 

 In (\ref{eq:psi_k}), let $\rho \equiv \gamma^2-v^2$ and write $\sum_{k=1}^\infty \phi_{k+1}\rho^k =: S$. Then we have 
 \be
 \sum_{k=0}^\infty \psi_k\rho^k =(1+S)^{-1/2} = \sum_{n=0}^\infty \left(\frac{-1}{2}\right)^n\frac{(2n-1)!!}{n!}S^n. \label{eq:Szego-Alt-1} \ee
 Let $D$ be the differential operator $\displaystyle{\frac{d}{d\rho}}$. Then 
 \begin{eqnarray}
     \psi_k &=& \left.\frac{1}{k!}D^k\left(\sum_{n=0}^\infty \psi_n\rho^n\right)\right|_{\rho=0}\nonumber \\
     &=& \frac{1}{k!} \sum_{n=0}^\infty \left(\frac{-1}{2}\right)^n\frac{(2n-1)!!}{n!}\left.D^k(S^n)\right|_{\rho=0}. \label{eq:Szego-Alt-2} 
 \end{eqnarray}
 Applying Fa\`{a} di Bruno's theorem for higher order derivatives of compositions (the generalised chain rule; see e.g.\ \cite{krantz2002primer}), we have, for $n\geq 1$,
 \be \left.D^k(S^n)\right|_{\rho=0}=
 \left\{ \begin{array}{cl}
      0, & k<n; \\
      k!\hat{B}_{k,n}(\phi_2,\dots,\phi_{k-n+2}),& k\geq n
 \end{array}
 \right.
 \label{eq:psi-Bell} 
 \ee
 where 
  \be 
 \hat{B}_{k,n}(x_1,x_2,\dots,x_{k-n+1}) = \sum \frac{n!}{m_1!m_2!\cdots m_{k-n+1}!}x_1^{m_1}x_2^{m_2}\cdots x_{k-n+1}^{m_{k-n+1}},\label{eq:Bell-def} \ee
 with the sum taken over all finite sequences of non-negative integers $(m_1,m_2,\dots,m_{k-n+1})$ such that 
 \be m_1+m_2+\cdots+m_{k-n+1} = n,\label{eq:Bell-index1} \ee
 and
 \be m_1+2m_2+\cdots+(k-n+1)m_{k-n+1} = k.\label{eq:Bell-index2} 
 \ee
The functions $\hat{B}_{k,n}$ are Bell polynomials, and they may be calculated either directly using (\ref{eq:Bell-def}), or via recursion formulae \cite{comtet1974advanced}. It follows that $\psi_0=1$, and 
\be \psi_k = \sum_{n=1}^k \left(\frac{-1}{2}\right)^n\frac{(2n-1)!!}{n!}\hat{B}_{k,n}(\phi_2,\dots,\phi_{k-n+2}), \label{eq:psi-Bell} \ee
which (along with (\ref{eq:coeffs a})) gives an explicit formula for Szego's coefficients $V_k$, and hence for the Hadamard coefficients $U_k$. 

Finally, we note a connection between the Hadamard coefficients considered here, and the Hadamard coefficients of the 4-dimensional Pleba{\'n}ski-Hacyan (PH) spacetime considered in Sec.~\ref{sec:PH} above. In  PH spacetime, the direct product of 2-dimensional Minkowski spacetime $\mathbb{M}_2$ and the (unit) 2-sphere $\mathbb{S}_2$,  the world function decomposes as 
\be \sigma_{PH}= \sigma_{\mathbb{M}_2}+\sigma_{\mathbb{S}_2}. \label{eq:sigma-PH} \ee
It is straightforward to show that 
\be \Box_{PH}\sigma_{PH}  = 3+\gamma\cos\gamma, \label{eq:BoxSigmaPH} \ee
where $\gamma$ is geodesic distance on $\mathbb{S}_2$ {and $\Box_{PH}$ is the D'Alembertian in PH}. Writing the tail term of the retarded Green function of the wave operator  $P = \Box_{PH} -\frac14$ in the form of a Hadamard series (see Eq.~(\ref{udef}))
\be V_{PH} = {\sum_{n=0}^\infty \nu_n}\sigma_{PH}^n, \label{eq:VPH} \ee
we can show that the Hadamard coefficients {$\nu_n$} depend only on $\gamma$. (We note that this operator is the one considered in Sec.~\ref{sec:PH} above: $P=\Box_{PH}-(m^2+2\xi)$ with $m^2+2\xi=1/4$.) Defining $\bar{V}_n$ by
\be \nu_n(\gamma) = \frac{2^n}{n!}\bar{V}_n(\gamma), \label{eq:VbarPH} \ee
we can show that the $\bar{V}_n$ satisfy the sequence of transport equations 
\be 2\gamma\bar{V}_k'+(2k+1+\gamma\cot\gamma)\bar{V}_k = -\frac12\left(\bar{V}_{k-1}''+\cot\gamma\bar{V}_{k-1}'-\frac14 V_{k-1}\right),\quad k\geq 1. \label{eq:VkPhrecur} \ee
These are precisely the transport equations (\ref{eq:Uk recur}) for the Hadamard coefficients of $AdS_2$ - but with a shift of index. We calculate that $\bar{V}_0(\gamma) = U_1(\gamma)$ (note the change of argument in $U_1$), and we
can conclude that, in this case of $M=0$,
\be 
{
\frac{k!}{2^k}
 \nu_k(\eta)=}
\bar{V}_k(\eta) = U_{k+1}(\eta),\quad k\geq 0. \label{eq:VkPH=Uk}\ee
Thus the Hadamard coefficients $ \nu_k$ of 4-dimensional PH spacetime are essentially the Hadamard coefficients {$ U_k$} of $AdS_2$, and can therefore be calculated using Bell polynomials {(see Eqs.~\eqref{eq:coeffs a} and \eqref{eq:psi-Bell})}, via Szego's theorem. 
 

\section{Regularized Self-field}\label{sec:SF}

We here show how
our results could be used to calculate the regularized self-field $\Phi_R$ (that is, the regularized value of the scalar field created by a scalar point charge, evaluated on the world-line of the charge itself) in Schwarzschild spacetime - see \cite{CNOW:2019}. As noted earlier, the self-field is relevant to self-force calculations of radiation reaction: the self-force is the covariant derivative of the regularized self-field. A practical calculation of the regularized self-field can be achieved by
integrating what we refer to as the \textit{non-direct} GF $\Gnd(x,x')$ over the worldline of the scalar charge\footnote{Another expression for the regularized self-field (see, e.g.,~\cite{Poisson:2011nh}) is like Eq.~\eqref{eq:SF} but with  the integrand being  the
full GF $G_R$ instead of its non-direct part $\Gnd$ and the  upper limit of integration being $\tau^-$ instead of $\tau$; in that case, the upper limit  
$\tau^-$ would effectively remove the sharp divergence of the integrand $G_R(z(\tau),z(\tau'))$ at coincidence which is alternatively removed in \eqref{eq:SF} by having subtracted from the GF its divergence at coincidence (see \eqref{eq:Gnd Had}).}:
\be\label{eq:SF}
\Phi_R(\tau)=\int_{-\infty}^
\tau
d\tau' \Gnd(z(\tau),z(\tau'))
\ee
where $\tau$ is the proper time along the world-line {$z(\tau)$}  of the charge and
\begin{align}\label{eq:Gnd Had}
\Gnd(x,x')\equiv
\left\{\begin{array}{l l}
\Gret(x,x')-
{\Gd(x,x')}
=
\V(x,x')\theta(-\s)\theta(\dt), 
& x'\in \mathcal{N}(x), \\
\displaystyle
\Gret(x,x'),&x'\notin \mathcal{N}(x).
\end{array}
\right.
\end{align}
{where
\be\label{eq:Gd}
\Gd(x,x')\equiv \U(x,x')\delta(\s)\theta(\dt)
\ee
is the direct part of the GF. For obvious reasons, the integral in \eqref{eq:SF} is usually called the tail integral.}
One may calculate the non-direct {part of the} GF by carrying out an $\ell$-mode decomposition as
\begin{align} \label{eq:Green nd}
&
\Gnd(x,x')\equiv
\frac{1}{4\pi r\, r'}
\sum_{\ell=0}^{\infty}(2\ell+1)P_{\ell}(\cos\gamma)\left(\Glret(r,r';\dt)-\Gld(r,r';\dt)\right)
\end{align}
where
\begin{equation} \label{eq:Gld}
  \Gld(r,r';\dt)=
2\pi\theta(\dt)
   \theta(\pi-\eta) U_0(x^A,x^{A'}) P_{\ell}(\cos\eta)\left(\frac{\sin\eta}{\eta}\right)^{1/2}
\end{equation}
are the modes of the direct part {$\Gd$} as obtained in Eq.~20~\cite{CNOW:2019} {and $U_0$ is the first Hadamard series coefficient in \eqref{eq:Had series 2d}}.
In Ref.~\cite{CNOW:2019},  $\Glret$ as well as $U_0$, $\eta$ and, with them,  $\Gld$, were all evaluated numerically to obtain the regularized self-field $\Phi_R$. In here, we instead use results in the main part of our paper  in order to give an analytic expression for the integrand in the regularized self-field expression \eqref{eq:SF}.
Using Eq.~\eqref{legendre-bessel} for $P_{\ell}$ in the modes \eqref{eq:Gld}  of the direct part and Eq.~\eqref{bessel-exp-cgl} for the modes $\Glret$ of the full GF, we notice a remarkable similarity between the representations for the two sets of modes, which allows us to write:
\be 
\Glret(r,r';\dt)-\Gld(r,r';\dt)=
 2\pi \theta(\Delta t) 
 \sum_{k=0}^\infty 
\left( 
\theta(-\sigma)  U_k\left(2\eta\right)^k-
 \theta(\pi-\eta)U_0\, {V_k}(\eta) 
\right)
\frac{J_k(\lam\eta)}{\rlam^k}.
\ee
We note that the expressions for $G_R$ in Section \ref{sec:Sum Hads} \textit{et seq.} would also apply to the non-direct part $\Gnd$ of the GF merely by applying the replacement $U_k \to U_k-U_0V_k(\eta)/(2\eta)^k$ in (\ref{alpha-hat-def1}) for $\eta<\pi$, and with no replacement for $\eta\geq \pi$.
The regularized self-field \eqref{eq:SF} can now be written in terms of the non-direct GF as
\be
\Phi_R(\tau)=
\frac{1}{r}
\sum_{\ell=0}^{\infty}\left(\ell+\frac{1}{2}\right)
\int_{-\infty}^{\tau^-}d\tau'
\frac{P_{\ell}(\cos\gamma)}{r'}
\sum_{k=0}^\infty 
\left( 
U_k\left(2\eta\right)^k-
 \theta(\pi-\eta)U_0\, {V_k}(\eta) 
\right)
\frac{J_k(\lam\eta)}{\rlam^k},
\ee
where we have used the fact that $\theta(-\sigma)=1$ along the worldline.  
This integral can be split as 
\be\label{eq:self-field ND}
\Phi_R(\tau)=
\int_{-\infty}^{\tau_0}d\tau' \Gret(z(\tau),z(\tau'))+
\frac{1}{r}
\sum_{\ell=0}^{\infty}\left(\ell+\frac{1}{2}\right)
\int_{\tau_0}^{\tau}d\tau'
\frac{P_{\ell}(\cos\gamma)}{r'}
\sum_{k=1}^\infty 
\left( 
U_k\left(2\eta\right)^k-
U_0\, {V_k}(\eta) 
\right)
\frac{J_k(\lam\eta)}{\rlam^k},
\ee
where $\tau_0<\tau$ is the value of the proper time corresponding to $\eta=\pi$.
The nice feature here is that we have subtracted the direct part from the GF in an exact analytical manner (note the starting value of the sum at $k=1$, as there has been an exact cancellation of the $k=0$ term).
{We note that, in the massless case $M=0$ and using Eq.~\eqref{eq:Uk-Vk,M=0}, the $k$-sum inside the second partial tail integral in \eqref{eq:self-field ND} vanishes (in agreement with the fact that the massless scalar field in flat spacetime has no tail) {\it term-by-term}.
}



\begin{thebibliography}{36}
\expandafter\ifx\csname natexlab\endcsname\relax\def\natexlab#1{#1}\fi
\expandafter\ifx\csname bibnamefont\endcsname\relax
  \def\bibnamefont#1{#1}\fi
\expandafter\ifx\csname bibfnamefont\endcsname\relax
  \def\bibfnamefont#1{#1}\fi
\expandafter\ifx\csname citenamefont\endcsname\relax
  \def\citenamefont#1{#1}\fi
\expandafter\ifx\csname url\endcsname\relax
  \def\url#1{\texttt{#1}}\fi
\expandafter\ifx\csname urlprefix\endcsname\relax\def\urlprefix{URL }\fi
\providecommand{\bibinfo}[2]{#2}
\providecommand{\eprint}[2][]{\url{#2}}

  \bibitem[{\citenamefont{OToole}(2019)}]{OToole:2021}
\bibinfo{author}{\bibfnamefont{C.} \bibnamefont{{O'Toole}}},
\bibinfo{author}{\bibfnamefont{A.C.} \bibnamefont{Ottewill}}
\bibnamefont{and}
\bibinfo{author}{\bibfnamefont{B.} \bibnamefont{Wardell}},
    \bibinfo{journal}{Phys. Rev.} \textbf{\bibinfo{volume}{D103}},
  \bibinfo{pages}{124022} (\bibinfo{year}{2021}),
  \eprint{2010.15818}.



\bibitem[{\citenamefont{Casals et~al.}(2019)}]{CNOW:2019}
\bibinfo{author}{\bibfnamefont{M.} \bibnamefont{Casals}}, 
\bibinfo{author}{\bibfnamefont{B.} \bibnamefont{Nolan}},
\bibinfo{author}{\bibfnamefont{A.C.} \bibnamefont{Ottewill}}
\bibnamefont{and}
\bibinfo{author}{\bibfnamefont{B.} \bibnamefont{Wardell}},
    \bibinfo{journal}{Phys. Rev.} \textbf{\bibinfo{volume}{D100}},
  \bibinfo{pages}{104037} (\bibinfo{year}{2019}),
  \eprint{1910.02567}.

\bibitem[{\citenamefont{Jonsson et~al.}(2019)}]{JACMK2019}
\bibinfo{author}{\bibfnamefont{R.~H.} \bibnamefont{Jonsson}}, 
\bibinfo{author}{\bibfnamefont{D.~Q.} \bibnamefont{Aruquipa}}, 
\bibinfo{author}{\bibfnamefont{M.} \bibnamefont{Casals}}, 
\bibinfo{author}{\bibfnamefont{A.} \bibnamefont{Kempf}},
\bibnamefont{and}
\bibinfo{author}{\bibfnamefont{E.} \bibnamefont{Mart{\'\i}n-Mart{\'\i}nez}}
    \bibinfo{journal}{Phys. Rev.} \textbf{\bibinfo{volume}{D101}},
  \bibinfo{pages}{125005} (\bibinfo{year}{2020}),
  \eprint{2002.05482}.


\bibitem[{\citenamefont{Leaver}(1986)}]{Leaver:1986}
\bibinfo{author}{\bibfnamefont{E.~W.} \bibnamefont{Leaver}},
  \bibinfo{journal}{Phys. Rev. D} \textbf{\bibinfo{volume}{34}},
  \bibinfo{pages}{384} (\bibinfo{year}{1986}).

\bibitem[{\citenamefont{Poisson et~al.}(2011)\citenamefont{Poisson, Pound, and
  Vega}}]{Poisson:2011nh}
\bibinfo{author}{\bibfnamefont{E.}~\bibnamefont{Poisson}},
  \bibinfo{author}{\bibfnamefont{A.}~\bibnamefont{Pound}} \bibnamefont{and}
  \bibinfo{author}{\bibfnamefont{I.}~\bibnamefont{Vega}},
  \bibinfo{journal}{Living Rev. Rel.} \textbf{\bibinfo{volume}{14}},
  \bibinfo{pages}{7} (\bibinfo{year}{2011}), \eprint{1102.0529}.
  
  
  \bibitem[{\citenamefont{Blasco et~al.}(2015)\citenamefont{Blasco, Garay,
  Mart{\'\i}n-Benito, and Mart{\'\i}n-Mart{\'\i}nez}}]{blasco2015violation}
\bibinfo{author}{\bibfnamefont{A.}~\bibnamefont{Blasco}},
  \bibinfo{author}{\bibfnamefont{L.~J.} \bibnamefont{Garay}},
  \bibinfo{author}{\bibfnamefont{M.}~\bibnamefont{Mart{\'\i}n-Benito}},
  \bibnamefont{and}
  \bibinfo{author}{\bibfnamefont{E.}~\bibnamefont{Mart{\'\i}n-Mart{\'\i}nez}},
  \bibinfo{journal}{Physical review letters} \textbf{\bibinfo{volume}{114}},
  \bibinfo{pages}{141103} (\bibinfo{year}{2015}).


\bibitem[{\citenamefont{Birrell and Davies}(1984)}]{Birrell:Davies}
\bibinfo{author}{\bibfnamefont{N.}~\bibnamefont{Birrell}} \bibnamefont{and}
  \bibinfo{author}{\bibfnamefont{P.}~\bibnamefont{Davies}},
  \emph{\bibinfo{title}{Quantum Fields in Curved Space}}
  (\bibinfo{publisher}{Cambridge University Press},
  \bibinfo{address}{Cambridge}, \bibinfo{year}{1984}).
  


\bibitem[{\citenamefont{Hadamard}(1923)}]{Hadamard}
\bibinfo{author}{\bibfnamefont{J.}~\bibnamefont{Hadamard}},
  \emph{\bibinfo{title}{Lectures on Cauchy's Problem in Linear Partial
  Differential Equations}} (\bibinfo{publisher}{Dover Publications},
  \bibinfo{year}{1923}), ISBN \bibinfo{isbn}{978-0486495491}.

\bibitem[{\citenamefont{Garabedian}(1998)}]{Garabedian}
\bibinfo{author}{\bibfnamefont{P.~R.} \bibnamefont{Garabedian}},
  \emph{\bibinfo{title}{Partial Differential Equations}}
  (\bibinfo{publisher}{Chelsea Pub Co}, \bibinfo{address}{New York},
  \bibinfo{year}{1998}), ISBN \bibinfo{isbn}{9780821813775}.

  \bibitem[{\citenamefont{K-R-W}(1997)}]{KRW:1997}
\bibinfo{author}{\bibfnamefont{B.} \bibnamefont{{Kay}}},
\bibinfo{author}{\bibfnamefont{M.} \bibnamefont{Radzikowski}}
\bibnamefont{and}
\bibinfo{author}{\bibfnamefont{R.} \bibnamefont{Wald}},
    \bibinfo{journal}{Commun. Math. Phys.} \textbf{\bibinfo{volume}{183}},
  \bibinfo{pages}{533-556} (\bibinfo{year}{1997}).

\bibitem[{\citenamefont{Ikawa}(2000)}]{Ikawa}
\bibinfo{author}{\bibfnamefont{M.}~\bibnamefont{Ikawa}},
  \emph{\bibinfo{title}{Hyperbolic partial differential equations and wave
  phenomena. Iwanami series in modern mathematics. Translations of mathematical
  monographs}} (\bibinfo{publisher}{American Mathematical Soc.},
  \bibinfo{address}{Providence}, \bibinfo{year}{2000}), ISBN
  \bibinfo{isbn}{9780821810217}.

\bibitem[{\citenamefont{Ori}()}]{Ori1short}
\bibinfo{author}{\bibfnamefont{A.}~\bibnamefont{Ori}}, \bibinfo{note}{{private
  communication (2008) and report (2009)} available at
  \url{http://physics.technion.ac.il/~amos/acoustic.pdf}}.

\bibitem[{\citenamefont{Casals et~al.}(2009{\natexlab{a}})\citenamefont{Casals,
  Dolan, Ottewill, and Wardell}}]{CDOWa}
\bibinfo{author}{\bibfnamefont{M.}~\bibnamefont{Casals}},
  \bibinfo{author}{\bibfnamefont{S.}~\bibnamefont{Dolan}},
  \bibinfo{author}{\bibfnamefont{A.~C.} \bibnamefont{Ottewill}},
  \bibnamefont{and} \bibinfo{author}{\bibfnamefont{B.}~\bibnamefont{Wardell}},
  \bibinfo{journal}{Phys. Rev.} \textbf{\bibinfo{volume}{D79}},
  \bibinfo{pages}{124043} (\bibinfo{year}{2009}),
  \eprint{0903.0395}.

\bibitem[{\citenamefont{Dolan and Ottewill}(2011)}]{Dolan:2011fh}
\bibinfo{author}{\bibfnamefont{S.~R.} \bibnamefont{Dolan}} \bibnamefont{and}
  \bibinfo{author}{\bibfnamefont{A.~C.} \bibnamefont{Ottewill}},
  \bibinfo{journal}{Phys. Rev.} \textbf{\bibinfo{volume}{D84}},
  \bibinfo{pages}{104002} (\bibinfo{year}{2011}), \eprint{1106.4318}.


\bibitem[{\citenamefont{Harte and Drivas}(2012)}]{harte2012caustics}
\bibinfo{author}{\bibfnamefont{A.~I.} \bibnamefont{Harte}} \bibnamefont{and}
  \bibinfo{author}{\bibfnamefont{T.~D.} \bibnamefont{Drivas}},
  \bibinfo{journal}{Physical Review D} \textbf{\bibinfo{volume}{85}},
  \bibinfo{pages}{124039} (\bibinfo{year}{2012}).

\bibitem[{\citenamefont{Casals and Nolan}(2012)}]{Casals:2012px}
\bibinfo{author}{\bibfnamefont{M.}~\bibnamefont{Casals}} \bibnamefont{and}
  \bibinfo{author}{\bibfnamefont{B.~C.} \bibnamefont{Nolan}},
  \bibinfo{journal}{Phys.Rev.} \textbf{\bibinfo{volume}{D86}},
  \bibinfo{pages}{024038} (\bibinfo{year}{2012}), \eprint{1204.0407}.

\bibitem[{\citenamefont{$\rm{Zengino\breve{g}lu}$ and
  Galley}(2012)}]{Zenginoglu:2012xe}
\bibinfo{author}{\bibfnamefont{A.}~\bibnamefont{$\rm{Zengino\breve{g}lu}$}}
  \bibnamefont{and} \bibinfo{author}{\bibfnamefont{C.~R.}
  \bibnamefont{Galley}}, \bibinfo{journal}{Phys. Rev. D 86,}
  \textbf{\bibinfo{volume}{064030}} (\bibinfo{year}{2012}), \eprint{1206.1109}.

\bibitem[{\citenamefont{Yang et~al.}(2014)\citenamefont{Yang, Zhang, Zimmerman,
  and Chen}}]{Yang:2013shb}
\bibinfo{author}{\bibfnamefont{H.}~\bibnamefont{Yang}},
  \bibinfo{author}{\bibfnamefont{F.}~\bibnamefont{Zhang}},
  \bibinfo{author}{\bibfnamefont{A.}~\bibnamefont{Zimmerman}},
  \bibnamefont{and} \bibinfo{author}{\bibfnamefont{Y.}~\bibnamefont{Chen}},
  \bibinfo{journal}{Phys.Rev.} \textbf{\bibinfo{volume}{D89}},
  \bibinfo{pages}{064014} (\bibinfo{year}{2014}), \eprint{1311.3380}.

\bibitem[{\citenamefont{$\rm{Zengino\breve{g}lu}$}()}]{Video:CausticsSchw}
\bibinfo{author}{\bibfnamefont{A.}~\bibnamefont{$\rm{Zengino\breve{g}lu}$}},
  \bibinfo{howpublished}{\url{http://www.youtube.com/watch?v=Pe8sRjqtldQ}}.

\bibitem[{\citenamefont{Casals et~al.}(2013)\citenamefont{Casals, Dolan,
  Ottewill, and Wardell}}]{CDOW13}
\bibinfo{author}{\bibfnamefont{M.}~\bibnamefont{Casals}},
  \bibinfo{author}{\bibfnamefont{S.}~\bibnamefont{Dolan}},
  \bibinfo{author}{\bibfnamefont{A.~C.} \bibnamefont{Ottewill}},
  \bibnamefont{and} \bibinfo{author}{\bibfnamefont{B.}~\bibnamefont{Wardell}},
  \bibinfo{journal}{Phys. Rev. D} \textbf{\bibinfo{volume}{88}},
  \bibinfo{pages}{044022} (\bibinfo{year}{2013}),
  \urlprefix\url{http://link.aps.org/doi/10.1103/PhysRevD.88.044022}.

\bibitem[{\citenamefont{Wardell et~al.}(2014)\citenamefont{Wardell, Galley,
  Zengino\ifmmode~\breve{g}\else \u{g}\fi{}lu, Casals, Dolan, and
  Ottewill}}]{PhysRevD.89.084021}
\bibinfo{author}{\bibfnamefont{B.}~\bibnamefont{Wardell}},
  \bibinfo{author}{\bibfnamefont{C.~R.} \bibnamefont{Galley}},
  \bibinfo{author}{\bibfnamefont{A.}~\bibnamefont{Zengino\ifmmode~\breve{g}\else
  \u{g}\fi{}lu}}, \bibinfo{author}{\bibfnamefont{M.}~\bibnamefont{Casals}},
  \bibinfo{author}{\bibfnamefont{S.~R.} \bibnamefont{Dolan}}, \bibnamefont{and}
  \bibinfo{author}{\bibfnamefont{A.~C.} \bibnamefont{Ottewill}},
  \bibinfo{journal}{Phys. Rev. D} \textbf{\bibinfo{volume}{89}},
  \bibinfo{pages}{084021} (\bibinfo{year}{2014}),
  \urlprefix\url{http://link.aps.org/doi/10.1103/PhysRevD.89.084021}.

\bibitem[{\citenamefont{Casals and Nolan}(2015)}]{PhysRevD.92.104030}
\bibinfo{author}{\bibfnamefont{M.}~\bibnamefont{Casals}} \bibnamefont{and}
  \bibinfo{author}{\bibfnamefont{B.~C.} \bibnamefont{Nolan}},
  \bibinfo{journal}{Phys. Rev. D} \textbf{\bibinfo{volume}{92}},
  \bibinfo{pages}{104030} (\bibinfo{year}{2015}),
  \urlprefix\url{http://link.aps.org/doi/10.1103/PhysRevD.92.104030}.

\bibitem[{\citenamefont{Friedlander}(1975)}]{friedlander}
\bibinfo{author}{\bibfnamefont{F.~G.} \bibnamefont{Friedlander}},
  \emph{\bibinfo{title}{{The Wave Equation on a Curved spacetime}}}
  (\bibinfo{publisher}{Cambridge University Press},
  \bibinfo{address}{Cambridge}, \bibinfo{year}{1975}), ISBN
  \bibinfo{isbn}{978-0521205672}.

\bibitem[{\citenamefont{Casals and Ottewill}(2012)}]{PhysRevLett.109.111101}
\bibinfo{author}{\bibfnamefont{M.}~\bibnamefont{Casals}} \bibnamefont{and}
  \bibinfo{author}{\bibfnamefont{A.}~\bibnamefont{Ottewill}},
  \bibinfo{journal}{Phys. Rev. Lett.} \textbf{\bibinfo{volume}{109}},
  \bibinfo{pages}{111101} (\bibinfo{year}{2012}),
  \urlprefix\url{http://link.aps.org/doi/10.1103/PhysRevLett.109.111101}.

\bibitem[{\citenamefont{Zauderer}(1971)}]{zauderer1971modification}
\bibinfo{author}{\bibfnamefont{E.}~\bibnamefont{Zauderer}},
  \bibinfo{journal}{IMA Journal of Applied Mathematics}
  \textbf{\bibinfo{volume}{8}}, \bibinfo{pages}{8} (\bibinfo{year}{1971}).

\bibitem[{\citenamefont{DeWitt and Brehme}(1960)}]{DeWitt:1960}
\bibinfo{author}{\bibfnamefont{B.~S.} \bibnamefont{DeWitt}} \bibnamefont{and}
  \bibinfo{author}{\bibfnamefont{R.~W.} \bibnamefont{Brehme}},
  \bibinfo{journal}{Ann. Phys.} \textbf{\bibinfo{volume}{9}},
  \bibinfo{pages}{220} (\bibinfo{year}{1960}).

\bibitem[{\citenamefont{D\'ecanini and Folacci}(2008)}]{Decanini:Folacci:2008}
\bibinfo{author}{\bibfnamefont{Y.}~\bibnamefont{D\'ecanini}} \bibnamefont{and}
  \bibinfo{author}{\bibfnamefont{A.}~\bibnamefont{Folacci}},
  \bibinfo{journal}{Phys. Rev.} \textbf{\bibinfo{volume}{D78}},
  \bibinfo{pages}{044025} (\bibinfo{year}{2008}), \eprint{gr-qc/0512118}.

\bibitem[{\citenamefont{Ottewill and Wardell}(2011)}]{Ottewill:2009uj}
\bibinfo{author}{\bibfnamefont{A.~C.} \bibnamefont{Ottewill}} \bibnamefont{and}
  \bibinfo{author}{\bibfnamefont{B.}~\bibnamefont{Wardell}},
  \bibinfo{journal}{Phys.Rev.} \textbf{\bibinfo{volume}{D84}},
  \bibinfo{pages}{104039} (\bibinfo{year}{2011}), \eprint{0906.0005}.



\bibitem[{\citenamefont{Hardy}(1949)}]{Hardy}
\bibinfo{author}{\bibfnamefont{G.}~\bibnamefont{Hardy}},
  \emph{\bibinfo{title}{Divergent Series}} (\bibinfo{publisher}{Oxford
  Clarendon Press}, \bibinfo{year}{1949}), ISBN
  \bibinfo{isbn}{978-0-8218-2649-2}.

\bibitem[{\citenamefont{Casals et~al.}(2009{\natexlab{b}})\citenamefont{Casals,
  Dolan, Ottewill, and Wardell}}]{CDOWb}
\bibinfo{author}{\bibfnamefont{M.}~\bibnamefont{Casals}},
  \bibinfo{author}{\bibfnamefont{S.}~\bibnamefont{Dolan}},
  \bibinfo{author}{\bibfnamefont{A.~C.} \bibnamefont{Ottewill}},
  \bibnamefont{and} \bibinfo{author}{\bibfnamefont{B.}~\bibnamefont{Wardell}},
  \bibinfo{journal}{Phys. Rev.} \textbf{\bibinfo{volume}{D79}},
  \bibinfo{pages}{124044} (\bibinfo{year}{2009}{\natexlab{b}}),
  \eprint{0903.5319}.

\bibitem[{\citenamefont{Gradshteyn and Ryzhik}(2007)}]{GradRyz}
\bibinfo{author}{\bibfnamefont{I.}~\bibnamefont{Gradshteyn}} \bibnamefont{and}
  \bibinfo{author}{\bibfnamefont{I.}~\bibnamefont{Ryzhik}},
  \emph{\bibinfo{title}{Table of Integrals, Series, and Products}}
  (\bibinfo{publisher}{Academic Press}, \bibinfo{year}{2007}).

\bibitem[{\citenamefont{von Szeg\"o}(1934)}]{vonSzego01011934}
\bibinfo{author}{\bibfnamefont{G.}~\bibnamefont{von Szeg\"o}},
  \bibinfo{journal}{Proceedings of the London Mathematical Society}
  \textbf{\bibinfo{volume}{s2-36}}, \bibinfo{pages}{427}
  (\bibinfo{year}{1934}),
  \eprint{http://plms.oxfordjournals.org/content/s2-36/1/427.full.pdf+html},
  \urlprefix\url{http://plms.oxfordjournals.org/content/s2-36/1/427.short}.

\bibitem[{\citenamefont{Erdelyi et~al.}(1953)\citenamefont{Erdelyi, Magnus,
  Oberhettinger, and Tricomi}}]{Erdelyi:1953}
\bibinfo{author}{\bibfnamefont{A.}~\bibnamefont{Erdelyi}},
  \bibinfo{author}{\bibfnamefont{W.}~\bibnamefont{Magnus}},
  \bibinfo{author}{\bibfnamefont{F.}~\bibnamefont{Oberhettinger}},
  \bibnamefont{and} \bibinfo{author}{\bibfnamefont{F.}~\bibnamefont{Tricomi}},
  \emph{\bibinfo{title}{Higher Transcendental Functions}}
  (\bibinfo{publisher}{McGraw-Hill}, \bibinfo{address}{New York},
  \bibinfo{year}{1953}).

\bibitem[{{\relax DLMF}()}]{NIST:DLMF}
{\relax DLMF}, \emph{\bibinfo{title}{{NIST Digital Library of Mathematical
  Functions}}}, \bibinfo{howpublished}{http://dlmf.nist.gov/, Release 1.0.5 of
  2012-10-01}, \bibinfo{note}{online companion to \cite{Olver:2010:NHMF}},
  \urlprefix\url{http://dlmf.nist.gov/}.
  
    \bibitem[{\citenamefont{Aruquipa and Casals}(2022)}]{AruquipaCasals}
\bibinfo{author}{\bibfnamefont{D.~Q.~} \bibnamefont{Aruquipa}}
  \bibnamefont{and} \bibinfo{author}{\bibfnamefont{M.}~\bibnamefont{Casals}},
  \bibinfo{journal}{arXiv preprint arXiv:2205.13677}  (\bibinfo{year}{2022}).


  \bibitem[{\citenamefont{Wade}(2010)}]{wadeintroduction}
\bibinfo{author}{\bibfnamefont{W.~R.} \bibnamefont{Wade}},
  \emph{\bibinfo{title}{{An Introduction to Analysis (4th ed.)}}}
  (\bibinfo{publisher}{Pearson Prentice Hall},
  \bibinfo{year}{2010}).


  
  
  \bibitem[{\citenamefont{Buss and Casals}(2018)}]{BUSS2018168}
\bibinfo{author}{\bibfnamefont{C.} \bibnamefont{Buss}}
  \bibnamefont{and} \bibinfo{author}{\bibfnamefont{M.}~\bibnamefont{Casals}},
  \bibinfo{journal}{Phys. Lett.} \textbf{\bibinfo{volume}{B776}},
  \bibinfo{pages}{168} (\bibinfo{year}{2018}).

\bibitem[{\citenamefont{Carib{\'e} et al}(2023)}]{caribe2023lensing}
\bibinfo{author}{\bibfnamefont{J.~G.~A.~} \bibnamefont{Carib{\'e}}},
\bibinfo{author}{\bibfnamefont{R.~H.~} \bibnamefont{Jonsson}},
\bibinfo{author}{\bibfnamefont{M.~} \bibnamefont{Casals}},
\bibinfo{author}{\bibfnamefont{A.~} \bibnamefont{Kempf}},
  \bibnamefont{and} \bibinfo{author}{\bibfnamefont{E.}~\bibnamefont{Mart{\'\i}n-Mart{\'\i}nez}},
  \bibinfo{journal}{Phys. Rev.} \textbf{\bibinfo{volume}{D108}},
  \bibinfo{pages}{025016} (\bibinfo{year}{2023}).


\bibitem[{\citenamefont{Olver et~al.}(2010)\citenamefont{Olver, Lozier,
  Boisvert, and Clark}}]{Olver:2010:NHMF}
\bibinfo{editor}{\bibfnamefont{F.~W.~J.} \bibnamefont{Olver}},
  \bibinfo{editor}{\bibfnamefont{D.~W.} \bibnamefont{Lozier}},
  \bibinfo{editor}{\bibfnamefont{R.~F.} \bibnamefont{Boisvert}},
  \bibnamefont{and} \bibinfo{editor}{\bibfnamefont{C.~W.} \bibnamefont{Clark}},
  eds., \emph{\bibinfo{title}{{NIST Handbook of Mathematical Functions}}}
  (\bibinfo{publisher}{Cambridge University Press}, \bibinfo{address}{New York,
  NY}, \bibinfo{year}{2010}), \bibinfo{note}{print companion to
  \cite{NIST:DLMF}}.

\bibitem[{\citenamefont{Bertotti}(1959)}]{Bertotti}
\bibinfo{author}{\bibfnamefont{B.} \bibnamefont{Bertotti}},
  \bibinfo{journal}{Phys. Rev.} \textbf{\bibinfo{volume}{116}},
  \bibinfo{pages}{1331} (\bibinfo{year}{1959}).

\bibitem[{\citenamefont{Robinson}(1959)}]{robinson1959solution}
\bibinfo{author}{\bibfnamefont{I.} \bibnamefont{Robinson}},
  \bibinfo{journal}{Bull. Acad. Pol. Sci. Ser. Sci. Math. Astron. Phys} \textbf{\bibinfo{volume}{7}},
  \bibinfo{pages}{351} (\bibinfo{year}{1959}).


\bibitem[{\citenamefont{Allen and Jacobson}(1986)}]{allen1986vector}
\bibinfo{author}{\bibfnamefont{B.} \bibnamefont{Allen}}
  \bibnamefont{and} \bibinfo{author}{\bibfnamefont{T.}~\bibnamefont{Jacobson}},
  \bibinfo{journal}{Comm. Math. Phys.} \textbf{\bibinfo{volume}{103}},
  \bibinfo{pages}{669} (\bibinfo{year}{1986}).


\bibitem[{\citenamefont{Krantz and Parks}(2002)}]{krantz2002primer}
\bibinfo{author}{\bibfnamefont{S.~G.}~\bibnamefont{Krantz}}
\bibnamefont{and} \bibinfo{author}{\bibfnamefont{H.~R.}~\bibnamefont{Parks}}
  \emph{\bibinfo{title}{{A Primer of Real Analytic Functions}}}
  (\bibinfo{publisher}{Springer Science and Business Media},
  \bibinfo{year}{2002}).


  

\bibitem[{\citenamefont{Comtet}(1974)}]{comtet1974advanced}
\bibinfo{author}{\bibfnamefont{L.}~\bibnamefont{Comtet}}
  \emph{\bibinfo{title}{{Advanced Combinatorics: The art of finite and infinite expansions}}}
  (\bibinfo{publisher}{Springer Science and Business Media},
  \bibinfo{year}{1974}).
\end{thebibliography}

\bibliographystyle{apsrev}

\end{document}